%% file: main.tex
\documentclass[5p,twocolum,reversenotenum,authoryear]{myelsarticle}

\usepackage[dvipsnames]{xcolor}
\usepackage[colorlinks=True, citecolor=Blue, urlcolor=blue, pagebackref=True, breaklinks=false]{hyperref}
\renewcommand*{\backref}[1]{}
\renewcommand*{\backrefalt}[4]{[{%
    \ifcase #1 Not cited.%
          \or Cited on page~#2%
          \else Cited on pages #2%
    \fi%
   }].}

\usepackage{amsmath}

\usepackage{mathptmx}
\usepackage{txfonts}
\usepackage[T1]{fontenc}
\usepackage{graphicx}	
\usepackage{amssymb}	
\usepackage{newtxtext,newtxmath}
\usepackage{inputenc}
\usepackage{comment}
\usepackage{threeparttable}
\usepackage[switch]{lineno}
\usepackage{orcidlink}
\usepackage[dvipsnames]{xcolor}
\usepackage[export]{adjustbox}

\usepackage{natbib}
   
\journal{Astroparticle Physics}

\begin{document}

\newcommand{\HESS}{H.E.S.S.}
\newcommand{\degree}{$^{\circ}$ }

\newcommand{\change}[1]{#1}

\begin{frontmatter}




\input{authors_2025_WP}
\input{affiliations_WP}

\title{Science Prospects for the Southern Wide-field Gamma-ray Observatory: SWGO}


\begin{abstract}

Ground-based gamma-ray astronomy is now firmly established as a key observational approach to address 
critical topics at the frontiers of astroparticle physics and high-energy astrophysics. Whilst the field of TeV astronomy was once dominated by arrays of atmospheric Cherenkov Telescopes, ground-level particle detection has now been demonstrated to be an equally viable and strongly complementary approach. Ground-level particle detection provides continuous monitoring of the overhead sky, critical for the mapping of extended structures and capturing transient phenomena. As demonstrated by HAWC and LHAASO, the technique provides the best available sensitivity above a few tens of TeV and, for the first time, access to the PeV energy range. Despite the success of this approach, there is so far no major ground-level particle-based observatory with access to the Southern sky. H.E.S.S., located in Namibia, is the only major gamma-ray instrument in the Southern Hemisphere, and has shown the extraordinary richness of the inner galaxy in the TeV band, but is limited both in terms of field of view and energy reach. 

SWGO, the Southern Wide-field Gamma-ray Observatory, is an international effort to construct the first wide-field instrument in the south with deep sensitivity from 100s of GeV into the PeV domain. The project is now close to the end of its development phase and planning for construction of the array in Chile has begun. Here we describe the baseline design, expected sensitivity and resolution, and describe in detail the main scientific topics that will be addressed by this new facility. 

We show that SWGO will have a transformational impact on a wide range of topics from cosmic-ray acceleration and transport to the nature of dark matter. SWGO represents one of the key pieces of infrastructure for multi-messenger astronomy in the next decade, with strong scientific synergies with the nearby CTA Observatory.

Alongside its core scientific goals, SWGO is committed to work hand in hand with Atacamenean communities in the region, supporting local educational activities and seeking ways to integrate local knowledge and heritage into the observatory mission.


\end{abstract}



\end{frontmatter}

\clearpage
 \tableofcontents
\clearpage


\input{Introduction}

\input{Swgo}

\input{Design}

\input{Performance}

\input{Galactic}

\input{Transients}
\input{ParticleBSM}

\input{CosmicRay}

\input{Multimessenger}

\input{Outreach}

\input{Summary}


\section*{Acknowledgments}
\label{sec:Acknowledgements}

\input{acknowledgements.tex}


\def\aj{AJ}
\def\actaa{Acta Astron.}
\def\araa{ARA\&A}
\def\apj{ApJ}
\def\apjl{ApJ}
\def\apjs{ApJS}
\def\ao{Appl.~Opt.}
\def\apss{Ap\&SS}
\def\aap{A\&A}
\def\aapr{A\&A~Rev.}
\def\aaps{A\&AS}
\def\azh{AZh}
\def\baas{BAAS}
\def\bac{Bull. astr. Inst. Czechosl.}
\def\caa{Chinese Astron. Astrophys.}
\def\cjaa{Chinese J. Astron. Astrophys.}
\def\icarus{Icarus}
\def\jcap{J. Cosmology Astropart. Phys.}
\def\jrasc{JRASC}
\def\mnras{MNRAS}
\def\memras{MmRAS}
\def\na{New A}
\def\nar{New A Rev.}
\def\pasa{PASA}
\def\pra{Phys.~Rev.~A}
\def\prb{Phys.~Rev.~B}
\def\prc{Phys.~Rev.~C}
\def\prd{Phys.~Rev.~D}
\def\pre{Phys.~Rev.~E}
\def\prl{Phys.~Rev.~Lett.}
\def\pasp{PASP}
\def\pasj{PASJ}
\def\qjras{QJRAS}
\def\rmxaa{Rev. Mexicana Astron. Astrofis.}
\def\skytel{S\&T}
\def\solphys{Sol.~Phys.}
\def\sovast{Soviet~Ast.}
\def\ssr{Space~Sci.~Rev.}
\def\zap{ZAp}
\def\nat{Nature}
\def\iaucirc{IAU~Circ.}
\def\aplett{Astrophys.~Lett.}
\def\apspr{Astrophys.~Space~Phys.~Res.}
\def\bain{Bull.~Astron.~Inst.~Netherlands}
\def\fcp{Fund.~Cosmic~Phys.}
\def\gca{Geochim.~Cosmochim.~Acta}
\def\grl{Geophys.~Res.~Lett.}
\def\jcp{J.~Chem.~Phys.}
\def\jgr{J.~Geophys.~Res.}
\def\jqsrt{J.~Quant.~Spec.~Radiat.~Transf.}
\def\memsai{Mem.~Soc.~Astron.~Italiana}
\def\nphysa{Nucl.~Phys.~A}
\def\physrep{Phys.~Rep.}
\def\physscr{Phys.~Scr}
\def\planss{Planet.~Space~Sci.}
\def\procspie{Proc.~SPIE}
\def\eprint{arXiv}

\bibliographystyle{aa_url}
\bibliography{references}

\section*{Affiliations}
\footnotesize\itshape\elsaddress

\end{document}

%% file: authors_2025_WP.tex

\author[1,2]{P.~Abreu}
\author[3]{R.~Alfaro}
\author[4]{A.~Alfonso}
\author[5]{M.~Andrade}
\author[6]{E.~O.~Angüner}
\author[7]{E.~A.~Anita-Rangel}
\author[8]{O.~Aquines-Gutiérrez}
\author[9]{C.~Arcaro}
\author[10]{R.~Arceo}
\author[11]{J.~C.~Arteaga-Velázquez}
% ...

%% file: affiliations_WP.tex


\affiliation[1]{country={Laboratório de Instrumentação e Física Experimental de Partículas - LIP, Av. Prof. Gama Pinto, 2, 1649-003 Lisboa, Portugal}}
\affiliation[2]{country={Departamento de Física, Instituto Superior Técnico, Universidade de Lisboa, Av. Rovisco Pais 1, 1049-001 Lisboa, Portugal}}
\affiliation[3]{country={Instituto de Física, Universidad Nacional Autónoma de México, Circuito de la Investigación Científica, C.U., A. Postal 70-364, 04510 Cd. de México, México}}
\affiliation[4]{country={Universidad de La Serena, Chile}}
\affiliation[5]{country={Instituto de Física de São Carlos, Universidade de São Paulo, Av. Trabalhador São-carlense 400, São Carlos, Brasil}}
\affiliation[6]{country={TÜBİTAK Research Institute for Fundamental Sciences, 41470 Gebze, Turkey}}
\affiliation[7]{country={Instituto de Astronomía, Universidad Nacional Autónoma de México, Circuito Exterior, C.U., A. Postal 70-264, 04510 Cd. de México, México}}
\affiliation[8]{country={Departamento de Física y Matemáticas, Universidad de Monterrey, Av. Morones Prieto 4500, 66238, San Pedro Garza García NL, México}}
\affiliation[9]{country={INFN - Sezione di Padova, I-35131, Padova, Italy}}
\affiliation[10]{country={Facultad de Ciencias en Física y Matemáticas, Universidad Autónoma de Chiapas, C. P. 29050, Tuxtla Gutiérrez, Chiapas, México}}
\affiliation[11]{country={Instituto de Física y Matemáticas, Universidad Michoacana de San Nicolás de Hidalgo, Morelia, Michoacan, Mexico}}
\affiliation[12]{country={Temple University , Dept. of Physics, Philadelphia, PA, USA}}
\affiliation[13]{country={Institute of Physics of the Czech Academy of Sciences, Prague, Czech Republic}}
\affiliation[14]{country={Centro Brasileiro de Pesquisas Físicas (CBPF), Rua Dr. Xavier Sigaud 150, 22290-180 Rio de Janeiro, Brasil}}
\affiliation[15]{country={Universidade de São Paulo, Instituto de Astronomia, Geofísica e Ciências Atmosféricas, Departamento de Astronomia, Rua do Matão 1226, 05508-090 São Paulo, Brasil}}
\affiliation[16]{country={Friedrich-Alexander-Universität Erlangen-Nürnberg, Erlangen Centre for Astroparticle Physics, Nikolaus-Fiebiger-Str. 2, D 91058 Erlangen, Germany}}
\affiliation[17]{country={Università di Padova, I-35131, Padova, Italy}}
\affiliation[18]{country={Pontificia Universidad Católica del Perú, Av. Universitaria 1801, San Miguel, 15088, Lima, Perú}}
\affiliation[19]{country={Max-Planck-Institut für Kernphysik, Saupfercheckweg 1, 69117 Heidelberg, Germany}}
\affiliation[20]{country={Department of Physics and Astronomy, University of Rochester, Rochester, NY, USA}}
\affiliation[21]{country={Istituto Nazionale di Astrofisica - Istituto di Astrofisica e Planetologia Spaziali (INAF-IAPS), Via del Fosso del Cavaliere 100, 00133, Roma, Italy}}
\affiliation[22]{country={CCTVal, Universidad Tecnica Federico Santa Maria, Chile}}
\affiliation[23]{country={Astronomical Observatory Warsaw University, 00-478 Warsaw, Poland}}
\affiliation[24]{country={INFN, Roma Tor Vergata, Italy}}
\affiliation[25]{country={Department of Physics, University of Roma Tor Vergata, Viale della Ricerca Scientifica 1, I-00133 Roma, Italy}}

\affiliation[26]{country={Università degli Studi di Torino, I-10125 Torino, Italy}}
\affiliation[27]{country={INFN, Sezione di Torino, Torino, Italy}}
\affiliation[28]{country={Universidad Nacional de Ingeniería, Av. Túpac Amaru 210 - Rímac. Apartado 1301, Lima Perú}}
\affiliation[29]{country={Instituto de Astrofísica, Pontificia Universidad Católica de Chile (PUC), Av. Vicuña Mackenna 4860, 782-0436, Macul, Santiago, Chile}}
\affiliation[30]{country={Università di Napoli “Federico II”, Dipartimento di Fisica “Ettore Pancini”, Napoli, Italy}}
\affiliation[31]{country={INFN, Sezione di Napoli, Napoli, Italy}}
\affiliation[32]{country={Politecnico di Milano, Dipartimento di Scienze e Tecnologie Aerospaziali, Milano, Italy}}
\affiliation[33]{country={INFN, sezione di Milano, Milano, Italy}}
\affiliation[34]{country={Departamento de Física, Universidad de Santiago de Chile, Av. Bernardo O’Higgins 3363, 9170022 Santiago, Chile}}
\affiliation[34a]{country={University of Groningen, Wirdumerdijk 34, 8911 CE Leeuwarden, The Netherlands}}
\affiliation[35]{country={Universidade de São Paulo, São Paulo, Brasil}}
\affiliation[36]{country={Wisconsin IceCube Particle Astrophysics Center (WIPAC) \& Department of Physics}}
\affiliation[37]{country={Los Alamos National Laboratory, Los Alamos, NM, USA}}
\affiliation[38]{country={New Mexico Consortium, Los Alamos, NM, USA}}
\affiliation[39]{country={Departamento de Raios Cósmicos e Cronologia, Instituto de Física "Gleb Wataghin", Universidade Estadual de Campinas, C.P. 6165, 13083-970 Campinas, Brasil}}
\affiliation[40]{country={Centro Federal de Educação Tecnológica Celso Suckow da Fonseca (CEFET), Rio de Janeiro, Brasil}}
\affiliation[41]{country={Luleä University of Technology, Laboratorievägen 14, SE-971 87 LULEÅ, Sweden}}
\affiliation[42]{country={Universal Scientific Education and Research Network, Italy}}
\affiliation[43]{country={Technische Universität Dortmund, D-44221 Dortmund, Germany}}
\affiliation[44]{country={Department of Physics, University of Maryland, College Park, MD, USA}}
\affiliation[45]{country={Middle East Technical University, Northern Cyprus Campus, 99738 Kalkanli via Mersin 10, Turkey}}
\affiliation[46]{country={Department of Physics and Astronomy, Michigan State University, East Lansing, MI, USA}}
\affiliation[47]{country={Department of Physics, Washington University in St Louis, St Louis, MO 63130, USA}}
\affiliation[48]{country={Politecnico di Milano, Dipartimento di Energia, Milano, Italy}}
\affiliation[49]{country={Instituto Federal de Educação, Ciência e Tecnológia Fluminense 28030-130 Campos, RJ, Brazil}}
\affiliation[50]{country={Universidad Finis Terrae, Chile}}
\affiliation[51]{country={Universidad Nacional de San Antonio Abad del Cusco, Av. de la Cultura, Nro. 733, Cusco - Perú}}
\affiliation[52]{country={Université Paris Cité, CNRS, AstroParticule et Cosmologie, Paris, France}}
\affiliation[53]{country={Dept. of Engineering Physics, Tsinghua University, 1 Tsinghua Yuan, Haidian District, Beijing 100084, China}}
\affiliation[54]{country={Scuola Superiore Meridionale, Napoli, Italy}}
\affiliation[55]{country={Universidade Estadual de Feira de Santana, Feira de Santana, Brazil}}
\affiliation[56]{country={Department of Physics, Sungkyunkwan University, Suwon, South Korea}}
\affiliation[57]{country={IFLP, Universidad Nacional de La Plata and CONICET, La Plata, Argentina}}
\affiliation[58]{country={Michigan Technological University, Houghton, Michigan, 49931, USA}}
\affiliation[59]{country={Department of Physics, Durham University, South Rd, Durham DH1 3LE, United Kingdom}}
\affiliation[60]{country={University of Rijeka, Faculty of Physics, 51000 Rijeka, Croatia}}
\affiliation[61]{country={Univ. Bordeaux, CNRS, LP2i Bordeaux, UMR 5797, F-33170 Gradignan, France}}
\affiliation[62]{country={French-Chilean Laboratory for Astronomy, IRL 3386, CNRS and Universidad de Chile, Casilla 36-D, Santiago, Chile}}
\affiliation[63]{country={Departamento de Física, Universidad Católica del Norte, Avenida Angamos 0610, Casilla 1280, 1270709 Antofagasta, Chile.}}
\affiliation[64]{country={Université Paris Cité, CNRS, Astroparticule et Cosmologie, F-75013 Paris, France}}
\affiliation[65]{country={Department of Physics and Astronomy, University of Utah, Salt Lake City, UT, USA}}
\affiliation[66]{country={INAF - Osservatorio Astronomico di Cagliari, Via della Scienza 5, 09047 Selargius, Italy}}
\affiliation[67]{country={University of Leicester, School of Physics and Astronomy, University Road, Leicester, LE1 7RH, United Kingdom}}
\affiliation[68]{country={Facultad de Ciencias Exactas, Universidad Nacional de Salta, Avda. Bolivia 5150, A4408FVY, Salta, Argentina}}
\affiliation[69]{country={University of Seoul, Seoul, Rep. of Korea}}
\affiliation[70]{country={Dipartimento di Fisica, Università degli Studi di Trieste, via Valerio 2, I-34127 Trieste, Italy}}
\affiliation[71]{country={INFN - Sezione di Trieste, via Valerio 2, I-34127 Trieste, Italy}}
\affiliation[72]{country={Instituto de Tecnologías en Detección y Astropartículas (CNEA, CONICET, UNSAM), Buenos Aires, Argentina}}
\affiliation[73]{country={Comisión Nacional de Investigación y Desarrollo Aeroespacial, Calle Luis Felipe Villarán 1069, 15046, Lima, Perú}}
\affiliation[74]{country={Universidad Nacional de Moquegua, Calle Ancash SN, 18001, Moquegua, Perú}}
\affiliation[74a]{country={Università degli Studi di Milano, Italy}}
\affiliation[75]{country={Instituto Nazionale Di Astrofisica (INAF), Torino, Italy}}
\affiliation[76]{country={Instituto de Ciencias Nucleares, Universidad Nacional Autónoma de México, Circuito Exterior, C.U., A. Postal 70-543, 04510 Cd. de México, México}}
\affiliation[77]{country={Stanford University, USA}}
\affiliation[77a]{country={Istituto Nazionale di Astrofisica (INAF), Padova, Italy}}
\affiliation[78]{country={Laboratori Nazionali di Frascati INFN, Italy}}
\affiliation[79]{country={Universidade Federal da Bahia, Brasil}}
\affiliation[80]{country={Universidad Nacional de San Agustin de Arequipa, Santa Catalina Nro. 117. Arequipa, Perú}}
\affiliation[81]{country={Instituto de Astronomía y Física del Espacio (IAFE (CONICET-UBA)), Ciudad Universitaria, CABA, Argentina}}
\affiliation[82]{country={Centro de Astronomía (CITEVA), Universidad de Antofagasta, Chile}}
\affiliation[83]{country={Università degli Studi di Bari Aldo Moro, Italy}}
\affiliation[84]{country={Centro di Ateneo di Studi e Attività Spaziali ”Giuseppe Colombo”, Via Venezia 15, I-35131 Padova, Italy.}}
\affiliation[85]{country={Departamento de Física, Facultad de Ciencias Básicas, Universidad Metropolitana de Ciencias de la Educación, Av. José Pedro Alessandri 774, Ñuñoa, Chile}}
\affiliation[85a]{country={Centro de Desarrollo de Investigaci\'on UMCE, Universidad Metropolitana de Ciencias de la Educaci\'on, Santiago, Chile}}
\affiliation[86]{country={Departamento de Física Teórica, Universidad de Zaragoza, Zaragoza 50009, Spain}}
\affiliation[87]{country={Universidad del Bío-Bío, Chile}}
\affiliation[88]{country={Universidad de Chile, Chile}}
\affiliation[89]{country={Université Savoie Mont Blanc, CNRS, Laboratoire d’Annecy de Physique des Particules—IN2P3, F-74000 Annecy, France}}
\affiliation[90]{country={Department of Physics and Astronomy, University of Alabama, Tuscaloosa, Alabama, 35487, USA}}
\affiliation[91]{country={IMAPP, Radboud University Nijmegen, Nijmegen, The Netherlands}}
\affiliation[92]{country={IRFU, CEA, Université Paris-Saclay, F-91191 Gif-sur-Yvette, France}}
\affiliation[93]{country={Unidade Acadêmica de Física, Universidade Federal de Campina Grande, Av. Aprígio Veloso 882, CY2, 58.429-900 Campina Grande, Brasil}}
\affiliation[94]{country={Bartol Research Institute, Department of Physics and Astronomy, University of Delaware, Newark, DE, USA}}
\affiliation[95]{country={Instituto Nacional de Astrofísica, Óptica y Electrónica, Puebla, Mexico}}
\affiliation[95a]{country={Facultad de Ciencias Sociales, Artes y Humanidades, Universidad de Antofagasta, Chile}}
\affiliation[95b]{country={Department of Physics, Missouri University of Science and Technology, Rolla, MO, USA }}
\affiliation[96]{country={RWTH Aachen University, Templergraben 55, 52062 Aachen, Germany}}
\affiliation[97]{country={Center for Astrophysics and Cosmology (CAC), University of Nova Gorica, Nova Gorica, Slovenia}}

%% file: Introduction.tex
\section{Introduction} 

The field of ground-based gamma-ray astronomy has grown rapidly since it was established in 1989~\citep{1989ApJ...342..379W}. Addressing themes in particle/astroparticle physics as well as high-energy astrophysics, astronomy with photon energies above about  $2\,\times\,10^{10}$~eV has emerged as a powerful tool within multi-messenger and multi-wavelength (MM/MWL) astrophysics.

Two complementary ground-based approaches exist: imaging atmospheric Cherenkov Telescope (IACT) arrays and ground-level particle detectors. The IACTs dominated the early development of the field, but ground-level particle detectors are now firmly established in the Northern Hemisphere. IACTs provide precision and access to the lowest energies, with excellent background rejection power at $\sim$TeV energies. However, they are pointed instruments with a modest field of view (FoV) that operate only during darkness. Ground-level particle detection offers a view of the whole overhead sky all of the time, albeit with more modest resolution and without access to the very lowest energies. 

The first developments for gamma-ray astronomy at \change{very} high altitude were associated to the Tibet Air Shower array \citep{Amenomori92} with the ARGO array implemented later at the same site~\citep{Aielli2012}. The pioneering Milagro detector first demonstrated the potential of the ground-particle approach with the first source detections using water as the target medium for producing Cherenkov radiation~\citep{MILAGRO}. It was clear though that a higher altitude detector was needed to bring the measurement level closer the maximum development depth of air-showers of the relevant energies \change{($\sim$400 g/cm$^{2}$ at 1 TeV)}.

In 2015, the HAWC array, a direct successor to Milagro, was completed at 4.1~km altitude in Mexico~\citep{2023NIMPA105268253A}. The sensitivity of ground-particle detectors reached a critical level in the HAWC era, where the complementarity of these instruments to the more established IACTs became very clear. According to TeVCat~\citep{tevcat}, HAWC discovered \change{many} new sources and added significantly to our knowledge of many more. Highlights of HAWC science include constraining the transport of cosmic rays (CRs) in the interstellar medium using nearby pulsars~\citep{Geminga_HAWC}, and establishing jetted microquasars as accelerators up to ultra-high energies~\citep[UHE, ][]{HAWCSS433}.

In 2021, the LHAASO facility was completed in Daocheng County, Sichuan Province, China, at 4.4 km above sea level. LHAASO contains two ground-level particle detection elements: the Water Cherenkov Detector Array (WCDA), which is a dense array $\sim$3.5 times larger than the dense central HAWC array, and the square-kilometer sparse array (KM2A). KM2A consists of buried muon detectors with a 4\% fill factor and surface scintillator detectors covering 0.5\% of the array footprint. With excellent background rejection power provided by its muon-tagging capabilities and its large area, KM2A has extended gamma-ray astronomy in to the PeV domain~\citep{lhaasocrab}.

Observations with H.E.S.S., an array of five IACTs in Namibia, have revealed the rich populations of particle accelerators in the inner galaxy~\citep{HESS_GPS}. The Galactic Plane Survey of H.E.S.S. and deep observations of the Galactic Center do, however, provide only limited sensitivity beyond $\sim$50~TeV, \change{with only the brightest sources detectable at these energies, see e.g. ~\citep{HESSCrab2024}}. The CTA Observatory (CTAO) is planned to include a southern site in Chile and will dramatically improve the survey depth~\citep{2024JCAP...10..081A} and resolution~\citep[see, e.g., ][]{2024APh...16303008S}. However, no major wide-field very-high-energy (VHE)/UHE instrument exists so far in the Southern Hemisphere, although developments are underway in Bolivia towards ALPACA, a UHE-focused instrument~\citep{2024arXiv241214550A}.

A wide-field observatory located in the Southern Hemisphere, with a broad energy range and a sensitivity comparable to that of LHAASO would complement CTAO, in particular in the areas of transient phenomena, large-scale emission, and UHE sources. Here we describe the expected scientific performance of a new ground-level particle detecting gamma-ray observatory planned for construction in the Southern Hemisphere and based on water Cherenkov detectors (WCDs): SWGO.

Section~\ref{sec:project} introduces the project, Section~\ref{sec:design} summarizes the site and the design, and Section~\ref{sec:performance} the expected performance. Subsequent sections focus on the expected scientific impact of SWGO.

%% file: Swgo.tex
\section{The SWGO Project} 
\label{sec:project}

The Southern Wide-field Gamma-ray Observatory (SWGO) Collaboration was created in 2019 as a convergence of several earlier concepts and efforts including LATTES~\citep{2018APh....99...34A} and the SGSO Alliance~\citep{2019arXiv190208429A}. The central concept for the observatory was defined as:
\begin{itemize} 
\item A gamma-ray observatory based on ground-level particle detection, with close to 100\% duty cycle and order steradian FoV.
\item Located in South America at a latitude of $-30$ to $-10$ degrees.
\item Placed at an altitude of 4.4~km or higher.
\item Covering an energy range from 100s of GeV up to the PeV scale.
\item Based primarily on WCD units.
\item Utilizing a high-fill-factor core detector with area considerably larger than HAWC and significantly better sensitivity, alongside a low density outer array.
\item Modular and scalable, with the possibility of extensions and/or enhancements.
\end{itemize}

Water Cherenkov Detector units are attractive in terms of cost per unit area of particle detection and were the approach of choice for Milagro, HAWC, and 
components of the Tibet and LHAASO arrays. 

The SWGO Collaboration has now grown to include over 100 institutes in 17 countries on four continents, and is close to completion of its {\it R\&D Phase}, 
with milestones defined at the beginning of the project reached over the last few years (see Table~\ref{tab:milestones}). Notable amongst these milestones was the establishment of a baseline approach based on extensive simulations covering a wide phase space \citep[M6: see][and Section~\ref{sec:design}]{2023arXiv230904577C}, as well as design development and cost estimation work. Another critical milestone was the identification of a preferred site for SWGO (M7), based on criteria of scientific performance (including altitude and latitude effects), cost (including construction and a nominal 5 years of operations), and risks. This milestone was reached in July 2024 (see Section~\ref{sec:site}) after a comprehensive process that considered 10 sites in four countries~\citep{Santander_2023}.

\begin{table}[]
    \centering
    \begin{tabular}{|cp{0.61\linewidth}c|} \hline
     & \bf Milestone & \bf Completed \\ \hline
        M1  &  R\&D Phase Plan Established & Q1 2020\\
    M2	& Science Benchmarks Defined & Q2 2020\\
M3	& Reference Configuration \& Options Defined & Q4 2020 \\
M4	& Site Shortlist Complete & Q3 2022 \\
M5	& Candidate Configurations Defined & Q1 2022 \\
M6	& Performance of Candidate Configurations Evaluated & Q1 2024 \\
M7	& Preferred Site Identified & Q3 2024 \\
M8	& Design Finalised & - \\
M9	& Construction \& Operation Proposal Complete  & - \\\hline
    \end{tabular}
    \caption{Milestones of the SWGO R\&D Phase, together with completion dates for Milestones already completed.}
    \label{tab:milestones}
\end{table}

The optimization process has been guided by science benchmarks associated with key science cases, established early in the project (M2) and allowing us to objectively optimize the instrument design based on our scientific objectives. The benchmarks cover the topics of transient sources (specifically GRBs, see Section~\ref{sec:GRB}), Galactic PeV particle accelerators (see Section~\ref{sec:PeVatron}), extended sources (with the example of PWNe and associated halos, see Section~\ref{sec:PWN}), large-scale diffuse emission (see Section~\ref{sec:diffuse}), dark matter detection (see Section~\ref{sec:DM}), and the mass-resolved anisotropy of CRs (see Section~\ref{sec:CR}).

\begin{figure*}[htbp]
    \centering
\includegraphics[trim={15cm 16cm 12cm 18cm},clip, width=\textwidth]{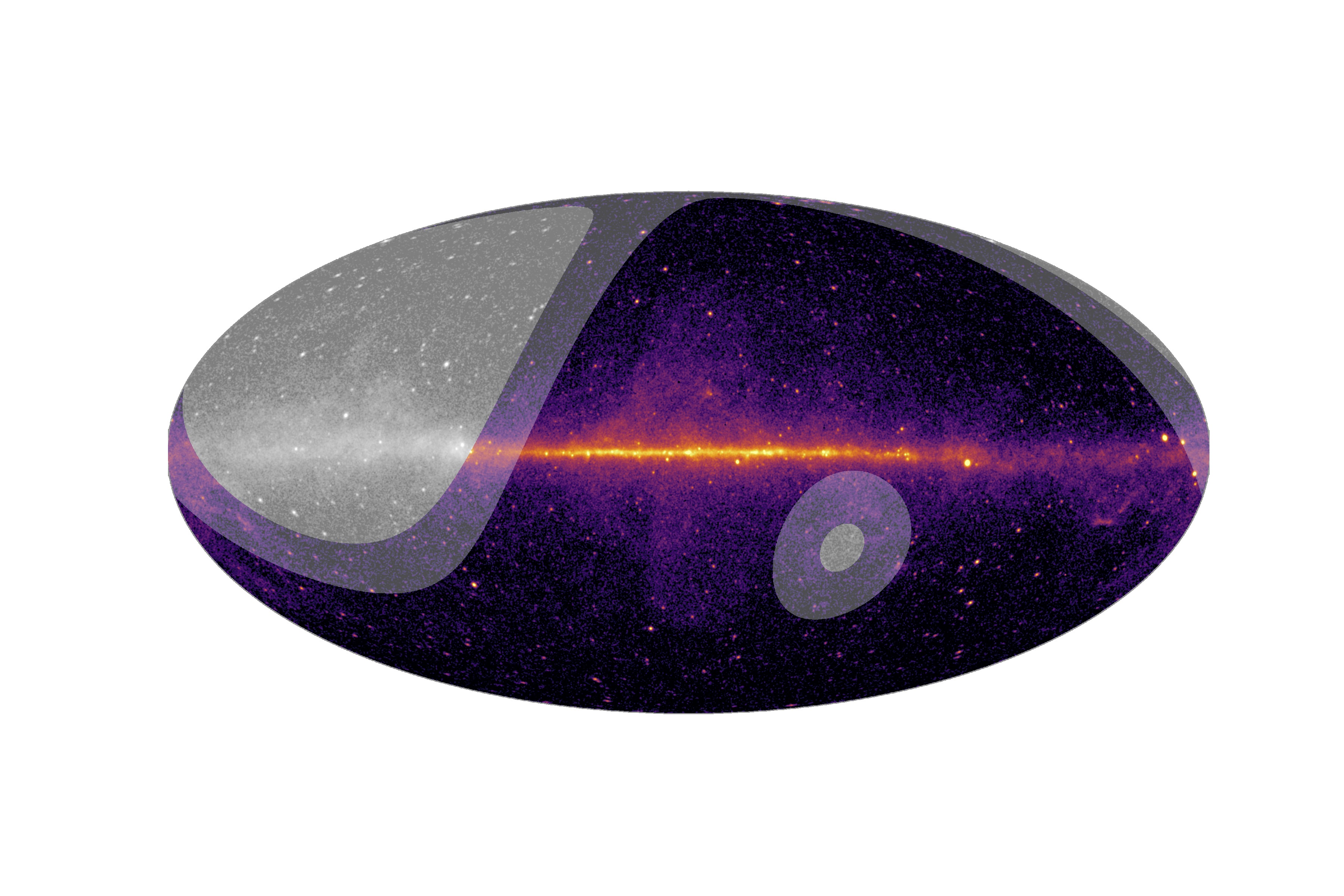}
    \caption{SWGO visibility map showing the sky observable to SWGO above 45$^{\circ}$ (full color) and 60$^{\circ}$ (color shaded) together with the \emph{Fermi}-LAT all-sky image from 3 to 300 GeV, available at \url{https://doi.org/10.26093/cds/aladin/3mva-x6}. Regions invisible to SWGO are shown in greyscale.}
    \label{fig:swgo_visibility}
\end{figure*}

Figure~\ref{fig:swgo_visibility} shows the sky visible to SWGO, including the Fermi Bubbles and the whole inner Galaxy. The Galactic Center passes directly overhead at the SWGO Site.

SWGO construction is anticipated to take place in stages. After a modest scale {\it Pathfinder}, \change{an option for the first science-capable phase is SWGO-A, focused} on the low-to-intermediate energy range (see Section~\ref{sec:performance}).

%% file: Design.tex
\section{Baseline Design} 
\label{sec:design}

The SWGO Collaboration is now approaching Milestone 8 --- the definition of the Baseline Design. The array site has been chosen, the overall layout has been defined,  and the detector unit design is fixed, at least for the inner part of the array. In this section, we give a brief overview of these aspects of the design, discuss areas where options still exist, and note the existence of already funded and possible future extensions to the project.

\subsection{The SWGO Site}
\label{sec:site}

\begin{figure*}
    \centering
    \includegraphics[width=1.0\linewidth]{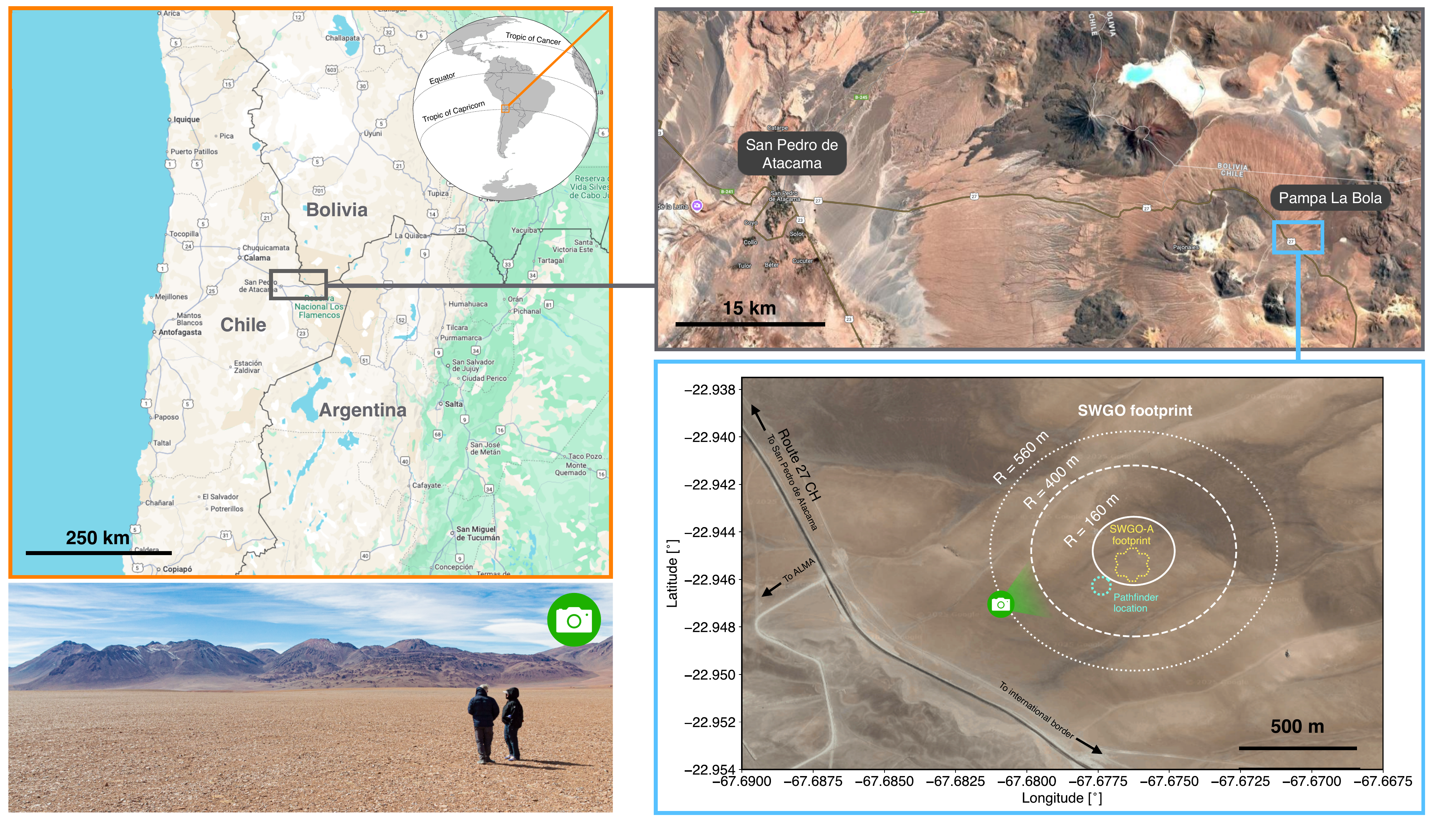}
    \caption{Location of the SWGO site in Pampa La Bola, Atacama Astronomical Park, Chile. The photo in the bottom left shows a view of the site. Map data: Google 2025. Satellite imagery: Landsat / Copernicus 2025}
    \label{fig:site_PL}
\end{figure*}

The SWGO primary site --- Pampa La Bola --- is located in the Calama commune in northern Chile at an altitude of 4,770 meters. The site lies on an extensive plateau within the Atacama Astronomical Park \citep[AAP;][]{10.1117/12.2561044} in close proximity to several major astronomical facilities, including the Atacama Large Millimeter/submillimeter Array (ALMA). The site has excellent access, being only a 45 minute drive from the town of San Pedro de Atacama, which lies at 2.4 km altitude. Figure~\ref{fig:site_PL} illustrates the location of the site within Chile, the local region, and the array layout (see Section~\ref{sec:layout}), as well as a photograph showing the excellent ground conditions. The site lies close to the international road Route 27, with the center of the array located at latitude 22.94480$^{\circ}$S, longitude 67.67625$^{\circ}$W. The site was selected from three shortlisted candidate sites in Peru, Argentina, and Chile in 2024, following an extensive search campaign~\citep{Santander_2023}. Key advantages of the Chilean site include its altitude, flatness/ease of construction, and the presence of a community of other observatories within the AAP. 
The back-up SWGO site is located close to the town of Imata, in the Arequipa district of Peru. 

The first activities planned at the site include building the Pathfinder (SWGO-PF), aimed to demonstrate the validity of the chosen technologies and assembly and operation procedures. The SWGO-PF will consist of 4--6 WCDs plus associated control and readout, water storage and purification, and power systems. The main purpose of the SWGO-PF is to gain experience and capture lessons learned for the installation of the tanks and for the water transport and quality, as well as to demonstrate system integration at high altitude and to validate our simulations.

\change{Beyond the Pathfinder the project moves towards implementation of a science-capable subset of the final array. One option for this is }
SWGO-A, a first component of the high-fill-factor inner array zone ($R<160$~m, see Figure~\ref{fig:site_PL}).

\subsection{SWGO Array Layout \& Simulation}
\label{sec:layout}

\begin{figure}
    \centering
    \includegraphics[width=0.95\linewidth]{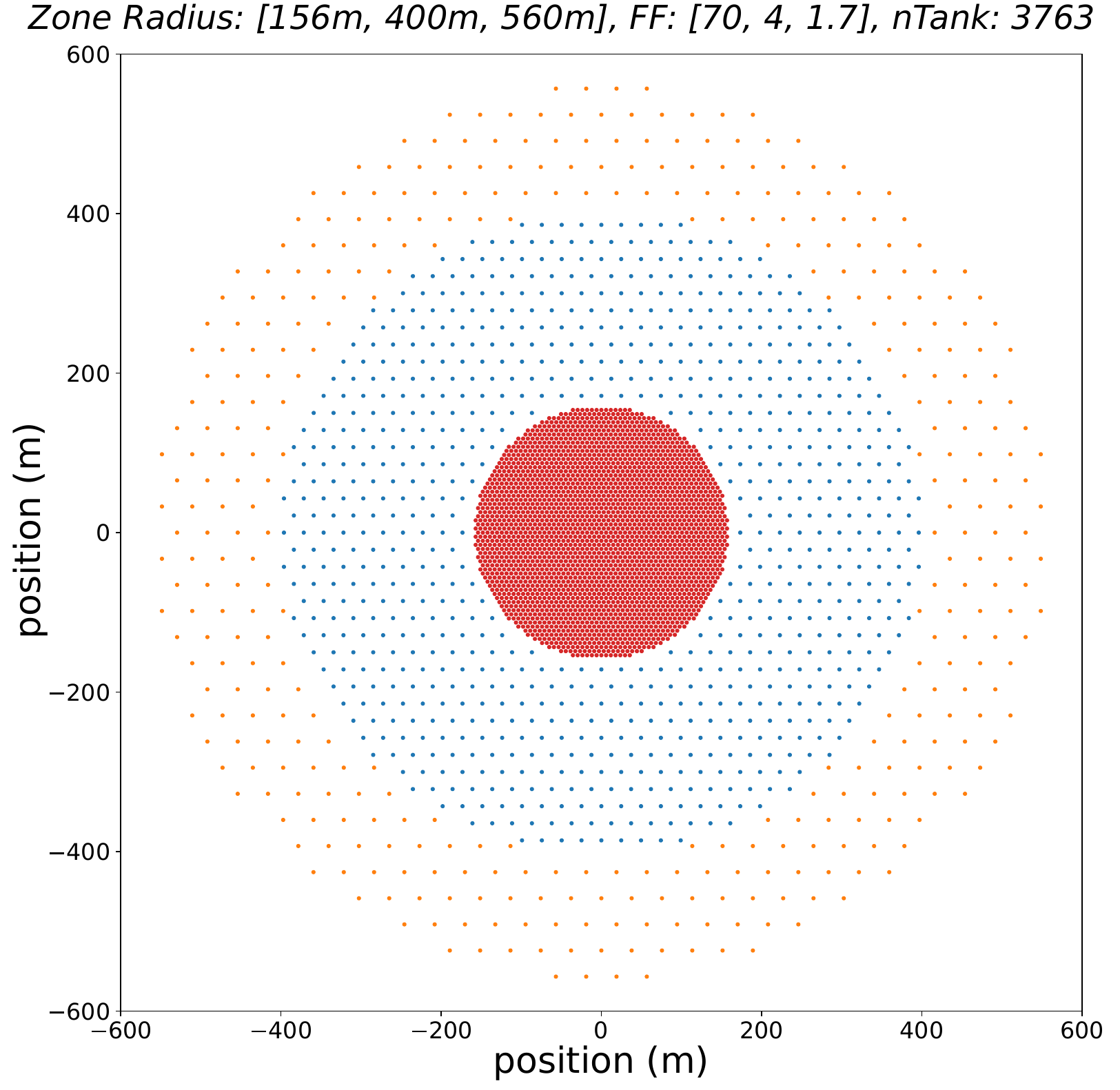}
    \caption{Design of SWGO featuring three zones with different fill factors (inner array in red, with the outer array in blue and orange) used to estimate its expected performance.}
    \label{fig:swgo_layout_D8}
\end{figure}

The SWGO detector is made up of an array of independent WCD units. These units are distributed in a 3-zone configuration (see Figure~\ref{fig:site_PL}).
The central zone consists of closely packed tanks with only enough separation to accommodate construction and maintenance. The fill factor of the central zone is 70\%.
This high-density region provides the instrument's sensitivity from hundreds of GeV to tens of TeV gamma rays and is fundamental to the sensitivity for the detection of transients and distant gamma-ray emitters.
Surrounding the central zone are two outer zones of gradually decreasing fill factor: 4\% for the intermediate zone and 1.7\% for the outermost zone.
These outer zones extend the total instrumented area to $\approx\,$1 km$^2$ and provide detection capability up to the PeV range.
In total, the configuration contains 3,763 WCDs.
The design is shown in Figure~\ref{fig:swgo_layout_D8}, where red markers denote the position of WCDs in the inner array.
The WCDs of the outer array are depicted as blue and orange markers.

\change{Whilst the inner array WCD design has been fully baselined and simulated, the final configuration for the outer part of the array is still being consolidated (see Section~\ref{sec:outer}).}

\subsection{Water Cherenkov Detector Unit Design}

\begin{figure*}
    \centering
      \includegraphics[width=0.32\linewidth, valign=b]{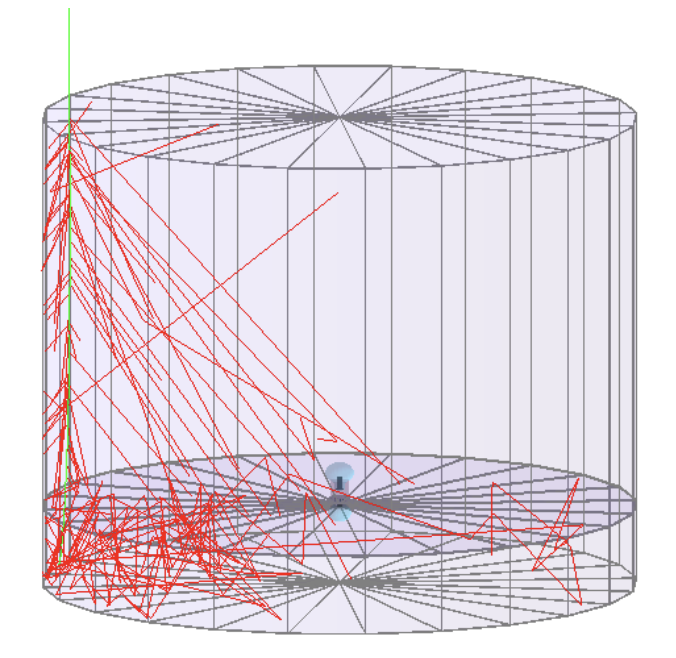}
    \includegraphics[trim=0cm 4cm 0cm 0cm,clip, width=0.32\linewidth, valign=b]{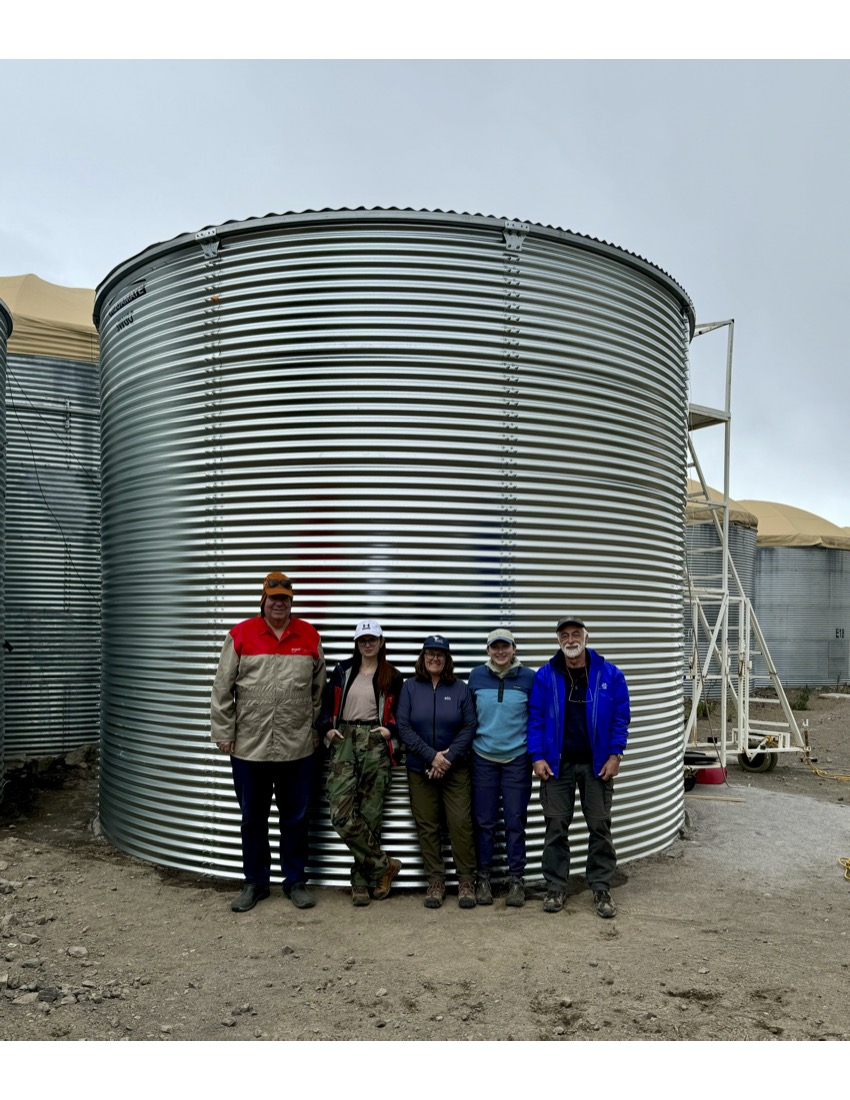}
  \includegraphics[trim=24cm 0cm 16cm 0cm, clip, width=0.35\linewidth, valign=b]{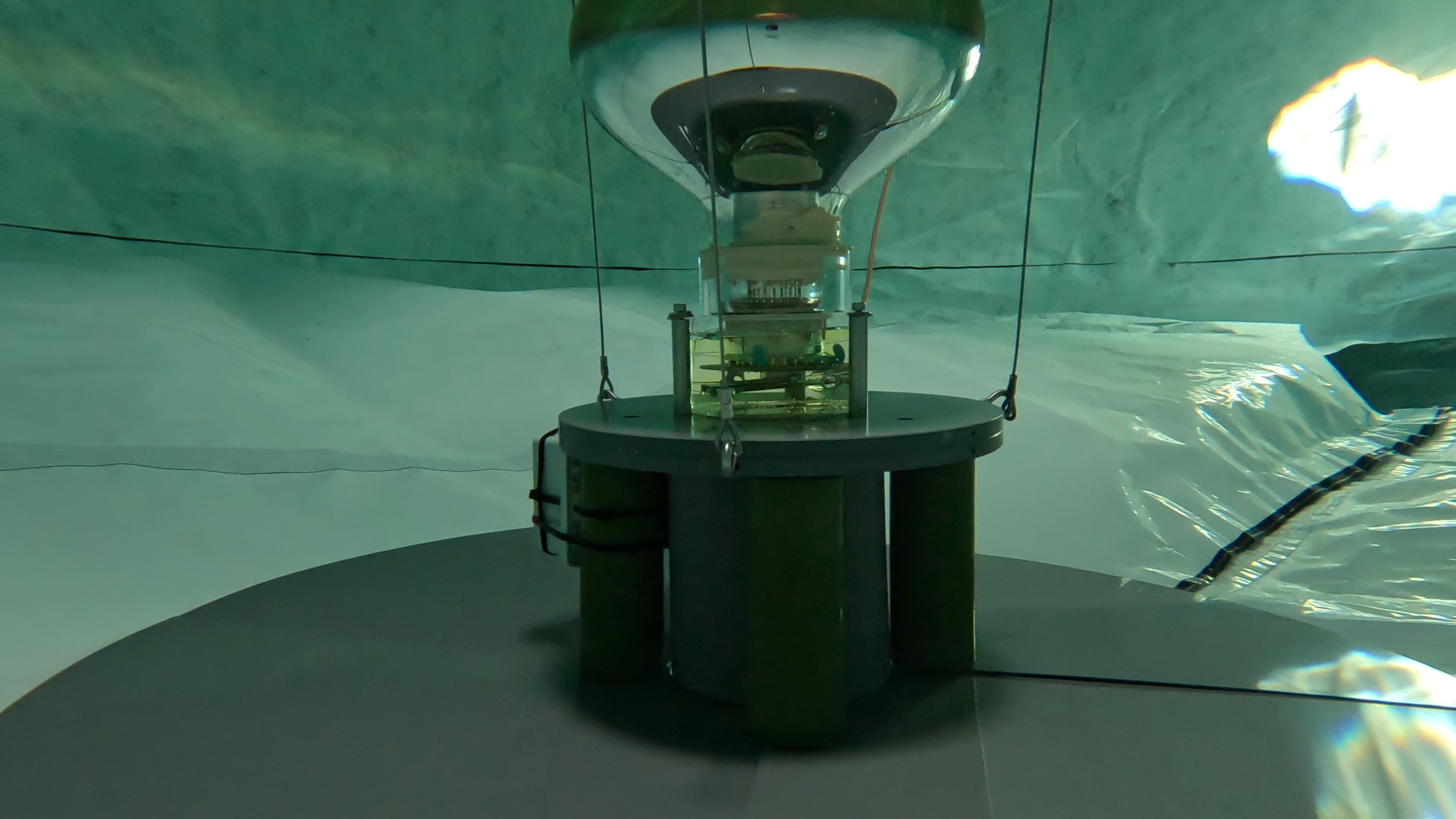}
    \caption{\textbf{Left:} simulation of a vertical muon (green) producing Cherenkov photons (red) in the dual-layer 5.2 $\times$ 4.0 m water volume of a SWGO WCD, showing reflections in the lower chamber. The two photo-multiplier tubes (PMTs) are shown in light blue. \textbf{Center:} photograph of the prototype tank of these dimensions assembled at the HAWC site for testing, together with some U.S. members of the SWGO Collaboration who aided with its installation. \textbf{Right:} Photograph of a deployed dual-PMT unit in a dual-layer 3.6~m bladder at the MPIK in Germany (as seen from the upper chamber), showing the 10'' PMT facing upwards.  }
    \label{fig:SWGOTankAlt}
\end{figure*}

The baseline for the inner array WCD unit is a cylindrical steel tank, 5.2~m in diameter and 4.1~m in height, assembled on site, with a bladder to contain the water. The bladder has a total height of 4.0~m and comprises two volumes, with the lower section measuring 80 cm in height, and a Tyvek surface layer for high reflectivity~\citep[see][for a study on the dual-layer approach and the impact of liner reflectivity]{Kunwar:2022iav}. We note that MILAGRO made use of two layers of photosensors and that a dual-layer design was also studied for the Pierre Auger Observatory~\citep{dual_layer_auger}. 
A photograph of an assembled prototype 5.2~m tank at the HAWC observatory can be seen in Figure~\ref{fig:SWGOTankAlt}, together with a photograph taken inside a deployed unit and a simulated muon passing through the WCD. The lower chamber provides the unit with muon-tagging capability to aid in rejection of the charged cosmic-ray background.

Once filled with filtered water, a double-PMT unit is lowered into a double-chamber bladder to position a downward-looking 8--10" PMT in the lower section and a 10" PMT looking up from the bottom of the upper section. We are currently evaluating two options for high-voltage (HV) supplies to the PMT: active (locally generated HV, similar to the active base used for IceCube) and passive (signal and HV carried on the same cable).


Signals from the PMTs in each WCD are routed to nearby {\it Field Nodes} (FNs), which serve as digitization and control units. Each FN nominally serves 55~WCDs, corresponding to 110~PMT channels. Housed in standalone, environmentally protected cabinets, the FNs include heating, cooling, insulation, and a programmable logic controller (PLC) for local control and monitoring. PMT signals are digitized using FlashCam electronics, a system developed for the CTAO~\citep{FlashCam}, which \change{samples at 250~MHz} with deep waveform buffers. Digitization takes place following a first-level trigger, based on the signal from any PMT exceeding a fraction of a photoelectron, thereby requiring a low level of electronic noise. 

Depending on the HV configuration, signal routing differs. In the case of passive HV distribution, a single coaxial cable carries both the PMT signal and HV from the WCD to the FN.
Inside the FN, dedicated PhantomHV boards~\citep{PhantomHV} generate the HV and separate it from the signal path. The resulting signal is routed to the FlashCam digitizers. The PhantomHV boards include a time-domain reflectometry system to measure the signal propagation delay along the coaxial cable to each PMT, allowing for per-channel cable transit time correction. This is used to compensate for temperature-dependent variations in cable delay and can also be used to detect faults or damage in the cable. The system achieves sub-nanosecond precision, supporting accurate event timing across the array, as required for precise direction reconstruction. 

In configurations with locally generated HV at the WCD, Cat6 \change{(Ethernet type)} cables carry the PMT signal, low-voltage power, and slow-control communication directly between the WCD and FN. In this case, electronics within the FN split the signal and auxiliary lines, with the signal again routed to FlashCam for digitization. 

Each FN includes a redundant White Rabbit (WR) timing system, which provides sub-nanosecond synchronization across the array. WR combines IEEE 1588 Precision Time Protocol 
with Synchronous Ethernet and phase-tracking techniques to distribute a common time and frequency base over dedicated fiber links. Each FN is connected to a central location via fiber for timing, control, and raw data readout. Power is supplied via a single-phase AC feed.

Centrally, the approximately 50~FNs needed for the inner array are connected to a computing rack for control and data acquisition. Several WR switches, with one acting as a master node, are connected to a GPS receiver to provide time synchronization across the array. An array-level trigger is formed in software using the digitized waveforms in real time. 

All elements of the inner array baseline have been prototyped, and the simulated performance presented below is based on this design. For the outer array, alternative solutions have also been prototyped; see Section~\ref{sec:outer} below.

\subsection{SWGO-A}
\label{sec:swgoa}

\begin{figure}[h]
    \centering
    \includegraphics[width=0.8\linewidth]{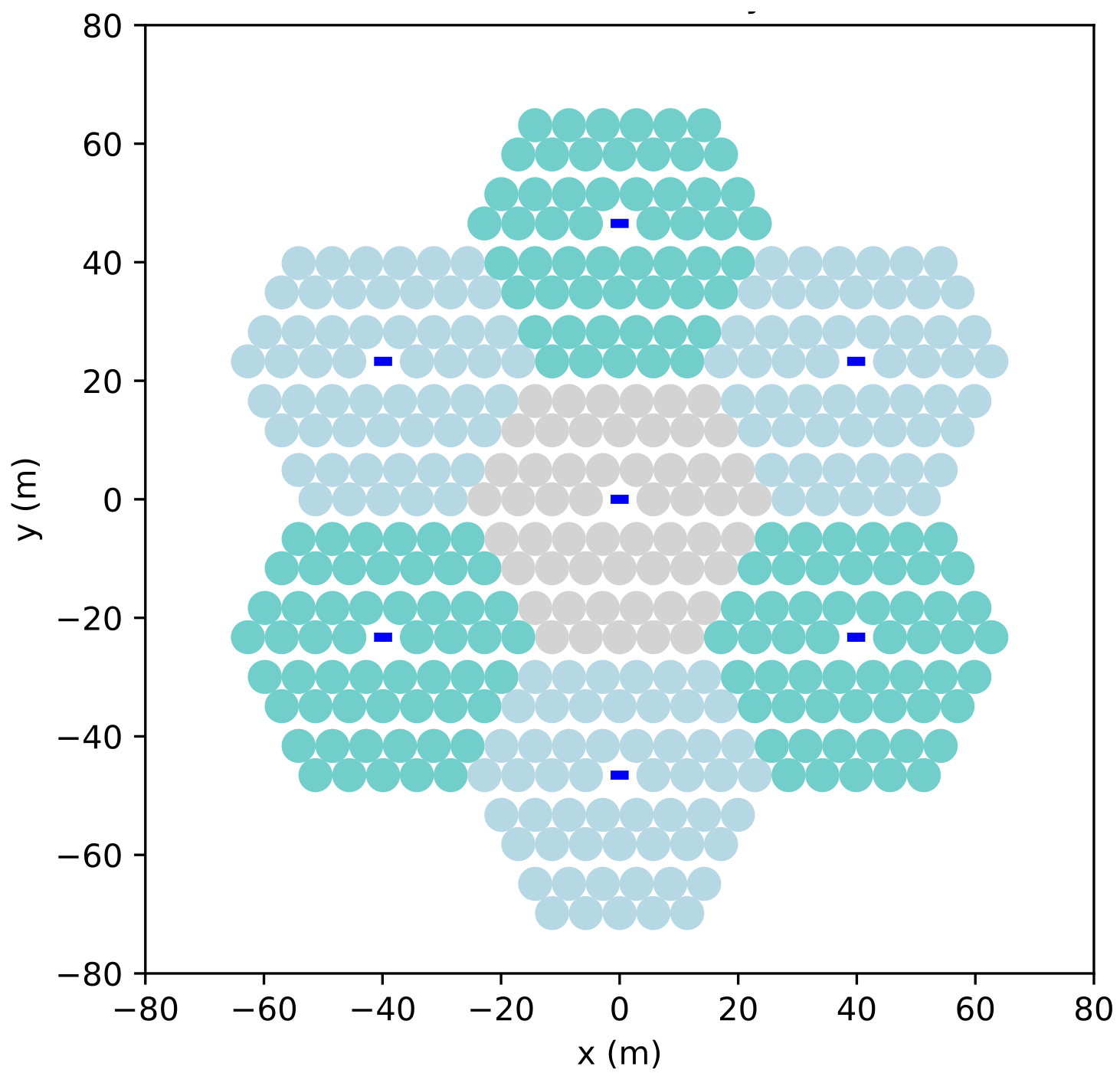}
    \caption{The planned SWGO-A layout showing seven clusters, each of 55 WCDs and surrounding one Field Node (blue rectangle).}
    \label{fig:SWGO-A}
\end{figure}

The SWGO-A layout corresponds to a 385 WCD-unit segment of the inner array, with a fill factor of \change{70}\%, arranged in seven independent clusters of 55 tanks. Each cluster will be electrically connected to a single FN as seen in Figure~\ref{fig:SWGO-A}. The WCDs will be separated by 50 cm, permitting access to all sides. Between every two rows of tanks, a 1.5 m wide aisle is planned for installation of the field node and access by small cranes or similar equipment as needed for deployment and maintenance. The overall fill factor is slightly larger than that of the central array of the HAWC observatory.

\subsection{Outer Array}
\label{sec:outer}

The outer array plays a vital role in increasing the effective detection area, enhancing background rejection, and boosting the overall performance of the observatory at higher energies. Due to the large footprint of this element of the array, alternative tank configurations \change{have been} explored. The final design is 
\change{undergoing an optimization process aimed at improving overall performance at fixed cost, but several aspects are already fixed as outlined below.} 

Smaller tanks offer advantages in terms of ease of deployment and reduced separation at constant fill factor. For smaller tanks an alternative technology becomes available: 
High Density Polyethylene (HDPE) roto-molded plastic tanks. 
Such tanks, measuring 3.6 meters in diameter, could be produced near the observatory site and transported by standard trucks. 
Prototype HDPE tanks have been produced for SWGO, based on a redesigned version of the tanks used at the Pierre Auger Observatory.
\change{Our studies indicate that a tank of this diameter and of depth 2.9~m is very promising in terms of performance/cost and has now been adopted as the outer array baseline.}

Compared to the original Pierre Auger design, several improvements are being considered: the use of enhanced resin materials for greater mechanical strength and UV resistance under harsh Atacama conditions; a potential double-wall structure for thermal insulation to prevent water freezing; and  the inclusion of a new Reflective Oxide Nano-coating (RON) on the inner surface. RON is an oxide-based nano-structured coating designed to enhance diffuse UV–visible reflectance and improve Cherenkov-light collection while remaining compatible with the polyethylene tank manufacturing process~\citep{mendes2026}. This could eliminate the need for a Tyvek-laminated polypropylene bladder, resulting in a WCD unit that may be more cost-effective, durable, and easier to deploy at high altitudes.

\begin{figure}
    \centering
    \includegraphics[width=0.95\linewidth]{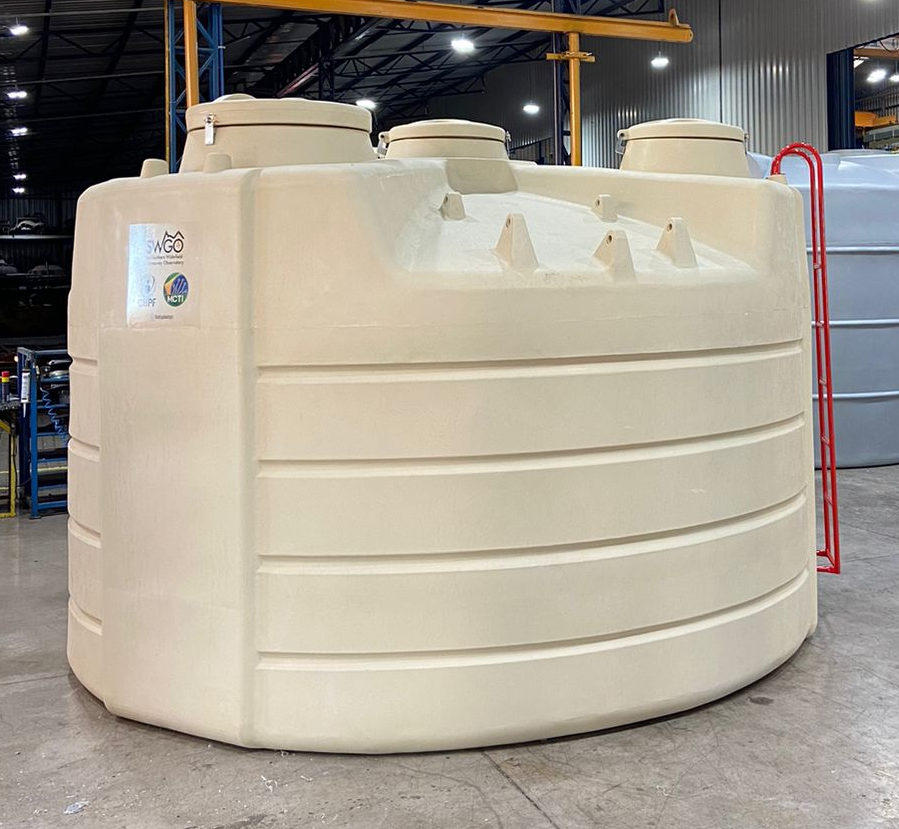}
    \caption{A prototype of a roto-molded HDPE tank under consideration for the outer (low-fill-factor) array. }
    \label{fig:HDPE}
\end{figure}

\begin{figure}
    \centering
    \includegraphics[width=0.95\linewidth]{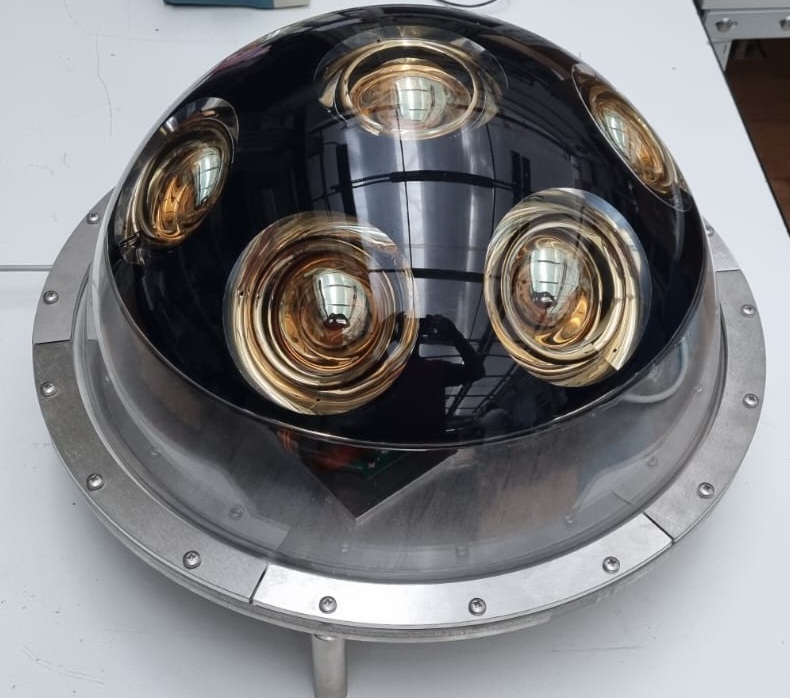}
    \caption{One of the multi-PMT modules to be deployed in the Pathfinder HDPE tanks. }
    \label{fig:multipmt}
\end{figure}

Some full-scale prototypes have been successfully produced and shipped to Chile for SWGO-PF (see Figure~\ref{fig:HDPE}), including one with a reflective nano-coating and one that is in operation equipped with a multi-PMT module (Figure~\ref{fig:multipmt}). This module is a development for SWGO, composed of seven 3” PMTs to maximize the angular acceptance of Cherenkov light produced by air-shower particles that penetrate the tank and to provide directionality information at each WCD unit, \change{and has now also been adapted as the baseline for the outer array WCDs.}

The outer array will make use of a WR time synchronization system similar to the one used for the inner array.

\subsection{The Radio Array SWGO-TURBO}
\label{sec:turbo}

An array of 800 radio antennas, named \emph{The Utility for Radio Beam-formed Observations} (TURBO), will be constructed as part of the SWGO Project, following a successful bid to the ERC Consolidator Grants. The dual-polarized antennas will observe the pulsed radio emission from air showers in the 50--200\,MHz frequency band, and data acquisition will be fully integrated within the observatory. Since SWGO-TURBO is already funded, it will be deployed during the early phase(s) of SWGO and therefore complement the particle observations for the full operational time of the observatory. The readout of the radio array will be initiated by a trigger from the particle detectors. By applying interferometric reconstruction~\citep{schoorlemmer-2021,schoorlemmer-2023}, it is anticipated that the influence of noise,  which is dominated by radio-wave emission from the Milky Way, can be significantly reduced compared to traditional radio observations of air showers.

The primary focus of TURBO is on air showers with high energy ($>$PeV) and large zenith angles (between $45^\circ$ and $70^\circ$). For these air showers, radio will complement the array of WCDs by providing accurate angular reconstruction, an alternative energy estimation, and reconstruction of the depth of the air shower maximum. These observations aim to increase the sensitivity and aperture of SWGO for photons with energy above PeV and enable more detailed air-shower observation of charged CRs. 

\subsection{Possible Ultra-High-Energy Extension in a Lake}
\label{sec:lake}

The deployment of WCDs inside a lake was considered as part of the SWGO design study~\citep{lake_paper, Goksu:2021Tw}, given that lake water provides good shielding for electromagnetic particles, with the potential to enhance muon tagging capabilities.

No suitable high-altitude lake was found, but investigations continue within the collaboration, targeting a somewhat lower altitude and a low-fill-factor (less than a few \%) array aimed at ultra-high energies. As well as searching for candidate lakes, the collaboration is developing lake deployment technologies and will compare the cost-effectiveness of a lake extension at a second site in comparison to expansion of the primary site.

%% file: Performance.tex
\section{Performance}
\label{sec:performance}

This section discusses the expected performance of SWGO using the reference layout described in Section~\ref{sec:layout}, which features a surface area of 1~km$^2$.
The areas of the three SWGO zones with fill factors of 70\%, 4\%, and 1.7\%, decreasing from the center, are shown in Figure~\ref{fig:swgo_layout_D8}.
In the following, we will refer to this layout as SWGO. The first stage of constructing this large instrument we, as described in Sections~\ref{sec:site} and \ref{sec:swgoa}, we refer to as SWGO-A. Figure~\ref{fig:SWGO-A} shows the layout of SWGO-A, which has a fill factor of \change{70}\%. The same WCD unit design was simulated for the three zones, as well as for SWGO-A. The performance of SWGO at an altitude of 4,770\,m is examined using simulations based on CORSIKA~\citep{corsika} and HAWCsim~\citep{HAWC_CRAB}. Air showers of secondary particles induced by gamma rays and protons are simulated with energies from 100\,GeV to 1\,PeV and utilized to derive Instrument Response Functions (IRFs) using the \texttt{pyswgo} software framework. The event reconstruction is partly carried out with \texttt{AERIE}~\citep{aerie, HAWC_CRAB}, originally developed for HAWC.

To ensure precise observations, only events with a multiplicity of at least 65 stations are accepted for SWGO\footnote{By the same standard, 30 triggered stations are required for SWGO-A.}.
Furthermore, events with reconstruction uncertainty that deviate by more than two standard deviations from the expected energy reconstruction or the angular reconstruction resolution are rejected.

\begin{figure}
    \includegraphics[width=\linewidth]{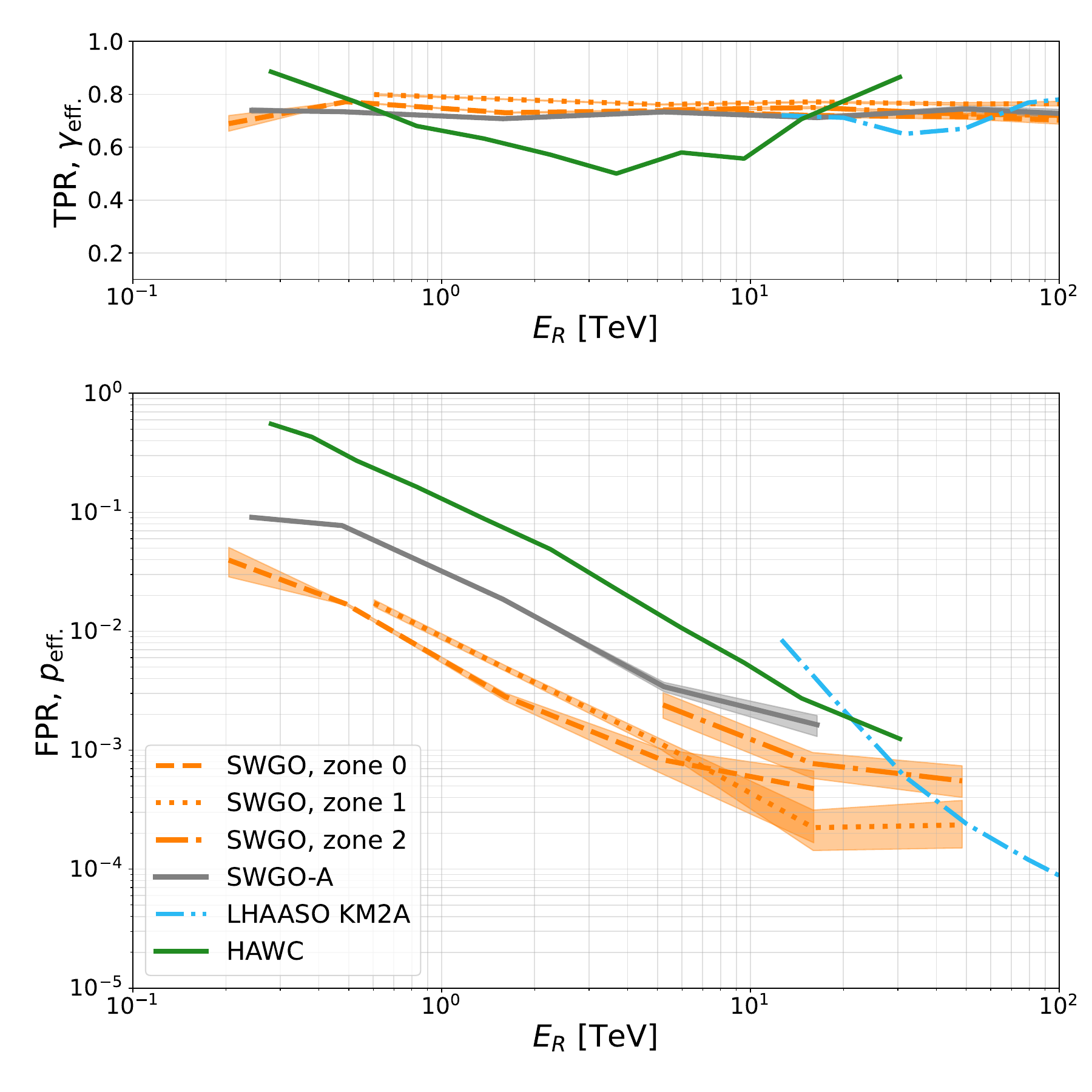}
    \caption{Performance of gamma-hadron separation as a function of energy for events with zenith angle between $0^\circ$ and $30^\circ$ for SWGO (orange) and SWGO-A (grey) compared to HAWC (green) and LHAASO KM2A (blue). Continuous, dashed, and dotted orange lines indicate showers reconstructed in different zones of SWGO. \textbf{Top}: Gamma-ray efficiency (true positive rate (TPR)). \textbf{Bottom}: Background efficiency (false positive rate (FPR)). At the highest energies, the background is limited by Monte Carlo statistics.}
    \label{fig:gamma-hadron}
\end{figure}

\subsection{Limitations}
\label{sec:limitations}
The current simulation setup does not consider the background from air showers (coinciding/overlapping showers, single muons) nor from dark counts.
The impact of these backgrounds on the analysis can be considerably reduced with dedicated filtering methods, as shown in the case for HAWC; however, they are not yet fully integrated into the SWGO software.
\change{The simulation used in this work is based on the well-tested and validated HAWC simulation, which accounts for noise from dark currents, out-of-time showers, single muons, neutrons/radioactivity, and after-pulsing.
To limit missing contributions on the quoted SWGO performance, we decided to set very conservative trigger thresholds.
The thresholds were derived by extrapolating from the HAWC 28-PMT trigger, scaled to SWGO's larger PMT count, while conservatively assuming the same Fano factor (i.e., signal-to-noise ratio) as HAWC.
This is quite a conservative estimate since HAWC is equipped with four PMTs compared to only a single one in the SWGO upper layer.
Therefore, HAWC has a higher Fano factor because a single charged-particle track in a single HAWC tank can produce four hits, whereas in SWGO it produces only a single hit.}
This yields a threshold of 65 for SWGO and 30 for SWGO-A.
\change{We also studied the performance in a more conservative high-threshold scenario, raising the hit threshold to 100 for SWGO and 50 for SWGO-A.}
When studying performance, we did not find significant differences across the performance metrics shown, besides a shift in the energy threshold\footnote{Compare Fig.~\ref{fig:sensitivity_cons} for more details.}.
Also note that the enormous size of SWGO will enable the exploitation of spatial triggering schemes or a directional coincidence approach, as used by LHAASO-WCDA.
Such improvements will further lower the SWGO threshold in the future.
A detailed noise model, comprising overlapping showers and background muons, PMTs, electronics, and digitization, is currently under development for future studies.

At higher energies, the current IRFs are limited by the strong rejection power of the gamma-hadron separation, causing the sensitivity estimates to feature a significant statistical uncertainty due to the very large number of required simulated Monte Carlo samples for obtaining enough surviving gamma events after separation.
\change{We assume in this region the background rejection to be constant, which leads to conservative estimates on the performance at energies larger than 100 TeV.}
This applies to all science benchmarks discussed in this document.
As the project continues to consider the layout and unit design of the outer array, adjustments in performance are anticipated.
Furthermore, we use state-of-the-art machine learning algorithms~\citep{dlfpr} during gamma-hadron separation.
Although their performance on simulations is exceptional and has been validated using measured data in CR physics~\citep{thepierreaugercollaboration2024measurement, thepierreaugercollaboration2024inference} and neutrino astronomy~\citep{gal_plane_2023}, a validation of these novel algorithms in gamma-ray astronomy is still lacking and will be crucial for achieving the performance shown here.
\change{Therefore, we note that the term `conservative' relates to the choice of trigger thresholds. 
Besides, due to the absence of data, our simulations describe an ideal detector, combined with the yet-to-be-provided validation of the novel reconstruction algorithms, the quoted results should consequently be understood as the expected performance under the outlined assumptions rather than as a general conservative estimate.}

\begin{figure}[t!]
    \includegraphics[width=\linewidth]{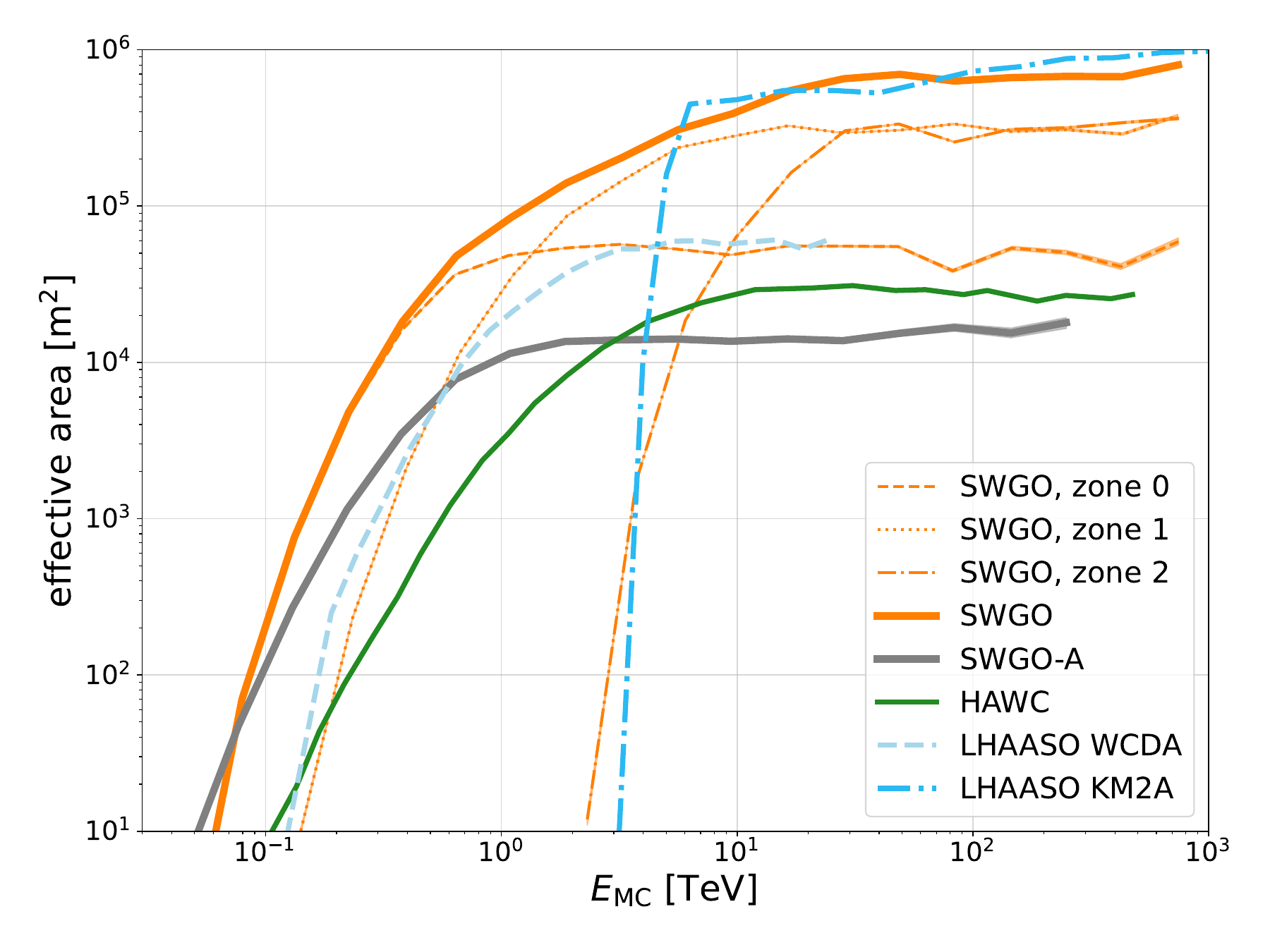}
    \caption{Effective area as a function of energy for SWGO and SWGO-A (grey) compared to HAWC (green) and LHAASO (blue). Different line styles denote the different zones of decreasing fill factors of the SWGO array.}
    \label{fig:eff_area}
\end{figure}

The derived quantities discussed hereafter provide an approximate performance estimate of SWGO and SWGO-A based on detailed simulations, dedicated reconstruction software, and reconstruction algorithms tailored to the SWGO design. Statistical uncertainties are shown as opaque regions and are obtained by bootstrapping.

\begin{figure*}[!th]
    \includegraphics[width=0.495\linewidth]{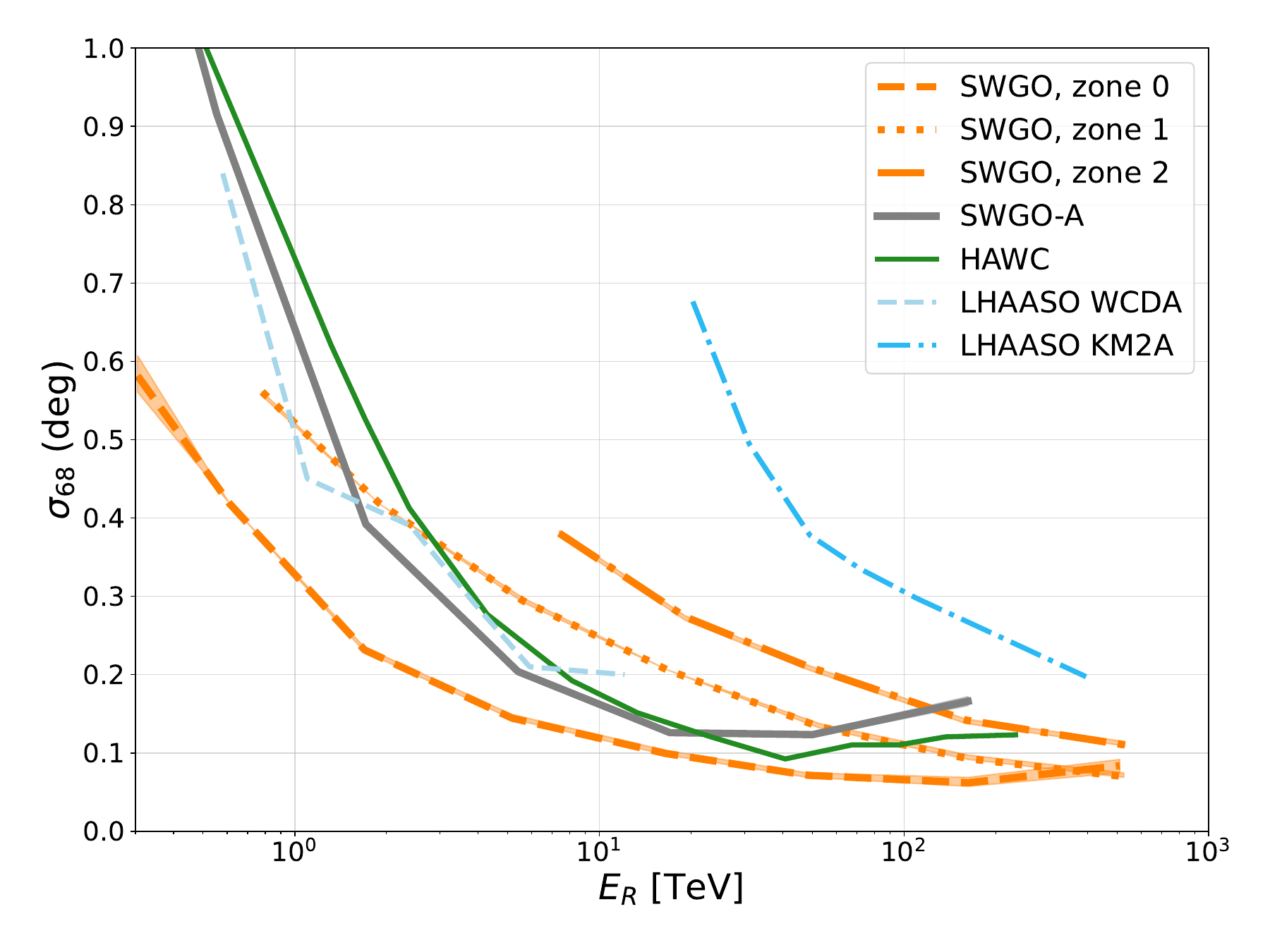}
    \hspace{0.25cm}
    \includegraphics[width=0.495\linewidth]{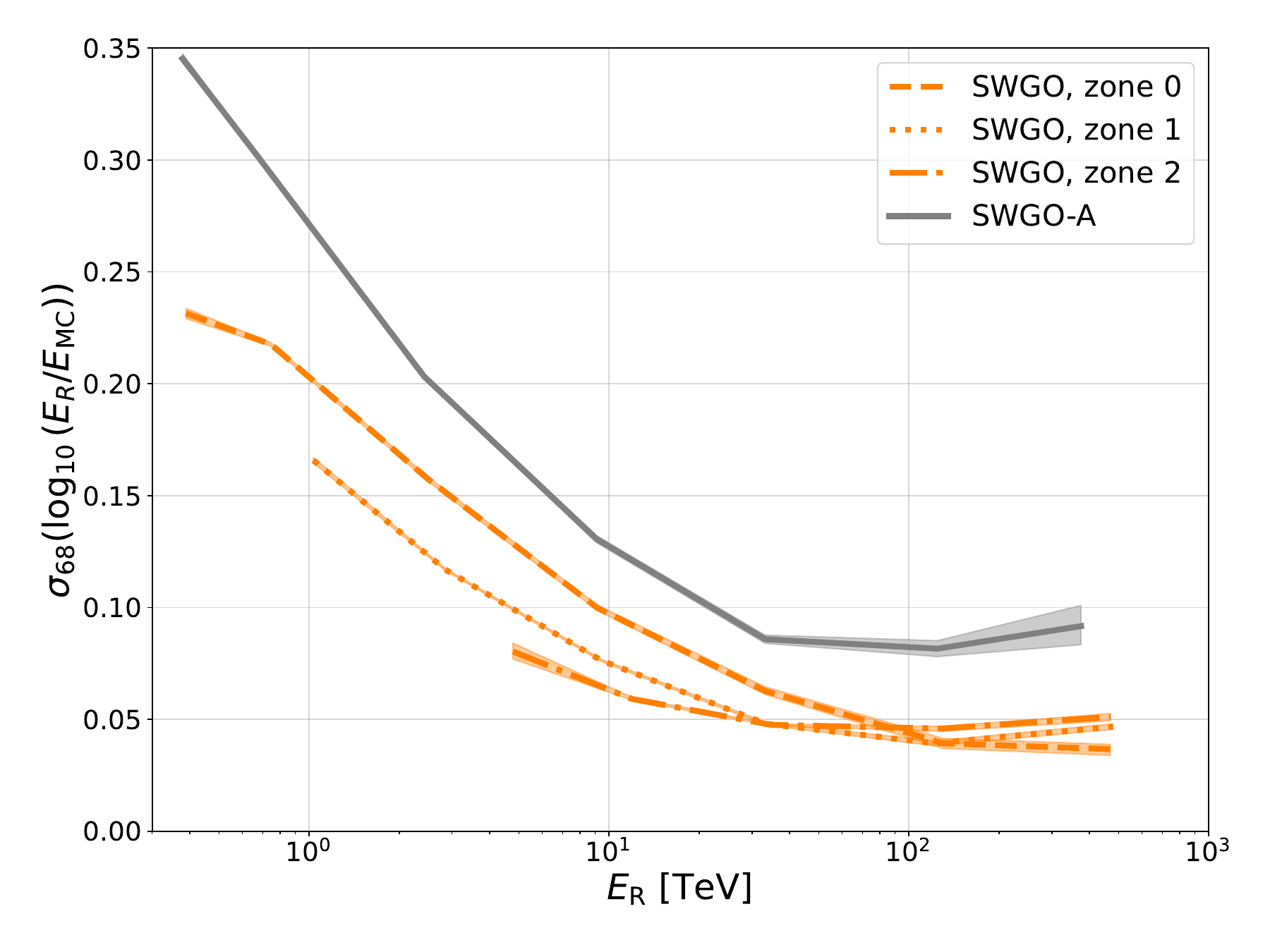}
    \caption{Reconstruction performance for gamma showers with zenith angles between $0^{\circ}$ and $30{^\circ}$ for SWGO (orange) and SWGO-A (grey). \textbf{Left:} Angular resolution as a function of energy compared to HAWC (green) and LHAASO (blue). \textbf{Right:} Energy resolution as a function of energy.}
    \label{fig:reco}
\end{figure*}

\subsection{Gamma-Hadron Separation}
Gamma-hadron separation is performed by training a shallow neural network given several separation observables in bins of shower energy, zenith, and core distances with respect to the array center, accounting for the graded design of SWGO.
In addition to low-level observables (e.g., zenith angle and number of triggered stations) and high-level observables \citep[such as PINCness and LDFChi$^2$, introduced by HAWC,][]{Alfaro_2022}, the analysis also makes use of novel deep-learning-based algorithms, such as graph neural networks~\citep{Glombitza_wcd_2025} and transformers~\citep{Watson:2023vx}, and uses the output scores of those algorithms in classification. The performance of the gamma-hadron separation is shown in Figure~\ref{fig:gamma-hadron} for events with zenith angle between $0^{\circ}$ and $30{^\circ}$ for SWGO (orange) and SWGO-A (grey), and compared to HAWC (green)~\citep{albert2024performancehawcobservatorytev} and LHAASO\footnote{At the time of writing, no detailed data on proton and gamma efficiency as a function of energy is available for the gamma-hadron separation performance of LHAASO WCDA.} KM2A (blue)~\citep{lhaasocollaboration2024optimizationperformancekm2aarray}.
Dashed, dotted, and dash-dotted lines indicate different zones of the SWGO layout, which are shown in Figure~\ref{fig:site_PL}, where zone~0 is the inner array. In the top panel, the gamma efficiency, i.e., the true positive rate (TPR), is shown and has been set to $80\%$ of the triggered events. Additional selection cuts slightly decrease these efficiencies.
The background rejection, i.e., the false positive rate (FPR), shown in the bottom panel, strongly decreases as a function of energy and reaches a suppression of $10^{-3}$ and lower for all zones, demonstrating the efficacy of the double-layer approach.
SWGO-A reaches a rejection of $10^{-1}$ at low energies and approaches $10^{-3}$ above 10~TeV, showing significant improvements compared to HAWC.
Compared to LHAASO KM2A, SWGO exhibits superior background rejection performance until 100\,TeV at which --- due to the strong rejection and the limited amount of Monte Carlo statistics --- no background remains in any of the investigated zones and layouts.

In Figure~\ref{fig:eff_area}, we show the effective area for detecting gamma rays after gamma-hadron separation using SWGO (orange) and SWGO-A (grey) as a function of energy for events with zenith angle between 0$^\circ$ and 30$^\circ$.
As a comparison, we also show HAWC (green)~\citep{albert2024performancehawcobservatorytev}, LHAASO WCDA (light blue)~\citep{lhaasocollaboration2021performancelhaasowcdaobservationcrab}, and LHAASO KM2A (blue)~\citep{lhaasocollaboration2024optimizationperformancekm2aarray}.
The thin orange lines denote the effective areas of events reconstructed with the shower core in the three different SWGO zones of decreasing fill factors (70\% (zone 0), 4\% (zone 1), and 1.7\% (zone~2)), which are depicted in Figure~\ref{fig:site_PL}. Whereas SWGO zone 0 and SWGO-A become fully efficient at 1\,TeV, zones 1 and 2 are efficient above 10\,TeV and 30\,TeV, respectively. The continuous orange line denotes the total effective area of the entire detector, which is comparable to LHAASO KM2A. With its $10^6$\,m$^2$ size and planned construction at 4,800~m altitude, SWGO's graded-zone design --- with the fill factor decreasing from a densely-instrumented inner core to a sparser outer array --- enables efficient gamma-ray detection from 300\,GeV to the PeV scale. Similar to HAWC, we included off-array events in this work to increase the effective area of SWGO-A, reaching almost $2\times 10^4$~m, close to HAWC. Due to its high-altitude location, SWGO-A already lowers the threshold compared to HAWC and LHAASO WCDA, providing {\bf significant} potential for improved observations at low energies.

\begin{figure*}[!ht]
\centering
\includegraphics[width=0.82\linewidth]{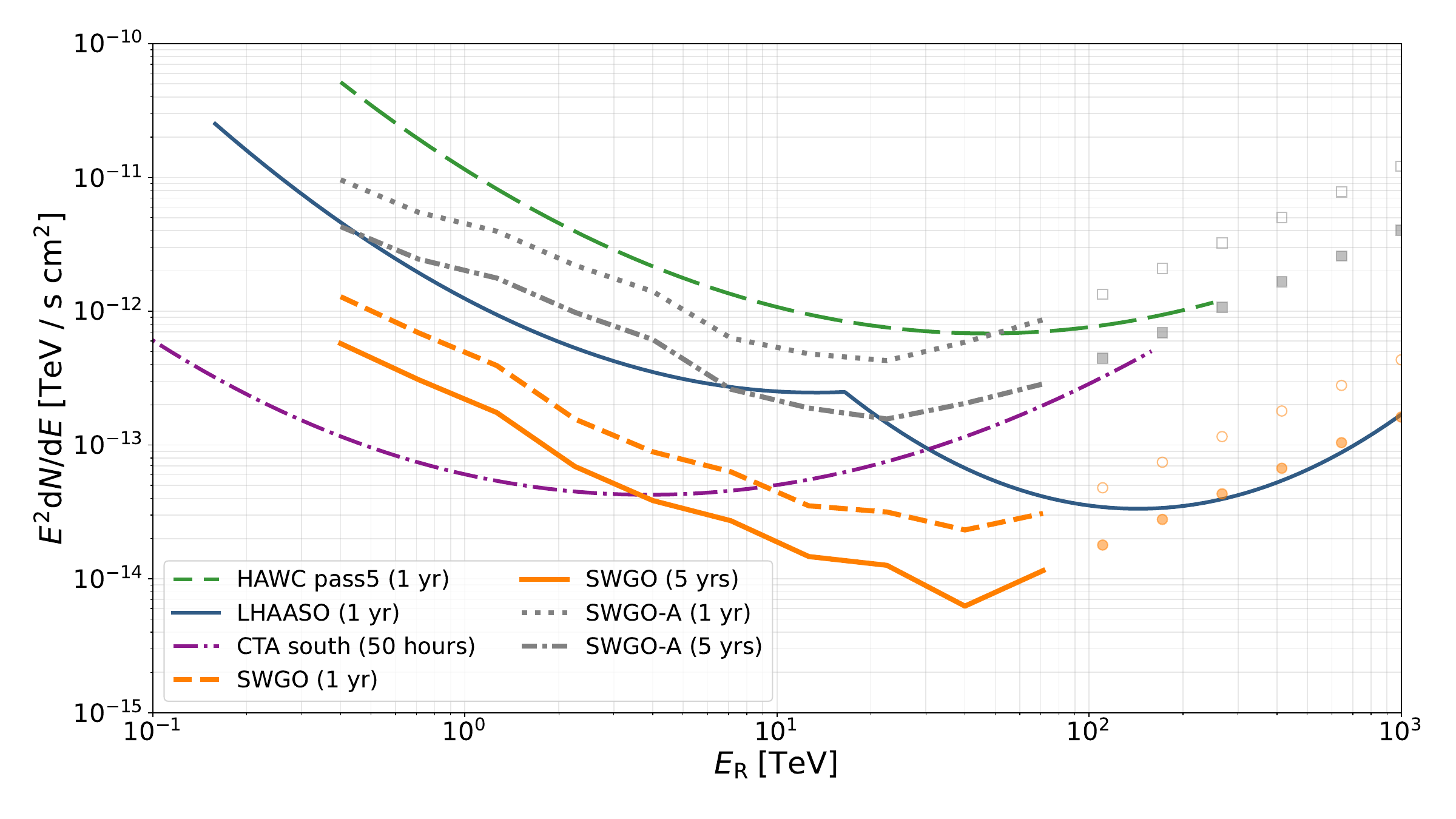}
    \caption{Differential point source sensitivity for SWGO and SWGO-A compared to CTA South, HAWC, and LHAASO. While the lines show the expected sensitivity calculated using simulations, the dotted regions denote extrapolations.}
    \label{fig:sensitivity}
\end{figure*}

\subsection{Event Reconstruction}
\label{sec:EventReco}
The utilized angular reconstruction is based on the Gaussian fitting procedure introduced by HAWC~\citep{HAWC_CRAB}, in which a plane with a second-order curvature model of a shower front is fit to arrival times. The angular resolution, defined as the $68\%$ quantile of the angular distance, is shown in the left panel of Figure~\ref{fig:reco} for SWGO-A (grey) and SWGO (orange) and compared to various operating ground-based gamma-ray observatories. While SWGO-A shows a resolution very similar to HAWC (green)~\citep{albert2024performancehawcobservatorytev} and LHAASO WCDA (light blue)~\citep{lhaasocollaboration2021performancelhaasowcdaobservationcrab}, SWGO demonstrates a very precise reconstruction with significant improvements over LHAASO and HAWC. After running the angular fit, the shower core and energy reconstruction are carried out using template-based methods~\citep{template_Parsons_2014, templates_vikas} adapted to the SWGO design~\citep{template_leitl}, with the arrival direction serving as seeding parameters.
These state-of-the-art template methods utilize likelihood models of the signals in tanks as a function of core distance in different bins of energy and arrival direction, which are derived using a comprehensive simulation library. By maximizing the likelihood of the observed event, template methods offer reliable performance across a large phase space, thereby significantly improving the reconstruction compared to previous approaches.
The quality of the energy reconstruction is summarized in the right panel of Figure~\ref{fig:reco} for SWGO-A (grey) and the three different zones\footnote{See Figure~\ref{fig:swgo_layout_D8} for details on the layout and the different zones.} of the SWGO array (orange). The energy resolution, defined as the $68\%$ quantile of the difference in log space, is improving significantly as a function of energy. Above 1\,TeV, SWGO (SWGO-A) reaches a resolution better than 0.2 (0.3) for all zones, improving to 0.05 (0.1) at the highest energies. The core resolution, defined as the 1D 68\% quantile of the distribution of the distance between the reconstructed and the true core position, improves from about 25\,m to around 2\,m and is below 10\,m for energies larger than 3\,TeV, providing accurate event reconstruction binned in different zones.

Based on the derived IRFs and the \texttt{Gammapy} software~\citep{Gammapy}, Figure~\ref{fig:sensitivity} shows the one-year (dashed) and five-year (continuous) differential sensitivity for SWGO (orange) and SWGO-A (grey) to detect a steady point source with an energy spectrum of $E^{-2}$ with significance at the level of $5\sigma$ for a continuously emitting source at $-20^\circ$ declination.
Whereas lines represent the expected performance estimated using the simulated IRFs, the dotted regions indicate extrapolations of the sensitivity in the phase space where no background events pass the gamma-hadron separation, assuming vanishing signals and constant rejection performance. Note that these extrapolations are currently not considered in the IRFs used for the science benchmark, limiting their expressiveness beyond 100 TeV.
For comparison, we show the sensitivity of LHAASO (blue)~\citep{Vernetto_2016} and HAWC (green)~\citep{albert2024performancehawcobservatorytev} with one-year observation time, as well as 50 hours of observation time using CTAO (purple)~\citep{CTA_ScienceTDR}.
Utilizing its innovative detector design at high altitudes, SWGO-A provides distinct improvements at low energies compared to HAWC. The expected SWGO sensitivity shows very promising performance over the entire energy range, with prospects for significant improvements over LHAASO at low and medium energies, demonstrating its strong potential to complement CTAO in the TeV range. As the layout and unit design of the outer array will be considered in the future, improvements are expected at the highest energies.

\begin{figure}[!ht]
\centering
\includegraphics[width=0.98\linewidth]{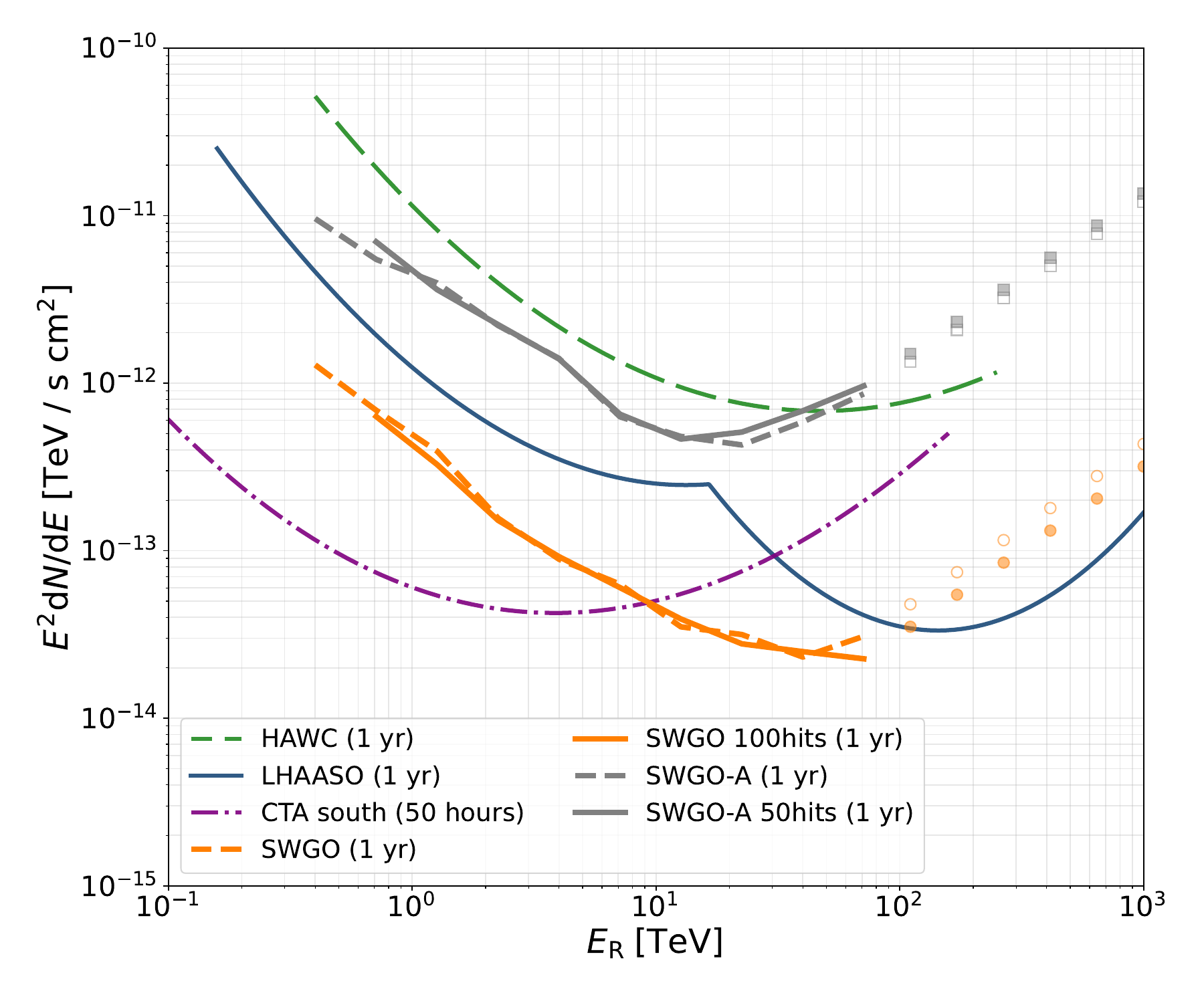}
    \caption{Differential point source sensitivity for SWGO and SWGO-A with different threshold cuts, compared to CTA South, HAWC, and LHAASO. While the continuous lines of SWGO show the expected sensitivity, the dashed lines denote the more conservative \emph{high-threshold} trigger scenario for SWGO (100 hits instead of 65) and for SWGO-A (50 instead of 30). Except for the higher energy threshold, the performance is similar. Open (filled) markers denote extrapolations of SWGO with 100 (65) hits.}
    \label{fig:sensitivity_cons}
\end{figure}

To further demonstrate the robustness of the quoted performance against the trigger threshold derived from HAWC as outlined in Section~\ref{sec:limitations}, we compare the nominal sensitivity with that obtained using substantially more conservative thresholds of 100 hits for SWGO and 50 for SWGO-A in Figure~\ref{fig:sensitivity_cons}.
As can be seen, the sensitivity above $\sim$700\,GeV is largely unaffected by this change.
Small differences at high energies are due to statistical uncertainties arising from retraining and recalibration of gamma-hadron separators.
However, the extremely high triggers cause a shift in the energy threshold, removing the first sensitivity bin.
This shows that the performance estimates throughout this work are robust against the specific choice of trigger threshold.
Thus, noise contributions from overlaying showers will not drive the quoted results at the energies most relevant to the science benchmarks discussed in this document.
For science tasks that require very high sensitivity at low energies, we performed a detailed evaluation using both cut criteria and discuss it in the text.
\change{We refer to these IRFs with very high thresholds as \emph{high-threshold} scenario.}

%% file: Galactic.tex
\section{Galactic Particle Acceleration and Transport}
\label{sec:galactic}

\subsection{Survey of the Galactic Plane}\label{gps}

\begin{figure*}
    \centering
    \includegraphics[width=\textwidth, trim={5.4cm 0.8cm 5.7cm 0.8cm},clip]{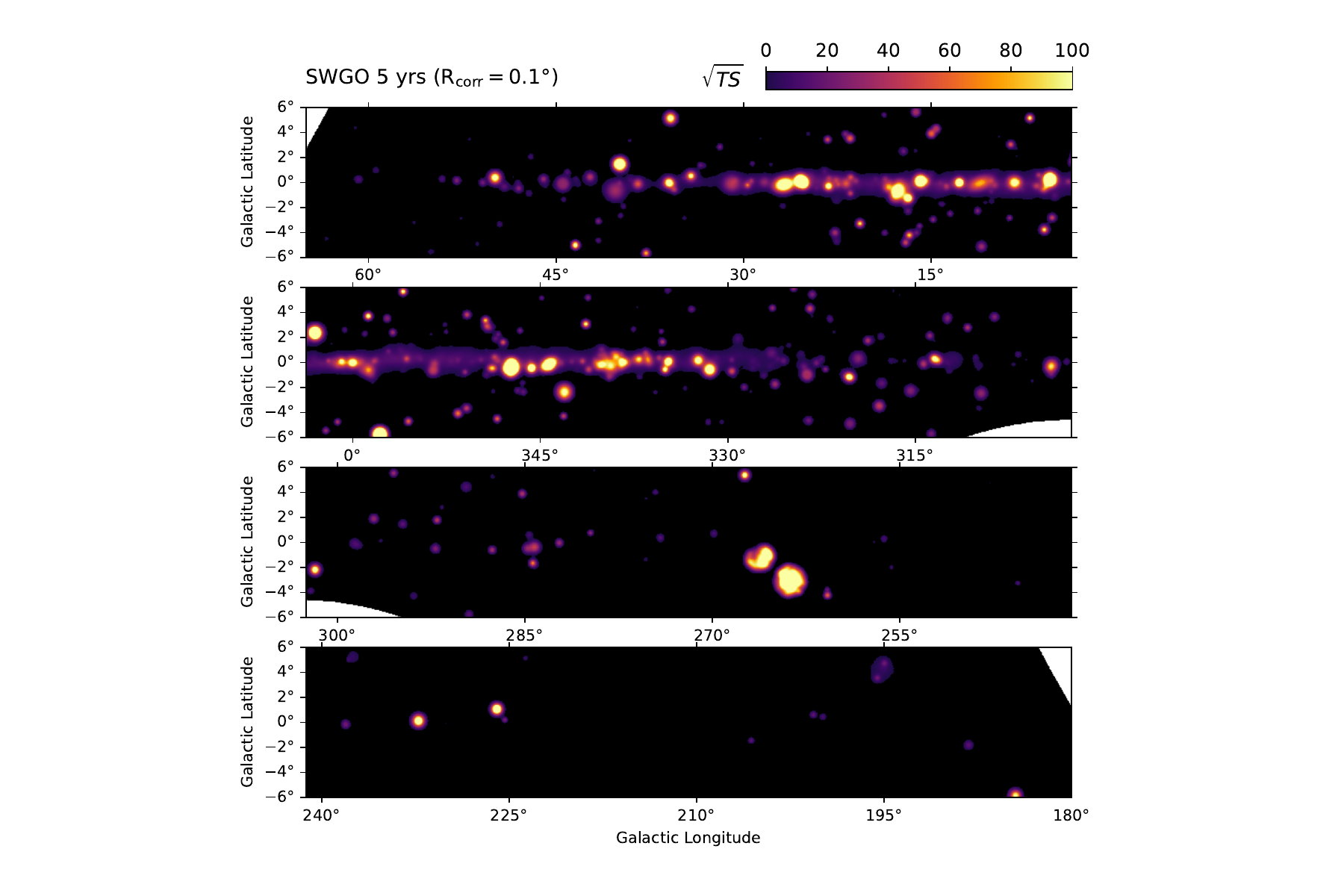}
    \caption{Significance map of the excess above the background integrated in a 0.1\degree correlation radius for energies larger than 1 TeV and after five years of SWGO observation. The simulation uses the same source population model as the CTA-GPS simulation and the interstellar emission model called "IEM varmin-rescaled" \citep[for details on these models, see][]{2024JCAP...10..081A}.}
    \label{fig:gal_signi_0p1deg}
\end{figure*}

Previous generations of instruments have detected several populations of Galactic sources, including pulsar wind nebulae (PWNe), supernova remnants (SNR), star-forming regions (SFRs), TeV halos, pulsars, gamma-ray binaries, and microquasars (see Section~\ref{sec:gal_pop} for dedicated discussions). A large fraction of new detections arose from surveys by HEGRA, H.E.S.S., VERITAS, MAGIC, Milagro, TIBET AS$\gamma$, HAWC, LHAASO, and the \emph{Fermi Gamma-ray Space Telescope}, highlighting the importance of unbiased observational strategies. By providing large and unbiased samples of gamma-ray sources, surveys help to uncover the mechanisms of particle acceleration, their relative efficiencies, the nature of their transport, and the contributions of different source populations to the Galactic CR population.

The SWGO design is exceptionally well-suited for surveying the Galactic Plane due to its wide FoV and high duty cycle. It will allow a large fraction of the sky to be continuously monitored, including most of the inner Galaxy where the majority of Galactic sources are concentrated.

To test the survey capabilities of SWGO, 1 to 10 years of Galactic Plane observations were simulated between Galactic longitude $-180^\circ<l<70^\circ$ and latitudes $|b|<6^\circ$.
The source population and interstellar emission models used are the same as for the recent CTA Galactic Plane Survey (CTA-GPS) simulation\footnote{These models are publicly available at \url{https://zenodo.org/records/10008527}} \citep{2024JCAP...10..081A}.
This includes both a compilation of known sources detected at GeV or TeV energies and synthetic populations for the main classes of Galactic VHE sources (SNRs, PWNe, and gamma-ray binaries). Section~\ref{sec:diffuse} discusses the specifics of the alternative interstellar emission models considered and the capabilities of SWGO to detect diffuse emission.
A potential view of the Galaxy above 1~TeV as seen by SWGO after five years of observation is given by Figure~\ref{fig:gal_signi_0p1deg}, which shows the significance of the excess above the CR background integrated in a 0.1\degree correlation radius.

The number of expected detections above a given test statistic (TS) value, $\rm{TS}>25$, is computed as a benchmark quantity. For simplicity, this number is estimated only for the Asimov data set \citep[i.e., counts equals predicted counts, see][]{2011EPJC...71.1554C}, which gives a limit on the expected detection for the true simulated model. For this benchmark a minimal interstellar emission model (IEM) is used as in the CTA-GPS simulation, which yields an upper estimate for the number of detectable sources. For comparison, the number of expected detections above the same TS threshold is also computed for the dataset of the CTA-GPS simulation. Moreover, as SWGO and CTAO datasets are independent, the TS values obtained for each observatory can be summed to obtain a joint TS and the corresponding expected detection for the joint datasets.

In Table~\ref{tab:gps_detections}, the expected detections with $\rm{TS}>25$ above 1 TeV toward $-180^\circ<l<70^\circ$ and $|b|<6^\circ$ from SWGO are compared to the ones expected for the CTA-GPS, as well as to the joint CTAO-GPS+SWGO datasets. Number of detections in the same longitude and latitude ranges are also reported for source catalogs of the current generation of instruments such as H.E.S.S., HAWC, and LHAASO.
This result clearly shows that SWGO will have a transformational 
impact on the field by multiplying the number of detected sources by up to a factor of 3 in a single year and up to a factor of 6 in 10 years.

Table~\ref{tab:gps_detections} also indicates that the expected number of detections of the CTA-GPS is matched by SWGO in about five years while, so far, the scheduling of the CTA-GPS is proposed to be spread over 10 years.  The presence of SWGO may therefore have a direct impact on CTAO's observation strategy.  Regarding the synergy between SWGO and CTAO, among the sources detectable by either SWGO or CTAO, about 70\% would be significantly seen by both instruments independently. As SWGO and CTAO are both committed to deliver open data, this would offer great opportunities for joint analyses to the community and even the possibility to build a joint catalog of gamma-ray sources over four decades in energy. This could even be extended to six decades in energy with the inclusion of \emph{Fermi} Large Area Telescope (LAT) data.

The expected detections quoted in Table~\ref{tab:gps_detections} are computed considering the true simulated model, which is an ideal limit. For the CTA-GPS simulation analysis, it was shown that the constructed catalog tends toward this ideal expectation \citep[see Section 4 and Appendix D of][]{2024JCAP...10..081A}, i.e., it properly models the sources expected to be detectable both in terms of number and flux. We did not attempt to construct a catalog for SWGO, but the same techniques could be applied. Above 10 TeV where the sensitivities and angular resolution are comparable, similar performances could be expected.
At lower energies, where both spectral and angular resolution are degraded, larger confusion effects are expected, so it will be more difficult to identify individual sources or to study energy-dependent morphology. However, these issues could be alleviated by performing joint analyses with CTAO. We defer the investigation of these questions to future works.

\begin{table}
\caption{Sources detectable \textbf{ with $\rm{TS}>25$ above 1 TeV} toward the Galactic Plane in SWGO's FoV 
 ($-180^\circ<l<70^\circ$ and $|b|<6^\circ$).}
\label{tab:gps_detections}
\begin{center}

\begin{tabular}{c|c}
\hline \hline
Catalogue & Detected sources  \\
\hline
HGPS$^{(a)}$ &  96 \\ 
3HWC$^{(b)}$ &  48 \\
1LHAASO$^{(c)}$  & 54 \\
\hline
Future surveys & Expected detections \\
\hline
CTA-GPS$^{(d)}$ & 461 \\
SWGO-A 1 yr & 70 \\
SWGO-A 5 yrs & 135 \\
SWGO 1 yr & 359 \\
SWGO 5 yrs & 487 \\
SWGO 10 yrs  & 536 \\
SWGO 10 yrs + CTA-GPS$^{(e)}$ & 603\\
\hline \hline

\end{tabular}
\end{center}

\begin{tablenotes}
\item \textbf{Notes.}
\item (a)  \cite{2018A&A...612A...1H}, components with $\rm{TS}>30$.
\item (b)  \cite{3HWC}
\item (c)  \cite{2024ApJS..271...25C}, sources with $\rm{TS}>37$ in WCDA or KM2A, or  with $25<\rm{TS}<37$ in both.
\item (d) using the same simulated dataset as in \cite{2024JCAP...10..081A}
\item (e) combining both simulated datasets in a joint likelihood.
\end{tablenotes}
\end{table}

\subsection{Extended Galactic Structures}

The study of astrophysical sources with extended VHE emission has become a central topic in modern astrophysics. In particular, morphological characterization provides critical insights into the physical mechanisms driving particle acceleration and propagation. SWGO is designed as a survey instrument with a large FoV that will map substantial portions of the Galactic disk and study extended sources. Its high sensitivity at VHE, along with its good spatial and energy resolution, will enable the investigation of extended sources and the diffuse emission of the Galactic Plane (see also Figures~\ref{fig:reco} and \ref{fig:sensitivity}). 

\subsubsection{Galactic Diffuse Emission}
\label{sec:diffuse}

While escaping the accelerating source and diffusing in the Galaxy, CR protons and nuclei interact inelastically with the atoms and molecules of the interstellar gas and produce gamma rays through the decay of neutral pions. Electrons emit gamma rays through inverse Compton (IC) scattering off the radiation fields and through bremsstrahlung \citep{1997ApJ...481..205H,1998A&A...338L..75M,Strong_2000,2000A&A...362..937A,2006PhRvD..74c4018K,2021A&A...653A..18D,2022ApJS..262...30P}. 
\begin{figure}
    \centering
    \includegraphics[width=\columnwidth, trim={0.3cm 0.3cm 1.cm 1.6cm},clip]{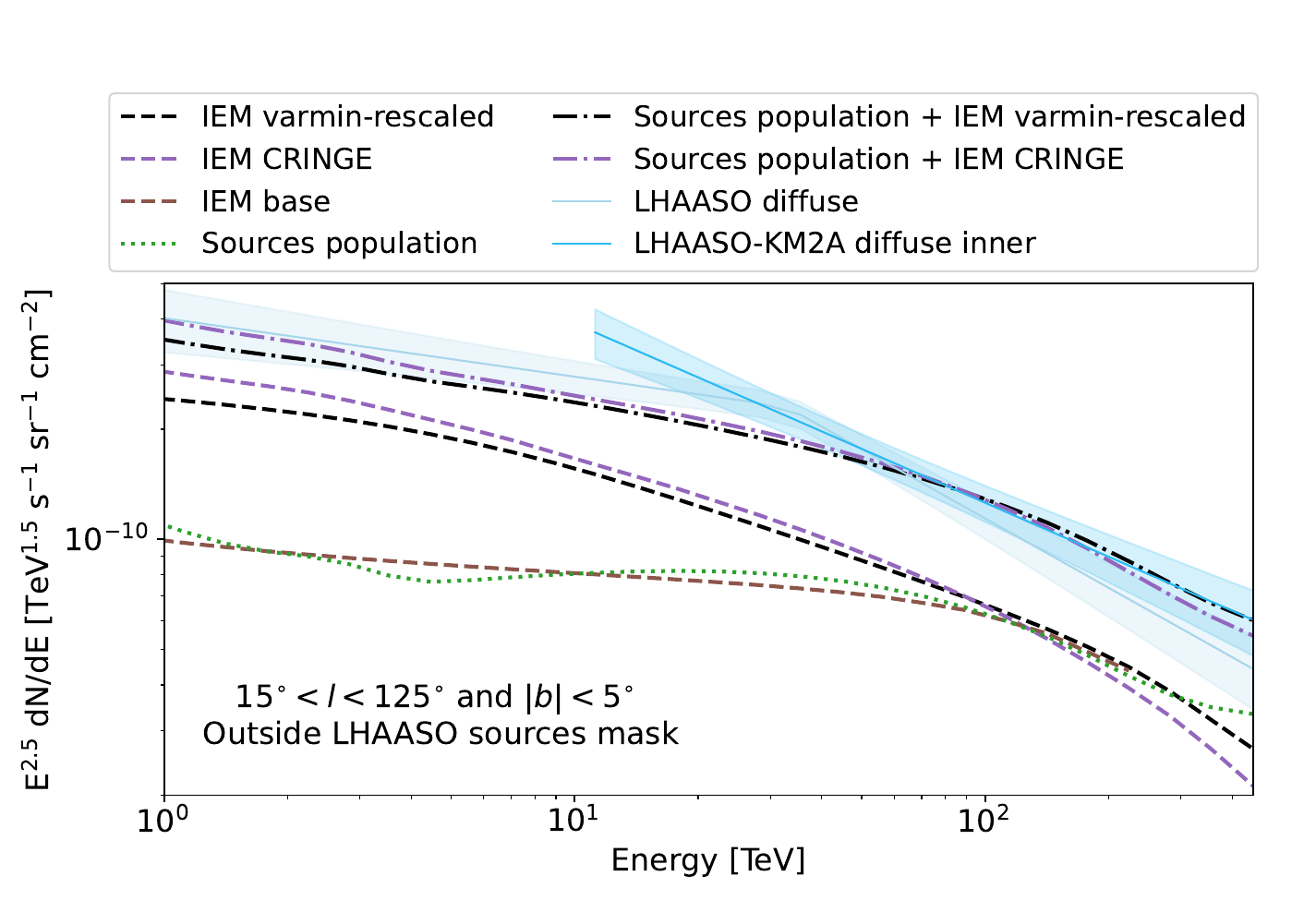}
    \caption{Differential flux of the diffuse emission models used in SWGO simulations compared to recent measurements by LHAASO-KM2A \citep[][in sky blue]{2023PhRvL.131o1001C} and LHAASO-WCDA \citep[][in light blue]{2024arXiv241116021L}. The diffuse emission can be modeled by different interstellar emission models (IEM) plus a contribution from unresolved sources. The ``IEM-CRINGE'' is taken from \cite{2023ApJ...949...16S}. The ``IEM-base'' and ``IEM-varmin-rescaled'', as well as the source population model, are taken from the CTA-GPS simulation \citep{2024JCAP...10..081A}. The LHAASO results correspond to a mean measurement toward $15^{\circ} < l < 125^{\circ}$ and $|b|<5^{\circ}$ outside of regions with a known source \citep[\textbf{this LHAASO sources mask is given in Figure~1 of }][]{2023PhRvL.131o1001C}. The emission of each model is averaged over the same region.}
    \label{fig:diffuse_lhaaso_mask}
\end{figure}

\begin{figure*}
    \centering
    \includegraphics[width=\textwidth, trim={5.4cm 0.8cm 5.7cm 0.8cm},clip]{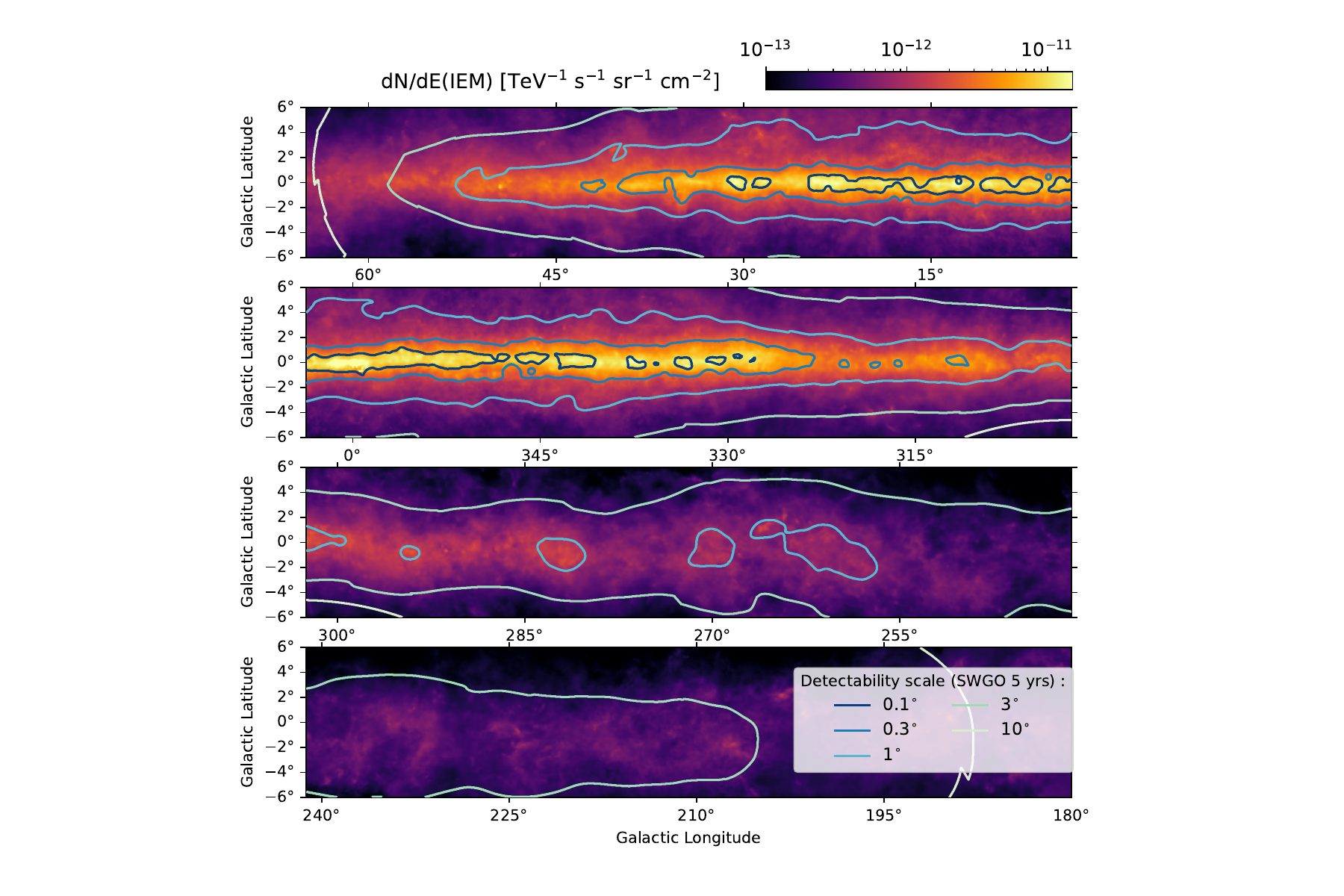}
    \caption{Detectability scale of the diffuse emission for five years of SWGO observation overlaid over the differential flux of the interstellar emission model considered \citep[IEM varmin-rescaled, see][]{2024JCAP...10..081A}. The contours correspond to the regions where the diffuse emission is detected at TS > 25 within the given correlation radius. At TeV energies, SWGO will have the capacity to detect diffuse emission at the scale of molecular clouds, down to $0.1^\circ$--$3^\circ$ in the inner Galaxy, like the \emph{Fermi}-LAT at GeV energies.}
    \label{fig:iem_detectability}
\end{figure*}

Although multiple studies of the Galactic diffuse emission have been performed at GeV energies using multiple missions and recently the \emph{Fermi}-LAT \citep{1997ApJ...481..205H,Ackermann_2012, 2016ApJS..223...26A}, measurements at TeV energies are more challenging because the CR spectrum and the resulting diffuse emission spectrum is steeper at higher energies \citep{pub.1136932664,2023PhRvL.131o1001C,pub.1169659179,2024ApJ...961..104A, 2024arXiv241116021L}. Furthermore, it is difficult to separate the faint large-scale Galactic diffuse emission from the residual hadronic background and possible contamination from unresolved sources.

While HAWC \citep{2024ApJ...961..104A} and LHAASO \citep{2025PhRvL.134h1002C} have deeply surveyed the Northern sky, including measurements of the diffuse emission, SWGO will extend the survey to the whole Southern sky. The SWGO design aims to achieve improved angular and energy resolution, as well as enhanced gamma–hadron discrimination relative to current water Cherenkov detectors, which is expected to translate into increased sensitivity for the detection and characterization of diffuse gamma-ray emission.

In Figure~\ref{fig:diffuse_lhaaso_mask}, the diffuse emission measurement by LHAASO is compared to different interstellar emission models (IEMs), in particular, ``CRINGE'' \citep{2023ApJ...949...16S} and ``varmin-rescaled'' \citep{2024JCAP...10..081A} that feature homogeneous and inhomogeneous CR transport, respectively. Interestingly, in this large region, which is far from the contamination of known sources, as considered in previous studies by LHAASO, the two models are barely distinguishable.
Once the contribution from the unresolved source is added \citep[considering the population model of][]{2024JCAP...10..081A}, both models can account for the diffuse emission flux measured by LHAASO within uncertainties. In order to distinguish between these two classes of models and further study the CR transport properties, it is necessary to probe the innermost region of the Galaxy at smaller scales, which should be achieved with SWGO.
 
\begin{figure}
    \centering
    \includegraphics[width=\columnwidth, trim={0.28cm 0.44cm 0.1cm 0.56cm},clip]{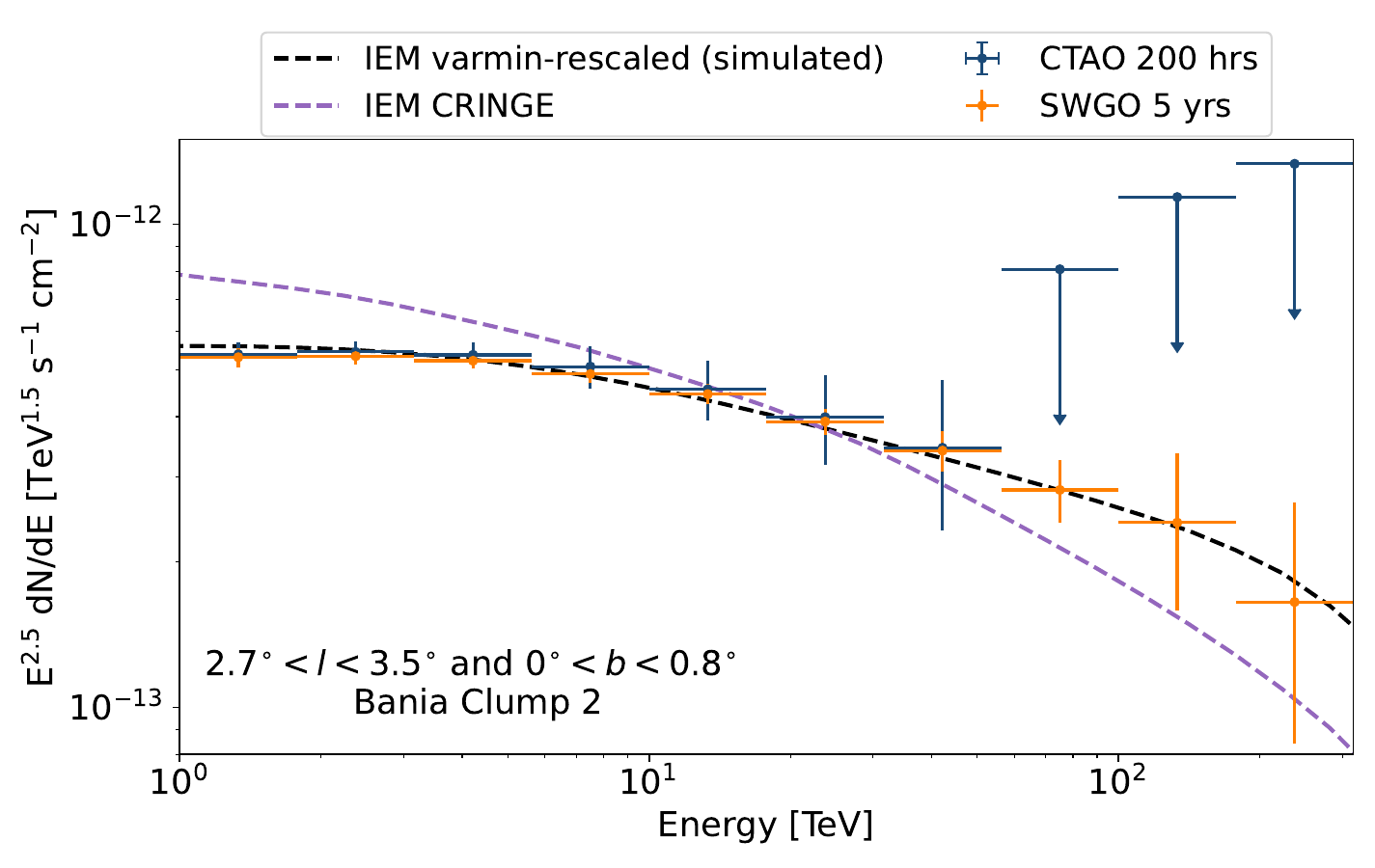}
    \caption{Differential flux of the diffuse emission toward Bania Clump~2 measured by SWGO in five years and 200 hours of CTAO observation. SWGO could significantly detect this molecular cloud up to 300~TeV. The two interstellar emission models considered featuring different CR transport properties could be discriminated even above 50~TeV with SWGO.}
    \label{fig:diffuse_bania_clump2}
\end{figure}

Indeed, SWGO will have the capacity to detect diffuse emission down to scales of \,$0.1^\circ$--$3^\circ$ in the inner Galaxy, and $1^\circ$--$10^\circ$ in the outer Galaxy. We define the detectability scale as the radius of the integration surface required to detect diffuse emission with TS > 25. It is derived by computing residual TS maps after subtraction of the CR background and simulated sources for a set of correlation radii (0.1, 0.3, 1, 3, and 10 degrees). In Figure~\ref{fig:iem_detectability}, the contour levels at TS~>~25 indicate the regions where the diffuse emission can be detected within each of the given correlation radius. It shows that SWGO can resolve individual molecular clouds (MCs) in the inner Galaxy at TeV energies, just as the \emph{Fermi}-LAT can at GeV energies. SWGO will detect gamma rays from giant MCs satisfying $A= M_5/d_{\rm kpc}^2 \gtrsim 0.5$, where $M_5$ is the cloud mass in 10$^5 M_{\odot}$ and $d_{\rm kpc}$ is its distance in kpc, assuming that the MC is illuminated by a CR spectrum similar to local measurements \citep{2022A&A...659A..57P}. Figure~\ref{fig:diffuse_bania_clump2} shows, as an example, that SWGO would be capable to detect the Bania Clump~2 \citep{1977ApJ...216..381B, 1986ApJ...306L..17S,2010ApJ...721..137B}, a well-known molecular cloud of the inner Galaxy highly deficient in star-formation activity, and probe its spectrum up to 300~TeV. We confirmed that there is no known source, or synthetic source, coincident with the cloud in our simulation. In regions where source contamination is negligible, as in this example, SWGO would not only detect individual MCs but also be able to discriminate between models featuring different CR transport properties.

\subsubsection{Fermi Bubbles} 

The Fermi Bubbles are among the most prominent large-scale gamma-ray structures in our galaxy. These massive lobes extend nearly 50$^{\circ}$ above and below the Galactic Center (GC), spanning a total diameter of approximately 10 kpc~\citep{su2010giant,ackermann2014spectrum}. 
Their characteristic hardness (around $E^{-2}$ in the gamma-ray range) challenges our understanding of particle acceleration processes occurring within the Milky Way, and suggests the presence of highly energetic particles that interact with interstellar matter and/or radiation fields~\citep{yang2018unveiling,crocker2011fermi,2024A&ARv..32....1S}.

Both the origin and the nature of the radiating particles in the Fermi Bubbles are important open questions in high-energy astrophysics. In fact these two questions are closely coupled. The competing scenarios for energizing the bubbles are associated with different timescales and hence different physical processes. 

In one scenario, the central supermassive black hole of our galaxy (Sgr A*) went through a period of much greater accretion a few million years ago, leading to Active Galactic Nuclei (AGN)-like jet activity that injected a significant amount of energy into the Milky Way’s halo~\citep{su2010giant,yang2018unveiling}. This scenario suggests that the Fermi Bubbles are relics of a past AGN phase, with accelerated particles producing gamma rays through IC scattering and hadronic interactions. Observations indicate that the Fermi Bubbles exhibit a surprisingly hard gamma-ray spectrum from 1 GeV to 100 GeV, with no detected spectral cutoff below 100 GeV. This spectrum remains relatively uniform across both northern and southern lobes, which is difficult to explain within standard Galactic CR propagation models.

The alternative scenario involves much slower inflation of the bubbles powered by ongoing star formation activity, and associated particle acceleration, over Gyr timescales. In this case, the bubbles are driven by large-scale winds originating from intense nuclear starburst activity at the GC~\citep{su2010giant,yang2018unveiling}. The very long radiative timescales for accelerated protons and nuclei in the bubbles are then matched to the overall lifetime of the system and become efficient in terms of gamma-ray production.

\change{SWGO is uniquely suited, thanks to its wide-field of view and continous coverage, to extend these studies to the multi-TeV regime, where the Fermi Bubbles have not yet been detected and where their spectrum would constrain the maximum energy of the radiating particles. At the SWGO site, the base of the bubbles culminates within $\sim$10$^\circ$ of zenith (compared with $z\gtrsim42^\circ$ at HAWC), and the southern bubble, never observed by a wide-field VHE instrument, at $z\lesssim35^\circ$. Scaling the published HAWC upper limits on the northern bubble (290~days;~\citealt{hawc2017fb}) with the SWGO effective area and gamma/hadron separation (Figs.~\ref{fig:gamma-hadron} and~\ref{fig:eff_area}), 5~years of SWGO data would reach limits of $\lesssim$$2\times10^{-8}$ and $\lesssim$$6\times10^{-9}$~GeV~cm$^{-2}$~s$^{-1}$~sr$^{-1}$ at 2 and 20~TeV, an order of magnitude below the published HAWC limits (a factor of $\sim$4--7 for equal exposure), neglecting systematic uncertainties in large-scale background estimation. This will allow SWGO to test the tentative $\sim$1~TeV cutoff at the base of the bubbles~\citep{hess2021fb}, provide the first VHE constraints on the southern bubble, and shed light on the activity history of the supermassive black hole at the center of our Galaxy.}

\subsubsection{The Galactic Center and the Central Molecular Zone}\label{sec:J1745} 

\begin{figure}
    \centering
    \includegraphics[width=\linewidth]{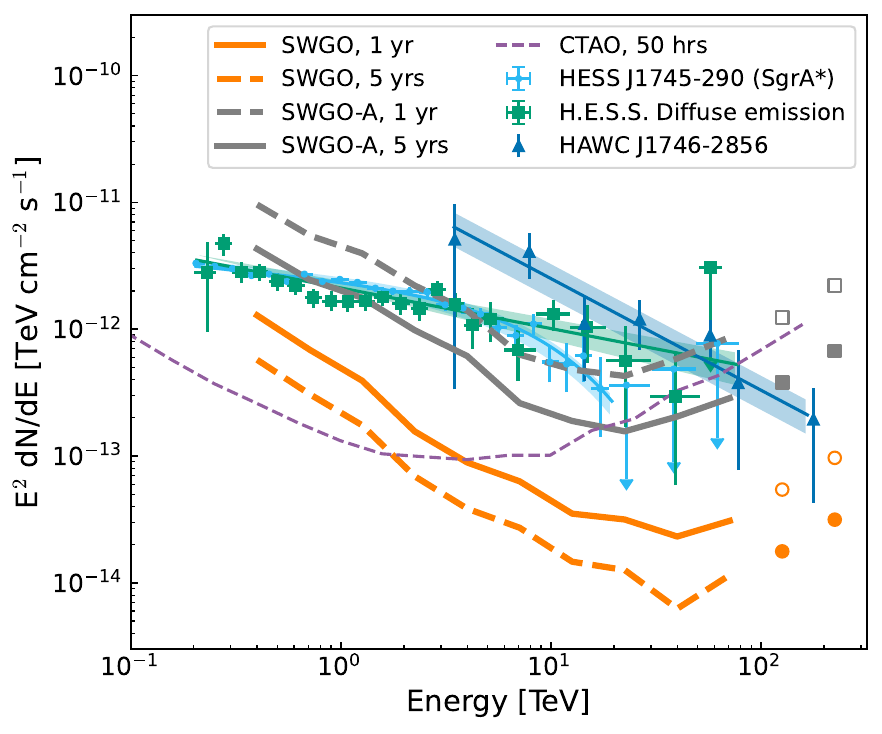}
    \caption{The measured gamma-ray flux and spectrum for the central source (Sgr A*) and the diffuse emission from the CMZ as seen by H.E.S.S., and the gamma-ray flux and spectrum of HAWC J1746-2856 measured by HAWC. Differential point-source sensitivity curves for SWGO for one year and five years of observation time are shown, as well as the sensitivity curves for CTAO (South) and HAWC for comparison.}
    \label{fig:gc}
\end{figure}

The center-most $\sim$200~pc of our Galaxy is a very rich region populated by dense MCs forming the central molecular zone (CMZ), SNRs like the composite SNR~G0.9+0.1, massive stellar clusters like the Quintuplet, the Arches, and the Nuclear clusters, where active star formation is ongoing, and the supermassive black hole Sgr~A$^{\star}$. Disentangling the sources and processes active in this complicated region remains a long-standing challenge in gamma-ray astronomy. 

Observations using H.E.S.S. revealed diffuse VHE gamma-ray emission correlated with the dense gas in the CMZ and provided evidence for the presence of a PeVatron in the GC region \citep{HESSGC}. In addition to the diffuse component, H.E.S.S. detected a bright central source spatially coincident with Sgr~A$^{\star}$ (HESS~J1745$-$290), whose spectrum is well described by an exponential cutoff power law with a cutoff energy of approximately 10~TeV \citep{HESSGC}. The origin of this source remains under debate, with possible contributions from the supermassive black hole itself and other accelerators located in the central parsecs.

Subsequent studies with larger H.E.S.S. datasets found evidence for a more complex spectral behavior of the diffuse CMZ emission, with indications of a spectral steepening when larger regions of the CMZ are considered \citep{2018A&A...612A...9H}. Similar evidence for spectral curvature has also been reported by MAGIC \citep{2020A&A...642A.190M}, while VERITAS observations remain compatible with a power-law description within uncertainties \citep{2021ApJ...913..115A}. As a result, the existence and energy scale of a spectral cutoff or break in the diffuse CMZ emission remain open questions.

Recently, HAWC reported the detection of the GC region after seven years of observations \citep{2024ApJ...973L..34A}. Although the limited angular resolution at the large zenith angles of these observations does not allow a clean separation of the various point-like and diffuse components, the detection extends the observed emission toward $\sim$100~TeV energies (Figure~\ref{fig:gc}). Taken together with previous measurements, these results suggest that a more detailed characterization of the spectral shape at the highest energies is needed.

With its location in the Southern Hemisphere, SWGO will be the optimal ground-based particle detector for observations of the GC region, which passes close to zenith and can be monitored up to energies of several hundred TeV. As shown in Figure~\ref{fig:gc}, SWGO will detect the GC region in less than one year of observations and provide an unprecedented characterization of the spectral shape at the highest energies, particularly for the diffuse emission from the CMZ. These measurements are crucial for understanding particle acceleration and transport in the GC, constraining the three-dimensional distribution of dense molecular clouds in the CMZ, and assessing the evidence for PeV cosmic-ray acceleration in this unique environment.

\subsection{Ultra-High-Energy Sources and PeVatrons}\label{sec:PeVatron} 

The concept of a ``PeVatron'' refers to astrophysical accelerators of particles — either electrons, positrons or protons — to energies of at least 10$^{15}$ electronvolts (PeV). This definition is fundamentally determined based on the largest rigidity ($\mathcal{R}=E/Z$ with $Z$ the atomic charge) that particles can attain at a given acceleration site. PeVatrons can be divided into two categories: ``hadronic PeVatrons'' and ``leptonic PeVatrons.'' In hadronic PeVatrons, gamma-ray emission predominantly results from proton-proton (pp) and proton-photon (p$\gamma$) interactions via the decay of neutral pions, with neutrino emission from charged pions \cite[collisions between heavier nuclei may be approximated as bundles of nucleons that equally share the kinetic energy. See, for example,][]{2022A&A...661A..72B}. Approximately 10$\%$ of the energy from the parent proton is transferred to these secondary particles, while the exact energy fraction varies with the spectral distribution of the primary protons \citep{2006PhRvD..74c4018K, kafexhiu2014}. The dominant gamma-ray emission mechanism in leptonic PeVatrons is IC scattering, whereby relativistic electrons up-scatter ambient photon fields, primarily diffuse infrared, and the 2.7~K Cosmic Microwave Background (CMB). For electrons or positrons in the Klein-Nishina regime, the up-scattered photon energy satisfies
$E_{\gamma}=0.37(E_{\text{e}}/1~\text{PeV})^{1.3}$~PeV \citep{crab_pev}, implying that PeV electrons can produce gamma rays with energies $\sim$370 TeV. 
UHE gamma-ray emission emerges as the key observational probe for exploring PeVatron sources \citep[see, for example,][for reviews.]{Fuente2023ThePA, cta_pevatrons, ozi_review, pev_annual_rev, deOnaWilhelmi2024}.

From the first report by the HAWC collaboration listing three confirmed UHE sources \citep{hawc_56TeV}, the latest catalog from the LHAASO collaboration now reports 43 UHE sources emitting above 100~TeV, all detected with statistical significance beyond 4$\sigma$ \citep{2024ApJS..271...25C}. Moreover, the firm detection of photons exceeding 1~PeV energies from the Crab Nebula \citep{crab_pev}, along with the Cygnus Cocoon, a superbubble enveloping a massive SFR \citep{cygnusfermi,2021NatAs...5..465A}, provides proof that PeV-energy particles are accelerated somewhere in these environments \citep{lhaaso2021}. Regardless of whether the emission is hadronic or leptonic in origin, these findings firmly establish the Crab Nebula and Cygnus Cocoon as Galactic PeVatron sources. A wide range of astrophysical objects, including SNRs \citep[e.g.,][]{bell2013, celli_SNR, 2022PhRvL.129g1101F}, massive star clusters (MSCs) and superbubbles \citep[e.g.,][]{aharonian2019, MorlinoSC,Vieu2022}, PWNe \citep[e.g.,][]{amato2003,Ohira2018,AmatoOlmi21}, and microquasars \citep{microquasars,  vg_2024,LHAASOmicroquasar}, are considered promising targets for future PeVatron searches. Additionally, all of these sources are underexplored in the UHE Southern sky, where SWGO will monitor them.

The search for PeVatrons is directly connected to the long-standing mystery of the origin of CRs. The CR spectrum follows a power-law (PL) distribution with an index of $\sim$$-2.7$ above $\sim$30~GeV, transitioning at $\sim$3~PeV to a steeper slope of $\sim$$-3.0$, forming the well-known ``knee'' feature \citep[e.g.,][]{blumer_rev}. 
Despite advances in measurements of CRs arriving at Earth up to and beyond the knee, it remains unclear which acceleration sites produce them.
This limitation is well known, as particles are randomly scattering in the Galactic magnetic field before arriving at Earth. In the PeV regime, it is thus necessary to look for neutral messengers such as UHE gamma rays and neutrinos.
Despite recent groundbreaking advancements in identifying PeVatrons, no Galactic source has yet been unambiguously confirmed as an accelerator of \emph{hadrons} to PeV energies, largely due to the difficulty in disentangling hadronic and leptonic emission scenarios. Recent studies suggest that UHE gamma-ray emission may be a universal feature of the environments of powerful pulsars, particularly those with high spin-down luminosity \citep[e.g.,][]{Albert_2021,2021ApJ...908L..49B, pevatron_pulsars}, indicating that the UHE source population is dominated by PWNe. Since IC scattering of $\sim$PeV electrons on the CMB can readily produce UHE gamma rays, detecting UHE emission up to 1~PeV alone cannot conclusively establish a hadronic origin \cite[see, for example,][]{2021ApJ...908L..49B}. Consequently, multi-wavelength and multi-messenger approaches become crucial for resolving this ambiguity. A clear correlation with dense target material, such as a MC, would support a hadronic scenario. However, an undeniable confirmation of hadronic PeVatron activity would come from the simultaneous co-detection of neutrinos and UHE gamma rays. 

Ultra-high-energy gamma rays with energies beyond 200–300~TeV are also subject to attenuation through $\gamma\gamma$ absorption, primarily due to interactions with interstellar radiation fields, particularly the CMB \citep[e.g.,][]{absorption_vernetto, absorption_Lhaaso}. This results in a reduction of UHE gamma-ray flux detected at Earth especially at these high energies, and potentially introducing artificial features into the observed spectra. Consequently, for distant sources such as the GC, these absorption effects must be corrected for an accurate interpretation of the observed spectra. The angular resolution of WCD experiments, typically around 0.2$^{\circ}$--0.3$^{\circ}$, is 3--4 times worse when compared to current IACT experiments. This reduced resolution means source confusion is an issue, making it difficult to identify the precise origin of emission through morphological analysis. Hence, combining the strengths of both methods becomes crucial when studying PeVatrons. While SWGO will be effective at identifying regions containing PeVatron activity, CTA South will offer precise localization of acceleration sites within these broader regions, allowing for detailed morphological studies.

\begin{figure*}[ht!] 
\centering 
\includegraphics[width=\linewidth]{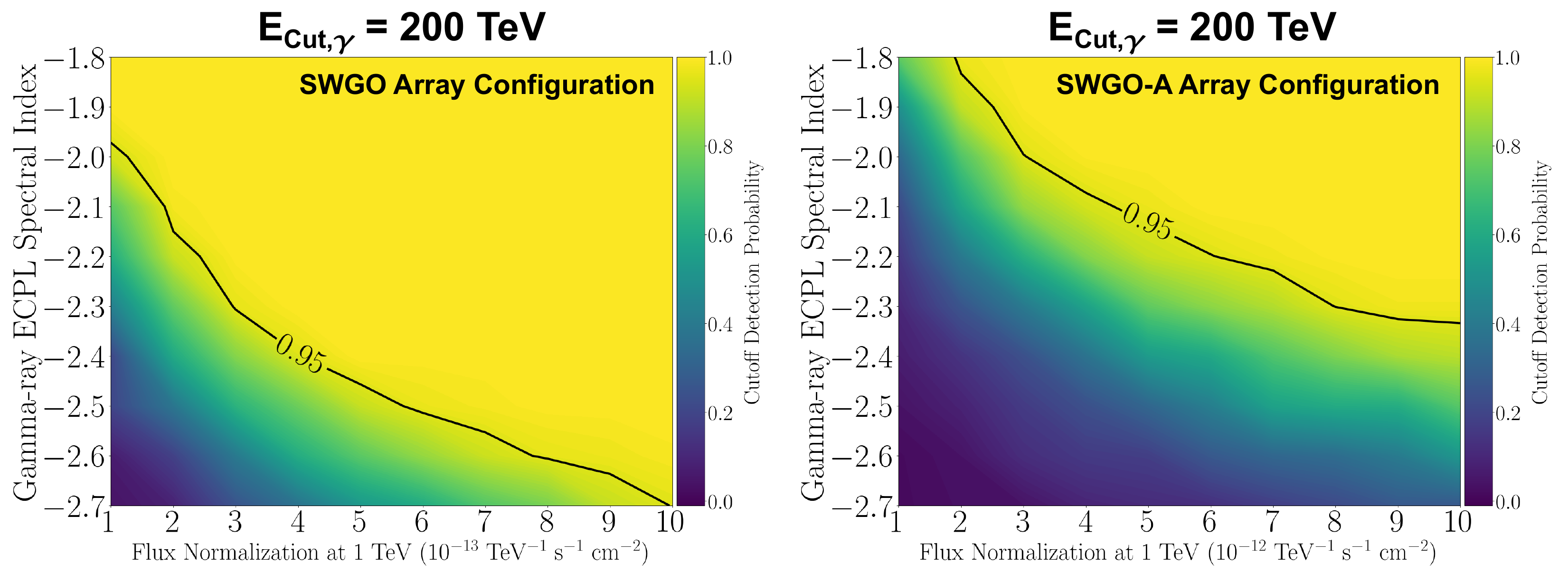} 
\caption{Spectral cutoff detection probability maps for five years of SWGO observations of point-like sources, assuming a spectral gamma-ray cutoff energy of E$_{\text{Cut},\gamma}=200$~TeV. The abscissa shows the true flux normalization, $\phi_0$, at 1 TeV and the ordinate shows the true spectral index, $\Gamma$, of an ECPL model. A TS$_{\lambda}$ cutoff detection threshold of 25 (corresponding to 5$\sigma$ detection level) is assumed. The color code shows the $\gamma$-ray spectral cutoff detection probability, while the black line shows the cutoff detection probability contours at 0.95 confidence level. The left and right panels show the spectral cutoff detection probability maps for the SWGO and SWGO-A array configurations, respectively. Note that the flux normalization factors have different scales for the left ($\times$10$^{-13}$) and right ($\times$10$^{-12}$) panels.} 
\label{cutoff_detmaps} 
\end{figure*}

SWGO promises significant advancements in future UHE observations \citep{swgo_pevatrons}, particularly in the Southern Hemisphere. SWGO is expected to surpass the sensitivity of 50-hour CTA observations in detecting point-like gamma-ray sources above 10~TeV, see Figure~\ref{fig:sensitivity}. In conjunction with existing LHAASO and HAWC observations, future SWGO data will enable the first complete UHE skymap of the Galactic Plane. Consequently, SWGO is set to play a crucial complementary role in Galactic Plane surveys and to work in synergy with LHAASO, HAWC, and the future CTA project.

Beyond the GC (see Section~\ref{sec:J1745} for detailed discussions), the southern sky also hosts other promising UHE sources, identified through either their high-energy fluxes or hard PL spectra without clear cutoffs, as observed by the H.E.S.S. telescopes. Among them, Westerlund~1 \citep{westerlund1_latest}, a young and massive stellar cluster where photons up to $\sim$81~TeV have been observed, stands out due to its extended, shell-like gamma-ray morphology spanning up to 2$^{\circ}$, making it an ideal target for SWGO observations. Additionally, unidentified hard gamma-ray sources, such as HESS~J1702$-$420A \citep{j1702}, with an exceptionally hard spectral index of $\gamma=1.53$, and HESS~J1641$-$463 \citep{j1641}, both exhibit hard spectra, have dense target material present along the line of sight at various distances, and are strong UHE source candidates potentially related to hadronic PeVatrons. Furthermore, the Southern sky may also host dark UHE sources, such as LHAASO~J2108$+$5157 \citep{j2108}, which emits gamma rays only above $\sim$20~TeV also showing significant E~>~100~TeV emission, yet remains undetected by current-generation IACTs, although a marginal ($\sim$3.7$\sigma$) point-like excess above 3~TeV has been reported by LST-1 \citep{lst1_j2108}.

To investigate SWGO’s sensitivity to spectral cutoffs from point-like PeVatron sources, Monte Carlo simulations were performed using IRFs corresponding to the SWGO and SWGO$-$A array configurations (see Figure~\ref{fig:sensitivity} and Section~\ref{sec:design}). Gamma-ray spectra were simulated assuming an Exponential Cutoff Power-Law (ECPL) model, expressed as:
\begin{equation} \label{ecpl_model} \Phi(E_{\gamma}) = \Phi_{0}\left(\frac{E_{\gamma}}{E_{0}}\right)^{-\Gamma_{\gamma}}\exp\left[-\left(\lambda_{\gamma} E_\mathrm{\gamma}\right)^{\beta_\mathrm{\gamma}}\right]\enspace, \end{equation}
where $\Phi_{0}$ is the normalization at reference gamma-ray energy of E$_{0}$, $\Gamma_{\gamma}$ is the gamma-ray spectral index, $\lambda_{\gamma}$ = (1/E$_\mathrm{\text{cut},\,\gamma}$) is the inverse cutoff energy, and $\beta_{\gamma}$ is the sharpness parameter describing the rate of exponential decay. This model is motivated by a parent proton spectrum that follows an ECPL form with a proton cutoff energy of E$_{\text{p},\mathrm{\text{cut}}}=3$~PeV and $\beta_{\text{p}}=1.0$, which characterizes a typical hadronic PeVatron source that contributes to the observed CR spectrum, especially at energies around the knee region. The resulting gamma-ray spectrum from the pp interactions follows the ECPL shape given in Eq.~\ref{ecpl_model} (see \cite{celli_2020} Eq.~17, Eq.~22, and Eq.~26). A log-likelihood ratio test, comparing a pure PL model to an ECPL model, is employed to quantify the statistical significance of gamma-ray spectral cutoffs, and defined as: 
\begin{equation}
\label{ts_lambda}
\mathrm{TS}_{\gamma}=-2\ln\frac{\hat{L}(\lambda_{\gamma}=0)}{\hat{L}(\lambda_{\gamma})}\enspace,
\end{equation}
where $\hat{L}(\lambda_{\gamma})$ and $\hat{L}(\lambda_{\gamma}=0)$ represent the maximum likelihoods over the full parameter spaces. Spectral cutoff detection maps, following the methodology introduced and detailed in \cite{cta_pevatrons}, have been produced for a broad range of ECPL flux normalization and spectral index parameters, with the cutoff energy fixed at E$_\mathrm{\text{cut},\,\gamma}=200$~TeV. These maps, demonstrating the sensitivity of SWGO to 200~TeV spectral cutoffs, are produced for SWGO and SWGO-A array configurations, and are shown in the left and right panel of Figure~\ref{cutoff_detmaps}, respectively. The figure demonstrates that the SWGO configuration exhibits strong sensitivity across a substantial range of PeVatron source parameters, even for soft sources. In contrast, the SWGO-A configuration shows a considerably lower capability for detecting a spectral cutoff at 200~TeV energies, performing approximately 25$-$30 times worse than the SWGO configuration throughout the investigated parameter space. These maps highlight the critical role of SWGO in identifying and characterizing PeVatron candidates, particularly in the Southern Hemisphere, where its capabilities will complement those of existing and future gamma-ray observatories. By bridging observational gaps and enhancing spectral cutoff detection at UHEs, SWGO is expected to play a key role in understanding PeV particle acceleration and significantly improve our understanding of Galactic PeVatrons.

\subsection{Galactic Populations}\label{sec:gal_pop}

\subsubsection{Pulsars, Pulsar Wind Nebulae, and TeV Halos}  
\label{sec:PWN} 

Pulsars and their environments are responsible for the largest fraction of identified Galactic gamma-ray sources throughout the GeV--TeV range \citep{tevcat} and are also the predominant source class associated with UHE Galactic objects observed at E~>~100~TeV by LHAASO \citep{Cao_2024_1lhaaso}. 

Due to pulsar's strong magnetic field and rapid rotation, charged particles extracted from the pulsar surface are accelerated and emit high energy photons, generating a pair-cascade process that produces copious amounts of electron-positron pairs. Under some circumstances, ions could potentially also escape the pulsar surface in non-trivial quantities \citep{2025arXiv250201318S}. Beyond the pulsar magnetosphere, 
the particles can escape and stream out into a magnetised pulsar wind \cite[see][for reviews]{2009ASSL..357..421K,2022ARA&A..60..495P}. There, they can be further accelerated at the wind termination shock (WTS), which is produced where the confining pressure of the (comparatively dense) surrounding medium reaches equilibrium with the pulsar wind's momentum flux \citep{1984ApJ...283..710K,nodes2008particle}.

Pulsed gamma-ray emission has been detected 
from four pulsars at energies from a few GeV up to TeV using ground-based IACTs. Notably, the MAGIC telescopes have detected pulsed emission from the Crab pulsar as a PL tail extending from the GeV bump observed by the \emph{Fermi}-LAT up to energies of a few TeV \citep{2016A&A...585A.133A}. In contrast, observations by H.E.S.S.~have revealed clear pulsed emission from the Vela pulsar up to 20~TeV \citep{vela_nature}, associated with the P2 pulse, characterized by a distinct spectral component separate from the GeV emission \citep{2023NatAs...7.1341HESS_Vela}. 

To evaluate the potential of SWGO to detect pulsar TeV emission, the observed pulsed emission from Vela based on the H.E.S.S. measurements was simulated, adopting an ECPL model with cutoff arbitrarily set at $40$~TeV, given that no spectral cutoff has yet been detected by H.E.S.S. \citep{2023NatAs...7.1341HESS_Vela}. The simulation results demonstrate that SWGO would achieve a clear detection of pulsations ($>5\sigma$) within one year of data collection. Moreover, a simulated five-year exposure would allow the identification and characterization of the spectral cutoff energy. Additionally, synthetic emission featuring a softer index of 1.9, a cutoff at 20~TeV, and a flux similar to the Vela pulsar was simulated. Depending on the On-phase to Off-phase ratio, such emission could be detected by SWGO in 1 to 5 years of observations.  

Given these findings, SWGO will provide important observations and constraints for pulsar astronomy in the TeV regime. Its advantageous location in the Southern Hemisphere grants excellent access to the Galactic Plane. In combination with SWGO's dedicated survey capabilities, this will yield good coverage and constrain the spectral cutoff for an increasing number of TeV-emitting pulsars. 

Beyond the WTS, a nebula of energetic particles forms, predominantly electrons and positrons\footnote{Hereafter referred to collectively as electrons, unless otherwise specified.}, that is gamma-ray bright due to IC scattering of these particles on the ambient radiation fields \citep{GaenslerSlane2006review}. PWNe consequently form one of the most common Galactic source classes known at TeV energies and will thus be a dominant component of the SWGO view of the Galactic Plane \citep{HESS_GPS,Cao_2024_1lhaaso}. 

While PWNe are believed to be leptonicly dominated, recent observations of the Crab PWN \citep{lhaasocrab} have suggested that there could be a subdominant hadronic component to the resulting gamma-ray emission at the highest energies for a small number of PWNe. 
SWGO's excellent sensitivity at high energies will help to distinguish between these possible scenarios for PWNe beyond the Crab.

Eventually, the expanding PWN may encounter a reverse shock from the progenitor SNR, causing the system to become disrupted. 
Particles escaping from the PWN following this stage may diffuse into the surrounding ISM, leading to the formation of a halo of energetic leptons, which is made visible on large scales due to further IC scattering. These pulsar halos (also known as TeV halos) were first identified in data from the wide-field HAWC observatory and have since been noted as an expected feature of PWN evolution \citep[e.g.,][]{giacintihalo}. The distinction between PWN and halo has been a matter of debate in recent years \citep[e.g.,][]{giacintihalo,sudohhalo}, yet extended gamma-ray emission around energetic pulsars appears to be commonplace \citep[e.g.,][]{2018A&A...612A...2HESSpwnpop,2024ApJS..271...25C,HAWC:2023jsq, Albert:2025gwm}. 

\begin{figure}
    \centering
    \includegraphics[width=\columnwidth, trim={0cm 0 0cm 0cm},clip]{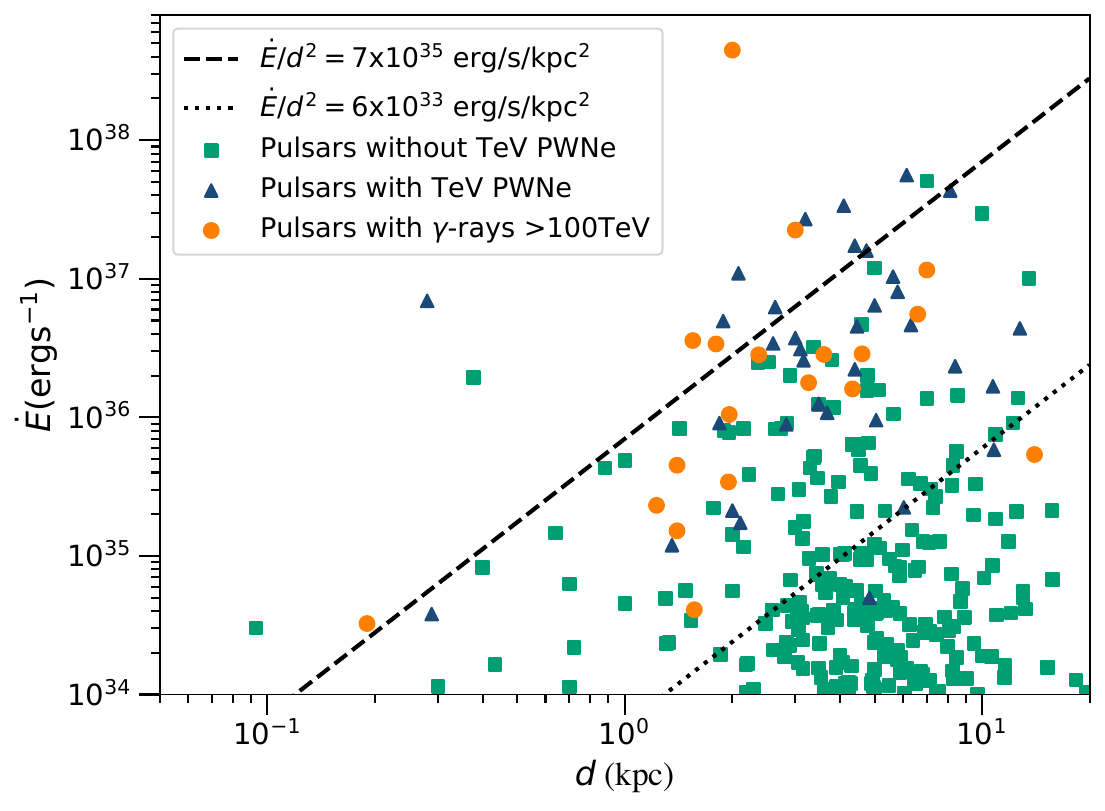}
    \caption{Population of pulsars in the ATNF, showing their spin-down luminosity as a function of distance \citep{ATNF}. Green squares indicate pulsars without known associated TeV emission, many of which are located in the Southern sky. Dashed and dotted lines indicate constant $\dot{E}/d^2$, roughly relating (under model assumptions) to the anticipated sensitivity after five years for SWGO-A and for the full SWGO, respectively. } 
    \label{fig:psrpop}
\end{figure}

Nearby halos, within $\sim$1\,kpc of Earth, exhibit degree-scale gamma-ray emission, enabling both the morphology and spectrum to be probed in detail. Wide-field instruments such as SWGO, and as demonstrated by HAWC, LHAASO, and Tibet-AS$\gamma$, are essential to capture the full extent of such emission. As electrons are transported away from the pulsar, they undergo energy losses, leading to distinct morphological and spectral profiles that can be used to infer properties of the particle transport, such as the diffusion coefficient normalization and the level of magnetic turbulence in the region \citep[e.g.,][]{Geminga_HAWC,LopezCoto_turbulence}. 

SWGO will contribute substantially to the study of PWNe and TeV halos by further surveying a wide variety of pulsar systems in the Southern sky in an unbiased manner. 
Using a synthetic population of pulsars, evolved using the COMPAS population synthesis code  
\citep{COMPASTeam:2021tbl}, a model of PWNe evolution based on \cite{2018A&A...612A...2HESSpwnpop} was applied to anticipate the likely gamma-ray emission along the Galactic Plane. Normalized in the Northern sky to HAWC data from the 3HWC catalog \citep{3HWC}, it is expected that SWGO will be able to detect no fewer than $\sim$30 PWN or TeV halo systems within $\sim$5 years of construction \citep{cozzolongoicrc}. 

Figure \ref{fig:psrpop} shows the currently known pulsar population as taken from the ATNF catalog \citep{ATNF}. The spin-down luminosity, $\dot{E}$, is a measure of the energy output by a pulsar over time due to loss of rotational energy and can be determined from the pulsar rotation via $\dot{E}=-4\pi^2I\dot{P}/P^3\,$,
where $P$ is the pulsar rotation period, $\dot{P}$ is the period's time derivative, and $I$ is the moment of inertia. 
Pulsars associated with TeV emission in the vicinity are indicated by blue triangles and those with emission reaching energies $\gtrsim $100\,TeV are indicated by orange circles. Green squares indicate pulsars for which no TeV gamma-ray emission has yet been identified. The sensitivity for a gamma-ray instrument is expected to scale as $\dot{E}/d^2$, where the relation between spin-down luminosity and anticipated gamma-ray flux is approximated using the model of \citet{2018A&A...612A...2HESSpwnpop}. This is a time-dependent, one-zone radiative model with a PL lepton-injection spectrum and spherically symmetric dynamical evolution including synchrotron, escape, and adiabatic losses \citep[see also][]{2012arXivMayerPWNmodel}.
Based on this model, lines in Figure~\ref{fig:psrpop} indicate the anticipated sensitivity of SWGO-A to pulsar $\dot{E}/d^2$ after 5 years (dashed) and for the full SWGO after five years (dotted). Green squares  above these lines therefore indicate the discovery potential for SWGO, as well as contributions to detailed studies of other known PWNe that lie above these thresholds. Whilst SWGO will be sufficiently sensitive to detect gamma-ray emission at the level anticipated by the PWN model of \citet{2018A&A...612A...2HESSpwnpop}, associating the emission unambiguously with these pulsars will require a careful joint analysis that takes other emission components, such as the Galactic diffuse emission, into account (see also Section~\ref{sec:diffuse}).

\subsubsection{Supernova Remnants}
\label{galsnrs}
 
Supernova Remnants are considered the most likely source of Galactic CRs, as their shocks are capable of accelerating charged particles, e.g., protons and heavier nuclei, and ejecting them into the surrounding medium. In some cases, these ejected CRs can interact with the ISM or radiation fields, leading to the production of gamma rays and neutrinos. 
Historically, it has been considered likely that SNRs could accelerate particles up to energies around the CR knee at $\sim$TeV, however, the lack of SNRs seen by the LHAASO experiment at UHEs is an encouraging consideration of alternative sources at the highest energies. 

Theoretical predictions based on the local supernova rate estimate the number of SNRs in the Galaxy to be more than 1,000. Currently, the number of detected SNRs is $\sim$300 \citep{snrcat}. The number of SNRs detected by radio observatories has grown significantly, with over 200 SNRs identified in the Southern Hemisphere through dedicated radio surveys, such as the SARAO MeerKAT Galactic Plane Survey (SMGPS) \citep[][respectively]{MeerKAT2024,MeerKATPlnSrv2025}.

The Southern Hemisphere offers a unique vantage point for observing certain SNRs, particularly those in the Galactic Plane. For example, young remnants like RX~J1713.7$-$3946 \citep{Acero_2017} and Vela Jr.~\citep{velajr} are more favorably positioned for observation from the Southern Hemisphere, allowing for detailed studies of their gamma-ray emission. SWGO and CTAO are both planned to be in the Southern Hemisphere, which will offer unprecedented opportunities to study UHE gamma-ray emissions from SNRs. These observatories will be capable of resolving spatial and spectral features of these sources with higher precision and energy sensitivity than any previous instruments.

\begin{figure}[!h] 
\centering 
\includegraphics[width=\columnwidth]{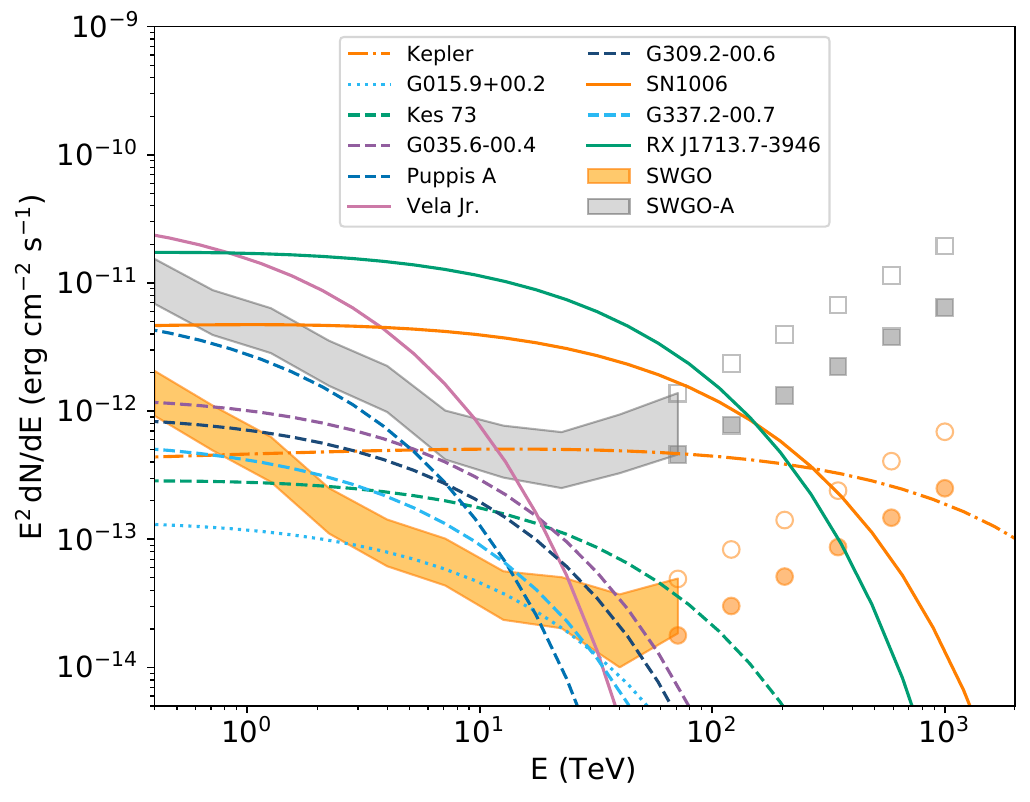} 
\caption{Predictions for the gamma-ray flux of SNRs within the observable sky for SWGO. SNRs detectable in one year with the SWGO-A array configuration are shown with solid lines and in five years by dot-dashed lines. Those SNRs detectable in only one year with the full array are shown with dashed lines, and those detectable with the full array in five years are shown with dotted lines. The upper limit of the shaded band shows the one-year point-source differential sensitivity, the lower bound the five-year sensitivity.} 
\label{fig:snrs} 
\end{figure}

To quantify the number of SNRs detectable to SWGO, the previous analysis from \cite{Scharrer2024} was repeated using the same simplistic model, but with updated sensitivity curves used throughout this work. The results of this are shown in Figure~\ref{fig:snrs}. With the full configuration, nine SNRs will likely have a flux detectable within one year and be within the SWGO FoV, with another one detectable in five years. Roughly half of these SNRs will likely have shell-like extended emission visible to SWGO \citep{Scharrer2024}.

SWGO, in particular, will observe a larger fraction of the sky compared to previous studies of the Galactic Plane in the TeV by H.E.S.S.~\citep{HESS_GPS}, which will likely lead to a greater number of SNRs detected. This will constrain the acceleration mechanisms, the identification of the particle acceleration regions within the remnants providing insights into the physical conditions of these sources, and, combined with multimessenger observations, offer a more complete picture of SNRs as particle accelerators.  

\subsubsection{Stellar Clusters and Associations}  

Most massive stars are formed in clusters and associations and these objects are of long-standing interest in the acceleration of cosmic rays \citep[e.g.,][]{1979ApJ...231...95M,1983SSRv...36..173Cesarsky}.
To date, only a few such sources have been clearly identified as TeV gamma-ray sources, such as Westerlund\,1, Westerlund\,2, and Cygnus OB2 \citep{2012A&AWesterlund1HESS,2007HESSwesterlund2}, whilst several more have been detected by the \textit{Fermi}-LAT \citep{2020ApJSFermi4FGL,2017A&AFermi_NGC3603}. However, they represent a significant scientific target for the next generation of gamma-ray observatories, especially SWGO. 

In addition to the CMZ (discussed Section~\ref{sec:J1745}), we consider two additional source types: OB associations and young massive star clusters (YMSCs). 
Considering their gamma-ray emission, these source types can be distinguished by their compactness. Compact clusters such as Westerlund~1 contain a large number of young, massive hot stars in a compact volume, whose winds interact and drive a collective large scale wind. OB association, such as Cygnus OB2, contain a loose association of hot stars, whose winds have sufficient space to create isolated shocks and redistribute the kinetic energy before a large scale collective wind can be formed \citep{2024MNRAS.532.2174V}. Both source types provide myriad particle acceleration sites, including the termination shocks of both individual stellar winds \cite[e.g.,][]{1982ApJ...253..188V} and large scale collective winds \cite[e.g.,][]{MorlinoSC}, as well as supernova remnants. Young clusters and associations hosting massive stars will have a large number of powerful supernova events in the first several Myrs of their lives, during which they may act as PeV CR sources \citep{2023MNRAS.519..136V}. LHAASO has reported multi-PeV photons coincident with the position of Cygnus OB2 \citep{2024SciBu..69..449L}, making such sources of great interest.

SWGO will provide an unprecedented chance to study both Galactic and extragalactic (in the Large Magellanic Cloud (LMC)) with improved sensitivity in the 100~TeV range. Natural candidates for observation include objects already detected in gamma rays, such as Westerlund~1 and W49A \citep{2012A&AWesterlund1HESS,Brun2010}. In this regard, SWGO has a unique advantage due to its large FoV. The study of Westerlund~1 and W49A with SWGO will enhance our understanding of other well known SFRs, such as the Carina nebula~\citep{etaCarHESS}, or star forming regions in the LMC \citep{2024ApJ...970L..21A}.

Projections for the gamma-ray emission from young massive stellar clusters, based on recent catalogs of data from the Gaia satellite, suggest that $\sim$10--20 stellar clusters in the Milky Way could be gamma-ray bright \citep{2026ApJ..1000..279Mitchell_SCs}. Such systems are good targets for wide-FoV instruments such as SWGO, due to the large anticipated size of the wind-blown bubble. Despite significant integrated gamma-ray fluxes, the degree-scale angular size of the emission will yield comparatively low surface brightness. SWGO is therefore ideal to search for such extended, diffuse structures at high energies. By detecting or placing constraints on a variety of these systems, SWGO will be able to clarify the nature of the TeV-emitting stellar cluster population, including which properties of these stellar cluster and their environments are most conducive to particle acceleration in their winds. 

\subsubsection{Binaries and Microquasars}

X-ray binaries are systems in which a compact object acquires matter from a nearby companion star. Broadly speaking, they can be classified into two types: (1) Low-mass X-ray binaries (LMXBs), where the companion star has a lower mass than the compact object, and (2) High-mass X-ray binaries (HMXBs), where the compact object is the less massive component. In general, accretion in LMXBs is interpreted to occur through an accretion disk, while in HMXBs it typically results from the strong stellar winds of the companion star. However, in some HMXBs, material transfer occurs via an accretion disk, especially when the companion star has a Be spectral type, as stellar winds in these stars are not particularly strong.

Occasionally, these systems can exhibit jets of material traveling close to the speed of light. In such cases, the system is referred to as a microquasar, a name given due to its resemblance to Active Galactic Nuclei (AGN) \citep{Mirabel_1998}. A correlation has been found between the luminosity in the radio and X-ray bands and the mass of the compact object \citep[e.g.,][]{Flacke_2004}, further strengthening the similarities between AGN and microquasars. Long-term radio observations of these sources have linked the presence of jets to specific accretion states. During a ``hard state''—when the spectrum is dominated by energy losses through IC scattering, peaking around 100~keV—continuous outflows, known as ``compact jets,'' are observed in the radio band. Conversely, discrete ejections, sometimes appearing ``superluminal,'' are associated with the ``soft state,'' where the spectrum is dominated by thermal emission from the accretion disk, peaking around 1~keV. \citep[see, e.g.,][for further details]{Remillard_2006, Belloni_2011, Corbel_2013}.

If microquasars share similar mechanisms of particle acceleration and jet formation with AGN, known to be highly efficient regions for particle acceleration and to produce broadband non-thermal emission (radio to VHE gamma-ray), it is reasonable to consider that microquasars could also act as sources of VHE gamma-ray emission. Emission in the GeV and TeV bands has been predicted for some time \citep[e.g.,][]{Aharonian_2004, Bosch_2009}, where it is attributed to particle acceleration within the jets. Currently, a little over 20 microquasars are known in the Galaxy \citep{Corral_2016}. Only three of these have been detected at GeV energies by the \emph{Fermi}-LAT: Cyg X-3 \citep[e.g.,][]{Fermi_2009, Zdziarski_2018}, where orbital variation in gamma-ray emission has been detected, and Cyg X-1, where gamma-ray emission coincides with a hard X-ray state \citep{Zanin_2016,Zdz_2017}, both classified as HMXBs; and SS 433 \citep{Xing_2019, Li_2020}, where variable flux was observed correlating with the precession period of the inner jet. However, the physical process driving this behavior remains unclear. In the VHE range, five microquasars have been detected: SS~433 \citep{Abeysekara_2018}, V4641~Sgr \citep{vg_2024}, GRS~1915+105, MAXI~J1820+070, and Cygnus~X-1~\citep{LHAASOmicroquasar}.

The detection of TeV emission from the jets of the microquasar SS~433 represents the first time astrophysical jets were resolved at TeV energies. TeV emission was first detected by HAWC~\citep{HAWCSS433}, which prompted a follow-up campaign with the H.E.S.S. telescopes~\citep{HESS_2024}. The gamma-ray emission is extended and closely follows the shape of previously known X-ray structures~\citep{geldz_1980}. Modeling of the X-ray and TeV emission suggests that it originates from efficiently accelerated electrons at the base of the extended X-ray jets~\citep{ss433_2024, HESS_2024}. More recently, the LHAASO collaboration reported emission up to 300 TeV~from the SS~433 system, which might require additional hadronic contributions to the emission~\citep{LHAASOmicroquasar}. 


Another example is the microquasar V4641 Sgr, which was identified as a TeV gamma-ray source by HAWC \citep{vg_2024}. The gamma-ray spectrum of this source is among the hardest observed, with photons detected above 200~TeV. Modeling suggests that this emission is the result of emission from high-energy protons, rather than electrons. More recently, the LHAASO collaboration reported emission up to 800~TeV from this microquasar, supporting the hadronic nature of the emission \citep{LHAASOmicroquasar}.

All the microquasar systems detected at TeV energies are located in regions of the sky visible from the Northern Hemisphere and were all first detected by wide-field instruments. Since the distribution of X-ray binaries does not show a preferred half of the Galaxy, it is safe to assume that this is simply because there is no unbiased gamma-ray instrument looking at the Southern sky. Consequently, SWGO would be expected to at least double the number of known TeV-emitting microquasars, providing a clearer view into the population. The SWGO sensitivity is expected to peak above tens of TeV, the ideal range to probe the nature of microquasars as sources of PeV cosmic rays. 

Binary stellar systems, which can comprise a variety of different combinations of physical components, often exhibit time-variable phenomena in their gamma-ray emission. Typically, this variation is either due to spectral modulation of the gamma-ray emission with orbital phase or gamma-ray emission only occurring during distinct phases of the binary orbital period. An example of the former is LS~5039, a binary system comprising a compact object and a massive O6.5V star that exhibits spectral modulation over its 3.9-day orbital period \citep{LS5039_hess}. At superior conjunction, the spectrum follows a PL shape, whereas at inferior conjunction, the flux at 1$-$10~TeV is considerably enhanced following a ECPL model. Whilst LS~5039 is observable from both the Northern and Southern Hemispheres, several well-studied binaries are located in the Southern sky. 

Beyond systems displaying accretion-powered jets, X-ray binaries also include a distinct class of gamma-ray emitters known as gamma-ray binaries, characterized by spectral energy distributions that peak above 1~MeV, powered by the interaction between a relativistic pulsar wind and the radiation field or stellar wind of a massive O- or Be-type companion star \citep{Dubus_2015}. The class of VHE gamma-ray binaries comprises only a handful of known objects in our Galaxy \citep{Abramowski_2013}, making every new detection of significant value for understanding the population. Several of these systems are located in the Southern sky and have been detected at VHE by H.E.S.S., including LS~5039, PSR~B1259$-$63/LS~2883, and 1FGL~J1018.6$-$5856 \citep{Aharonian_2005, Aharonian_2006, Abramowski_2013}.

Among these are PSR~B1259$-$63 / SS~2883, a binary system with a pulsar orbiting a massive Be star \citep{2003PASJ...55..473M}, and Eta Carinae, a colliding wind binary composed of two massive stars, likely a luminous blue variable and a Wolf-Rayet star \citep{1997DavidsonEtaCarLBV,2005IpingEtaCarWR}. The binary pulsar PSR~B1259$-$63 is the only gamma-ray binary in which the compact object is unambiguously identified as a pulsar, consisting of a $\sim$48~ms rotation-powered neutron star in a highly eccentric 3.4-year orbit. It exhibits gamma-ray emission around the periastron passage of its 3.4-year orbit \citep{psrb1259HESS}, with H.E.S.S. confirming a repetitive behavior across the 2004, 2007, and 2010/2011 periastron campaigns, with spectra extending up to $\sim$10~TeV \citep{Abramowski_2013}. H.E.S.S. has detected gamma-ray emission from the colliding wind binary Eta Carinae around periastron in two consecutive passages of the system with a 5.5-year orbital period \citep{etaCarHESS,etaCarHESS2}. In this system, the gamma-ray emission is understood to arise from interactions between a stellar wind comprising hadronic CRs and the surrounding Homunculus nebula that provides dense material for pp interactions during the periastron passage \citep{SteinmasslEtaCar}. 

A particularly well-studied example of a gamma-ray binary accessible only from the Northern hemisphere is LS~I~$+61^{\circ}303$, a Be/compact object system at declination $+61^{\circ}$. Its VHE emission exhibits both orbital modulation, with a period of $\sim$26.5~days, and a superorbital modulation, with a best-fit period of $1610 \pm 58$~days --- consistent with the $1667 \pm 8$~day radio period --- as established through a four-year MAGIC monitoring campaign \citep{Ahnen_2016}. Recent observations by LHAASO have extended the spectrum of this system up to $267 \pm 61$~TeV, detecting orbital modulation at energies between 25 and 100~TeV at $3.9\sigma$ confidence level, favoring a composite lepto-hadronic emission scenario and suggesting that this system operates as a Galactic PeVatron candidate \citep{LHAASO_2025}. The detection of emission above 100~TeV from LS~I~$+61^{\circ}303$ establishes it as the first confirmed gamma-ray binary operating as an UHE source, setting a new precedent for the class. Although its declination places it outside the field of view of SWGO, this result compellingly illustrates the scientific potential of searching for analogous systems in the Southern Hemisphere. SWGO, with its wide FoV and continuous coverage of the Southern sky, is ideally positioned to complement H.E.S.S. observations: it will enable uninterrupted orbital monitoring of known Southern systems such as LS~5039, PSR~B1259$-$63/LS~2883, and 1FGL~J1018.6$-$5856, overcoming the limitations imposed by the reduced duty cycle of pointed IACTs, while its sensitivity above tens of TeV makes it the ideal instrument to identify new gamma-ray binaries in the southern Galactic plane, and to determine whether particle acceleration to PeV energies is a common feature of this class of sources.

\subsubsection{Globular Clusters}   

Globular clusters are evolved, gravitationally bound stellar systems characterized by high densities, often exceeding 1,000 stars per cubic parsec near their cores \citep[e.g.,][]{sollima2017global}. This environment fosters the creation of compact binary systems and stellar remnants, including neutron stars, white dwarfs, LMXBs, and millisecond pulsars \citep[MSPs;][]{pooley2003dynamical,claireYe2019millisecond,deMenezes2023dynamical}. These populations are expected to produce two distinct gamma-ray spectral components: a $\sim$3~GeV peak resulting from curvature radiation in pulsar magnetospheres \citep[e.g.,][]{caraveo2014gamma,kalapotharakos2022fundamental}, and a $\sim$1~TeV peak generated by IC emission as relativistic leptons interact with background photon fields \citep[e.g.,][]{venter2009predictions,kopp2013multi}.

While the GeV component has been identified in approximately 40 Milky Way globular clusters using \textit{Fermi}-LAT data \citep[e.g.,][]{abdo2009detection47Tuc,hui2010fundamental,abdo2010population,deMenezes2019milky,abdollahi2022incremental,deMenezes2023dynamical}, the TeV component remains largely undetected, with the possible exception of emission near Terzan~5 reported by H.E.S.S. \citep{abramowski2011very}. In this context, SWGO emerges as a powerful instrument to probe the VHE physics of globular clusters. With approximately 85\% of the Milky Way's globular clusters situated in the Southern sky \citep{Harris1996_GC_catalog}, SWGO's strategic location in the Southern Hemisphere offers a distinct advantage over other major WCD gamma-ray instruments, which are all based in the Northern Hemisphere. Notably, the ten brightest globular clusters observed in the GeV range with the \emph{Fermi}-LAT are located in the Southern sky \citep{abdollahi2022incremental}, positioning these clusters as prime candidates for future VHE observations and detailed study.

\begin{figure}
    \centering
    \includegraphics[width=\linewidth]{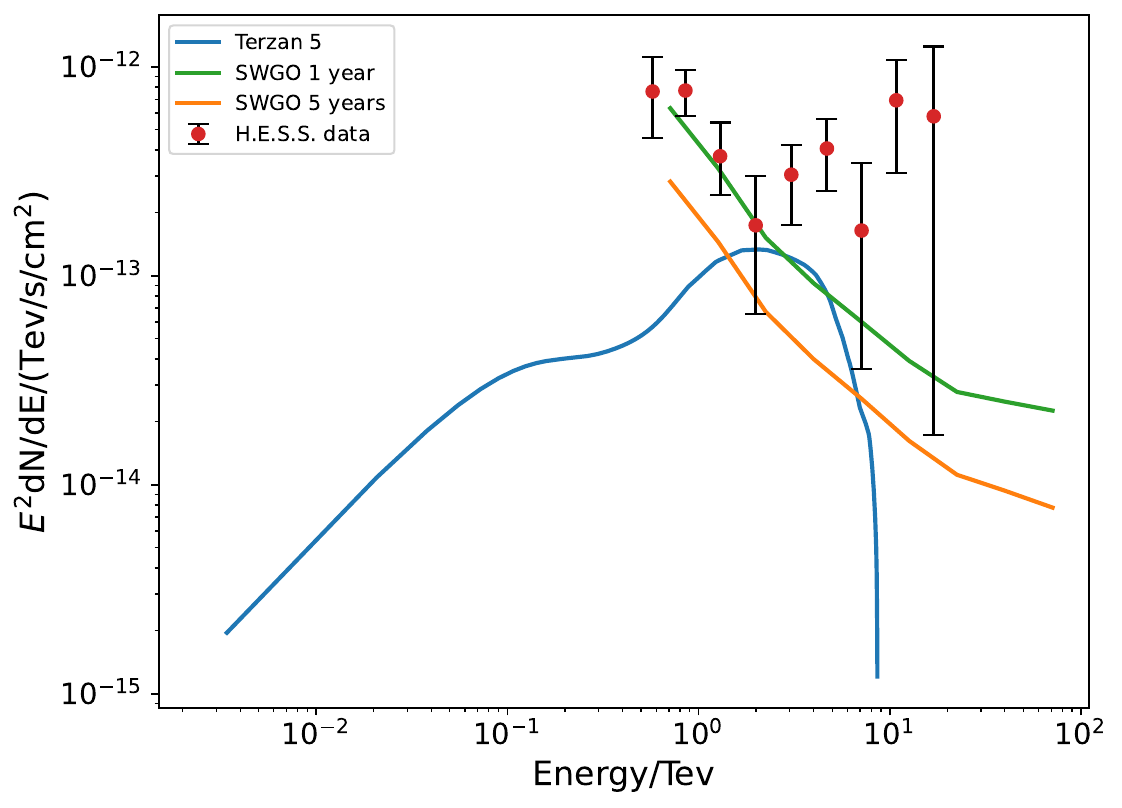}
    \caption{The predicted gamma-ray spectrum for Terzan~5, calculated under the assumption of a nebular magnetic field of $B = 10\mu$G \citep{venter2009predictions} and normalized based on the latest distance measurements for the cluster is shown alongside the full SWGO array sensitivity curves for one-year and five-year observation periods, as well as the H.E.S.S spectrum for a gamma-ray source observed in the direction of Terzan~5 (claimed to be the cluster). For the predicted model (blue line), the detection significances are $\sim$5$\sigma$ for one year (green line) and $\sim$20$\sigma$ for five years (orange line). For the H.E.S.S. source associated with Terzan~5, the significances are $\sim$29$\sigma$ for one year and $\sim$37$\sigma$ for five years.}
    \label{fig:terzan_spectrum}
\end{figure}

To evaluate the potential detectability of globular clusters with SWGO, Terzan~5 was modeled as a point-like source with a PL spectrum $dN/dE = k(E/E_0)^{-\Gamma}$, with normalization $k = (5.2\pm 1.1) \times 10^{-13}$ cm$^{-2}$ s$^{-1}$ TeV$^{-1}$, reference energy $E_0 = 1$ TeV, and spectral index $\Gamma = 2.5\pm 0.3$, i.e., consistent with the H.E.S.S. observations from the direction of this cluster \citep{abramowski2011very}. With this spectral model, the IRF for SWGO in the energy range from 0.5~TeV to 6~TeV was computed with the \textit{pyswgo} software package\footnote{\url{https://gitlab.com/swgo-collaboration/irf-production}}. The results 
demonstrate that over a one-year observation period, SWGO could achieve a detection significance of up to nearly $29\sigma$ (see Figure~\ref{fig:terzan_spectrum}). This is a significant improvement over the $7.5\sigma$ achieved by H.E.S.S. with 90 hours of observations using three- and four-telescope combinations. The superior performance of SWGO is primarily attributed to its enhanced sensitivity to photons with energies above 1~TeV. 

The main results expected to be achieved with SWGO on the topic of globular clusters are listed below.

\begin{itemize}
    \item Extended gamma-ray emission: in the GeV energy range, globular clusters are expected to appear as point-like sources due to the concentration of MSPs in the cluster cores, a consequence of dynamical friction \citep{deMenezes2023dynamical}. However, in the VHE domain, the situation changes. Relativistic leptons that escape the magnetospheres of MSPs can propagate within the cluster and scatter soft stellar photons to energies that exceed 100~GeV through the IC process \citep[e.g.,][]{bednarek2007high,bednarek2016tev}. Depending on the propagation behavior of these leptons, the associated IC emission could extend beyond a radius of $\sim$10~pc from the cluster centers \citep[e.g.,][]{cheng2010origin} and potentially produce extended gamma-ray sources detectable with SWGO. For a nearby cluster, such as NGC\,6752, this would imply an angular extension of $\sim$0.3$^{\circ}$, well within SWGO's angular resolution of $\sim$0.1$^{\circ}$.
    
    \item The correlation between gamma-ray luminosity ($L_{\gamma}$) and stellar encounter rate ($\Gamma$) is well established in the GeV range, where the gamma-ray luminosity of globular clusters scales linearly with their stellar encounter rate \citep{abdo2010population,hui2010fundamental,deMenezes2019milky}. Using this correlation, one can estimate the total number of MSPs in a cluster independently of radio observations. One major issue is that, in the GeV band, the gamma-ray emission of MSPs is beamed, leading to significant scatter in the $L_{\gamma} \times \Gamma$ correlation. \cite{deMenezes2023dynamical} show that this problem is mitigated in X-rays, where the emission from compact objects is expected to be roughly isotropic, and this should also be the case for the VHE range, where gamma-rays are roughly isotropically scattered following lepton propagation within the cluster. Consequently, we anticipate a much tighter $L_{\gamma} \times \Gamma$ correlation in the VHE domain compared to the GeV range, enabling more precise estimates of the total MSP population in globular clusters. Furthermore, \cite{deMenezes2023dynamical} shows that this linear correlation holds only for clusters with high stellar encounter rates ($\Gamma \gtrsim 100$, normalized with respect to 47~Tucanae, which has $\Gamma = 1000$), which suggests that channels other than stellar encounters can form MSPs in globular clusters. If a similar correlation were observed in the VHE regime, it would provide valuable insights into this phenomenon, and help to constrain the underlying mechanisms driving the relationship between $L_{\gamma}$ and $\Gamma$. The VHE emission model for the globular cluster Terzan~5 is shown in Figure~\ref{fig:terzan_spectrum}, alongside the SWGO sensitivity curves and the spectrum of the H.E.S.S. source associated with Terzan~5.
    
    \item Magnetic field, lepton injection spectra, and diffusion coefficient: the internal physical conditions within globular clusters remain poorly understood. A direct measurement of their VHE gamma-ray spectra could provide critical insights into key parameters, including the average magnetic field strength, the energy spectra of injected relativistic leptons, and the mechanisms governing CR propagation in these environments \citep[e.g.,][]{kopp2013multi,bednarek2016tev,ndiyavala2018identifying}. Such measurements would represent a significant step toward unraveling the complex interplay of astrophysical processes within globular clusters.
    \item Non-accreting white dwarfs: the escape of relativistic leptons from the magnetospheres of non-accreting white dwarfs in globular clusters may also upscatter surrounding soft photons to TeV energies via the IC process \citep[e.g.,][]{bednarek2012gamma}. While the spin-down luminosities of white dwarfs are typically about 1,000 times lower than those of MSPs, they may outnumber MSPs by a similar factor. Consequently, white dwarfs could contribute significantly to the observed TeV emission from globular clusters. Observations with SWGO may provide the means to disentangle the contributions of white dwarfs and MSPs, potentially via the identification of distinct spectral components in the VHE gamma-ray spectra of these clusters. Such a breakthrough would enable detailed studies of both source populations, and offer new perspectives on their roles in the astrophysical processes shaping globular clusters.
\end{itemize}

Beyond Terzan~5, several other globular clusters are promising targets for SWGO, with predicted VHE gamma-ray emissions comparable to, or potentially exceeding, that of Terzan~5. Notable examples include NGC~6388 and NGC~362, identified in previous studies \citep[see][for a comprehensive list]{ndiyavala2018identifying}.  

\subsubsection{Novae}
\label{subsec:novae}

Nova explosions occur 
when a white dwarf accreting matter from its companion star (i.e., a white-dwarf-star system), accumulates enough hydrogen on its surface to produce a thermonuclear explosion \citep[e.g.,][]{Chomiuk2021}. 
The 2021 eruption of the nova RS Ophiuchi (RS Oph) was detected by optical telescopes and the \emph{Fermi}-LAT (in GeV energies), as well as H.E.S.S. and MAGIC in TeV energies for the first time \citep{RSOph_Magic2022,RSOph_HESS2022}. Since a large portion of the Galactic Plane will be visible to SWGO, it is anticipated that a number of nova explosions may occur within the FoV of SWGO, which will enable prompt observations and/or constrain limits. 
 
Nova events are typically first identified due to a strong increase in optical brightness over a few hours. Whilst their optical light curves 
typically return to optical quiescence after $\sim$1~year \citep[e.g.,][]{2010Strope_novaLC}, the gamma-ray emission detected by the \emph{Fermi}-LAT generally ceases to be detectable within $\sim$30~days \citep[e.g.,][]{Fermi-LAT:2014hrt_novae}. Indeed, the only TeV-detected nova to date, RS\,Ophiuchi, remained detectable for  $\sim$2--4~weeks, however the emission was much reduced after the first few days \citep{RSOph_HESS2022}. 

The sensitivity of the full SWGO array over a period of 30 days was evaluated, and a conservative estimate of the integral sensitivity of $\sim$$2\times 10^{-12}\mathrm{cm}^{-2}\mathrm{s}^{-1}$ in the energy range 0.5 to 100~TeV was obtained. To compare this to the anticipated gamma-ray energy flux due to hadronic emission from a nova, the model of \citet{RSOph_Magic2022} was adopted, which enables an evaluation of the gamma-ray energy flux as a function of distance and shock velocity. For the closest likely novae such as T\,CrB, at a distance of $\sim$900~pc, a shock velocity of order $\gtrsim$5000~km/s is required for detection by SWGO, comparable to that reached by RS\,Oph during the 2021 outburst. However, at a distance more comparable to RS\,Oph (2.5~kpc), even higher velocities are required. Therefore only a small number of truly exceptional nova events are likely to be detectable with SWGO. Nevertheless, due to its large FoV and high duty cycle, SWGO remains poised to provide data coincident with novae more systematically than has  been feasible with IACTs. For example, while the full Moon caused the cessation of IACT observations during the 2021 RS\,Oph outburst, such an occurrence would have no effect on SWGO data acquisition \citep{RSOph_HESS2022,RSOph_Magic2022}.

%% file: Transients.tex
\section{Transients and Variable Sources}
\label{sec:transients}

SWGO will provide extensive sky coverage, which will enable the continuous monitoring and real-time detection of highly variable emission. This includes not only
AGN flares, but also transient phenomena such as Gamma-Ray Bursts (GRBs), which can occur anywhere in the sky without prior warning. With its wide FoV, SWGO will be capable of recording sudden increases in gamma-ray flux across multiple regions of the sky simultaneously, without the need to re-point at a specific source. This capability will allow for a rapid response to transient events, and offers a significant advantage for continuous sky monitoring. Alerts generated by SWGO are likely to be critical for pointed instruments to obtain information from the early phase of transient events.


The expected sensitivity of SWGO at low energies, from 100 to 300~GeV, will be better than current wide FoV detectors, which makes it a highly relevant instrument to locate and monitor extragalactic transient sources, as well as to provide rapid alerts to trigger complementary studies with other present and future VHE pointed instruments, such as CTAO.

\subsection{Active Galactic Nuclei} 
\label{sec:AGN} 
\label{subsec:AGN}

Active galaxies contain a nuclear region at least 100 times brighter than their integrated starlight due to the accretion of matter onto a supermassive black hole (SMBH) with masses up to $10^{10} M_{\odot}$ \citep[e.g.,][]{1978PhyS...17..265B, 1995PASP..107..803U}. 
There are several classification schemes for active galaxies, but only jetted AGN have been observed up to TeV energies. 
Regardless of orientation, most of these AGN show rapid flux variability at VHEs, suggesting that the emission occurs from small regions. The exact location of the VHE gamma-ray emission is uncertain, but it is thought to originate either along the jet \citep[e.g.,][]{2018Galax...6..116R} or at its base, near the SMBH \citep[e.g.,][]{2012JPhCS.355a2034R}. The nearby active galaxy Cen~A is a notable exception--- with TeV emission resolved along the kpc-scale jets~\citep{CenARef1}.

The mechanisms responsible for particle acceleration in AGN remain a significant open question. Fermi shock acceleration is often assumed to contribute, as it is a well established mechanism to produce non-thermal particle spectra over many decades in energy. Observational support is provided by the detection of shocks in the radio spectra, which occur further out along the jet.
Second-order Fermi acceleration due to magnetohydrodynamic (MHD) turbulence, as well as shear acceleration, have been proposed to explain the smooth evolution of the spectral index \citep[e.g.,][]{10.1007/978-1-4020-6118-9_19, 2021MNRAS.505.1334W}. Magnetic reconnection 
has also been suggested as a possible acceleration mechanism that may account for observed phenomena, especially rapid variability and large energy outputs \citep[e.g.,][]{2010MNRAS.408L..46G,2015MNRAS.450..183S,2016MNRAS.462.3325P}. 

The difference between blazars and radio galaxies is the angle between their jet axis and our line of sight. Blazars are observed along the jet axis, and dominate the VHE sky outside the Galactic Plane. They account for 70$\%$ of the extragalactic sources detected at VHE energies \citep{2020Galax...8...72C}. Radio galaxies are observed at larger angles, which allow for the study of the most inner region of the AGN. At the moment, there are four radio galaxies detected at VHE gamma-rays, including M87, Cen~A, NGC\,1275 and 3C\,264 \citep{2022Galax..10...61R}, with only Cen~A located in the Southern sky, and M87 visible from some Southern Hemisphere locations.

The broadband emission from blazars and radio galaxies is characterized by two dominant components in their spectral energy distribution (SED), which are often attributed to electron synchrotron at lower frequencies and IC at higher frequencies, although the identity of the high-frequency peak is a matter of some debate. Blazars can be further classified into flat spectrum radio quasars, Low-synchrotron-peaked BL Lacs (LBL), Intermediate-synchrotron-peaked BL Lacs (IBL), High-synchrotron-peaked BL Lacs (HBL), and Extreme high-synchrotron-peaked BL Lacs (EHBL), depending on the peak frequency of the electron synchrotron component of the SED. Some HBL sources can become EHBL when in a flare state, and these have been observed to reach up to tens of TeV.

\subsubsection{Emission Scenarios}

Active Galactic Nuclei can exhibit MWL emission from radio up to VHE gamma rays. As stated previously, the SED of AGN is composed of two primary components. The low-energy component is well understood as the result of synchrotron emission from relativistic electrons. The second component, responsible for gamma-ray emission from MeV energies up to TeV, is usually, but not universally, attributed to IC emission from the same electron population, either in external radiation fields~\citep[e.g.,][]{1992A&A...256L..27D,1994ApJ...421..153S} or via the synchrotron self-Compton (SSC) mechanism~\citep[e.g.,][]{Maraschi92,1996ApJ...461..657B}. Hadronic emission is generally invoked in combination with IC emission, but purely hadronic models have been propose to account for the entire high-energy peak \citep[e.g.,][]{1989A&A...221..211M,1995A&A...295..613M,2001APh....15..121M,2014MNRAS.441.1209F}. 

Whilst single-zone models have been very commonly applied in the past, it is clear that these are complex objects with multiple-zone models increasingly invoked~\citep[e.g.,][]{2018MNRAS.473.2542K, 2021ApJ...906...91S}. Note that whilst VHE gamma-ray activity in AGN is often correlated with that seen in the optical and X-ray, blazars have also been reported to exhibit {\it orphan flares} in the TeV band~\citep{krawczynski2004multiwavelength}. 

Continuous monitoring of AGN during both quiescent and flare states, across the electromagnetic spectrum, is crucial for constraining these models. Continuous monitoring at VHEs is one of the aims of SWGO, particularly for sources in the Southern Hemisphere.

The VHE horizon is limited by gamma-ray interactions with photons of the diffuse far-infrared to ultraviolet (FIR-UV) EBL, such that most AGN detected by SWGO are expected to be located at moderate redshift (see Figure~\ref{fig:Fermi_AGNs_SWGOsens}). Exceptions to this are powerful flaring episodes of more distant objects, anomalous optical depths due, e.g., to Axion-Like Particles (see Section~\ref{sec:ALPs}), or production of gamma-rays from secondary particles closer to the observer~\citep[e.g.,][]{2013ApJ...771L..32T}.

\subsubsection{Active Galactic Nuclei in the SWGO Field of View}

\begin{figure*}[ht!]
    \centering
    \includegraphics[width=\linewidth]{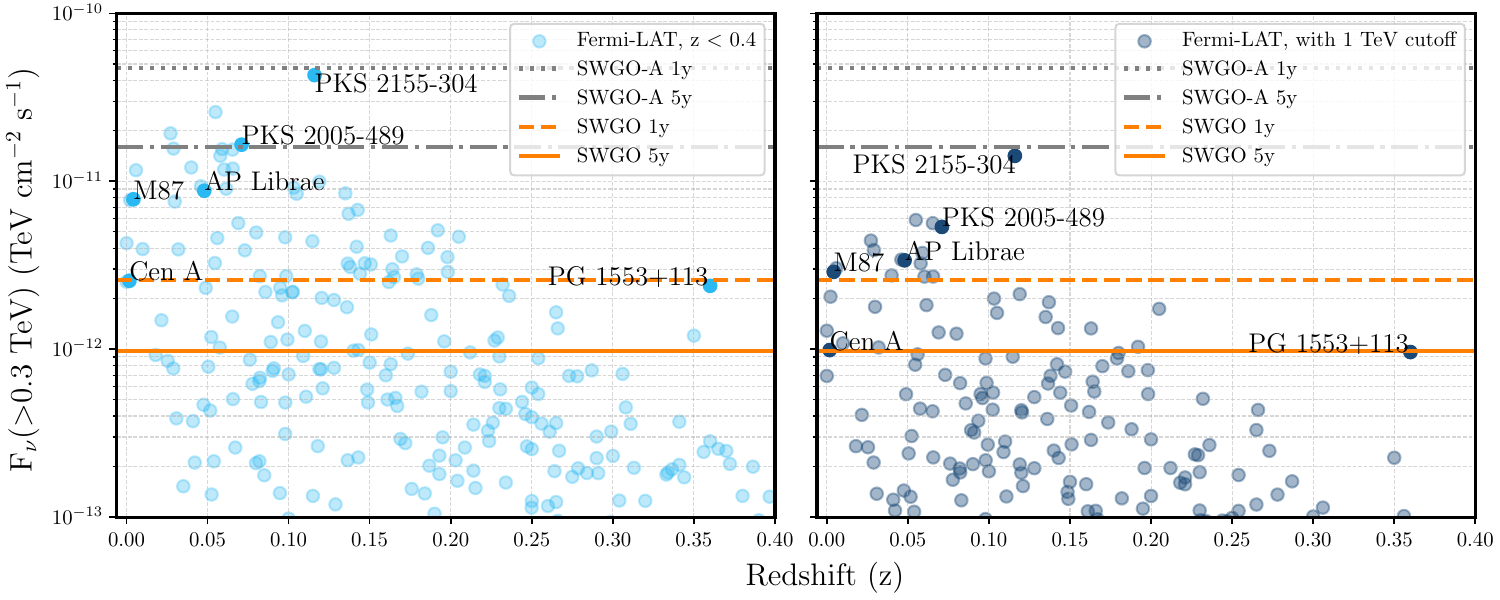}
    \caption{Both panels show sources from the Fourth Catalog of Active Galactic Nuclei detected by the \emph{Fermi}-LAT, based on 12 years of data (4LAC-DR3) \citep{2022ApJS..263...24A} in the SWGO FoV extrapolated from 300~GeV to 10~TeV, and EBL attenuated with the Dominguez model \citep{EBLDomin}. The left panel shows the extrapolated flux with the spectral model reported in the catalog, while the right panel presents the extrapolation with an added exponential cutoff beyond 1~TeV, in order to not overestimate the energy flux of each extrapolated source. The horizontal lines represent the SWGO-A and SWGO sensitivity for one and five years of observations.}
    \label{fig:Fermi_AGNs_SWGOsens}
\end{figure*}

Figure~\ref{fig:Fermi_AGNs_SWGOsens} shows the AGN that could be observed by SWGO and SWGO-A after one and five years of observation. The sources shown are taken from the \emph{Fermi}-LAT 4th AGN catalog \citep{2022ApJS..263...24A}, with their fluxes extrapolated to VHEs and limited to redshifts z~<~0.4. The left panel presents the extrapolated fluxes and accounts for attenuation due to the EBL, while the right panel includes both EBL attenuation and an additional exponential cutoff at 1~TeV. This cutoff is introduced to avoid an overestimate of the energy flux of each source and thus the number of sources potentially detectable.
Under the more optimistic scenario in the left panel, SWGO-A would be able to observe four sources after five years and SWGO would detect approximately 53 sources after one year. In the more conservative scenario, shown in the right panel, SWGO would detect about 15 sources over a one-year period. These estimates do not include the potential for flare activity, which may increase the total number of detectable AGN, but only transiently, rather than providing consistent, continuous coverage.

\begin{figure}
    \centering
    \includegraphics[width=1\linewidth]{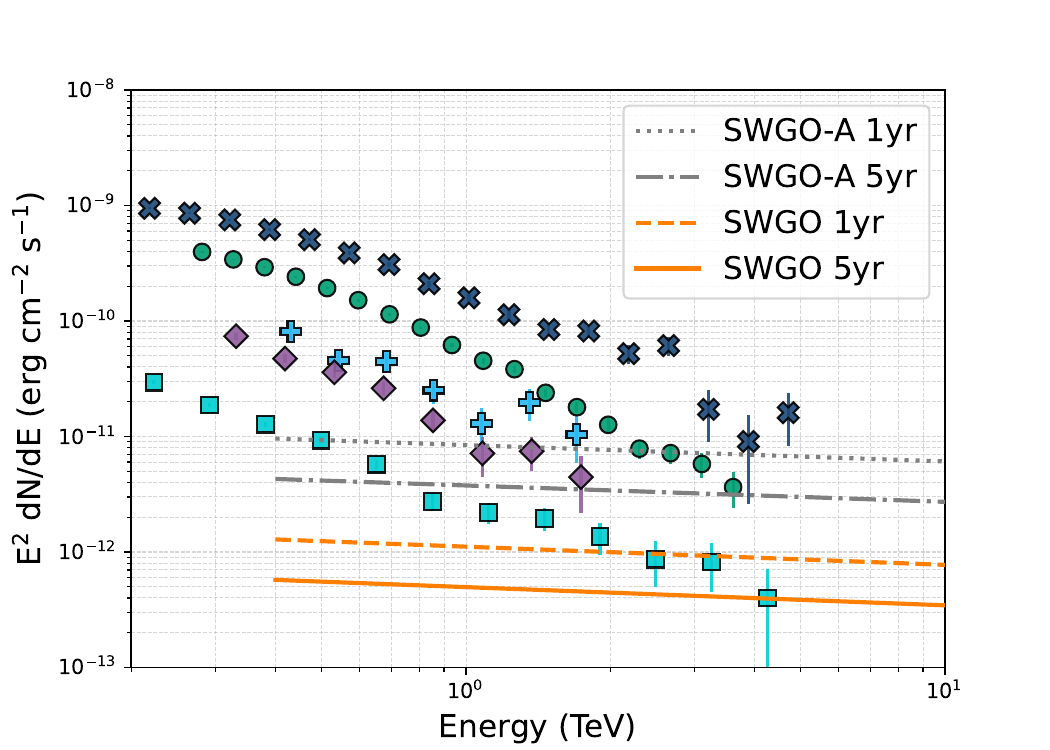}
    \caption{ Observed gamma-ray spectra of the blazar PKS 2155-304 (z~=~0.116), taken from \texttt{https://tools.ssdc.asi.it/SED/}. Different states of activity are shown: the low state in teal squared symbols \citep{2010A&A...520A..83H}; the intermediate states following the major flare of 2006, shown in light blue crosses and pink diamonds \citep{2012A&A...539A.149H}; and the spectra of the major 2006 flare and another active state shown in blue X symbols and green circles, respectively \citep{2013PhRvD..88j2003A}. The SWGO-A sensitivities (shown with dotted and dash-dotted lines) and SWGO sensitivities (shown with dashed and solid lines) correspond to one and five years of observation, respectively. SWGO will be able to observe quiescent and flare states in PKS 2155-304-like AGN.} 
    \label{fig:PKS2155_VHESpectra}
\end{figure}

SWGO will also be able to observe different activity states of AGN and potentially monitor their flare episodes. For example, in the case of the HBL blazar PKS\,2155$-$304, located at z~=~0.116, both SWGO-A and SWGO will be able to detect the source during its high and low states (see Figure~\ref{fig:PKS2155_VHESpectra}). 

Whilst the number of detected objects is modest in comparison to CTAO, the SWGO measurements will be important in terms of understanding the long-term average behavior of sources, probing highest-energy emission, and providing triggers to pointed instruments like CTAO.

\subsection{Gamma-Ray Bursts} 
\label{sec:GRB}

Gamma-Ray Bursts are among the brightest events in the Universe. Their emission is split into two phases: a highly variable prompt phase (from optical to MeV energies) and a longer-lasting afterglow across all wavelengths. Prompt emission durations follow a bimodal distribution, dividing GRBs into short ($<$2 s) and long ($>$2 s) types. Long GRBs are commonly associated with massive star collapses, while short GRBs are typically linked to compact object mergers~\citep[e.g.,][]{1992ApJ...392L...9D}, although exceptions and alternative scenarios remain under investigation.

The emission in both the prompt and afterglow phases are thought to arise from synchrotron and SSC processes: synchrotron can reach up to tens of GeV, while SSC dominates from tens of GeV to TeV energies.

Until recently, GRB afterglows were only observed up to approximately one hundred GeV. However, detections from GRBs 180720B, 190114C, 190829A, and the exceptionally bright GRB 221009A ({\it the Brightest Of All Time}, BOAT) have revealed photons up to nearly 4 TeV—and beyond 10 TeV in the last case---with VHE emission lasting thousands of seconds post-trigger \citep[e.g.,][]{2019Natur.575..464A,2019Natur.575..455M,doi:10.1126/science.abe8560,2023SciA....9J2778C}. These discoveries confirm that GRB afterglows can emit VHE gamma rays for several days, challenging current models.

So far, VHE emission has only been firmly detected from long GRBs, but a detection in the VHE range would place powerful constraints on the physics models of short GRBs from compact binary mergers. SWGO-A, with its broad sky coverage and rapid localization (<10~arcmin within seconds), can greatly enhance MWL follow-up, especially crucial for Gravitational-Wave (GW)-related events.

While prompt VHE emission remains undetected, including being remarkably absent in the BOAT, SWGO-A is well-positioned to investigate this open question. The next sections first focus on a detailed case study of GRB~190114C, with simulations of early-time emission with and without prompt VHE flux, to illustrate how SWGO-A could complement CTAO-South in specific, well-characterized events. We then explore SWGO-A’s detection prospects using a synthetic population of GRBs, which provides a statistical view of its sensitivity across various scenarios.  

\subsubsection{GRB 190114C as a Case Study}
\begin{figure}
    \centering
    \includegraphics[width=\linewidth]{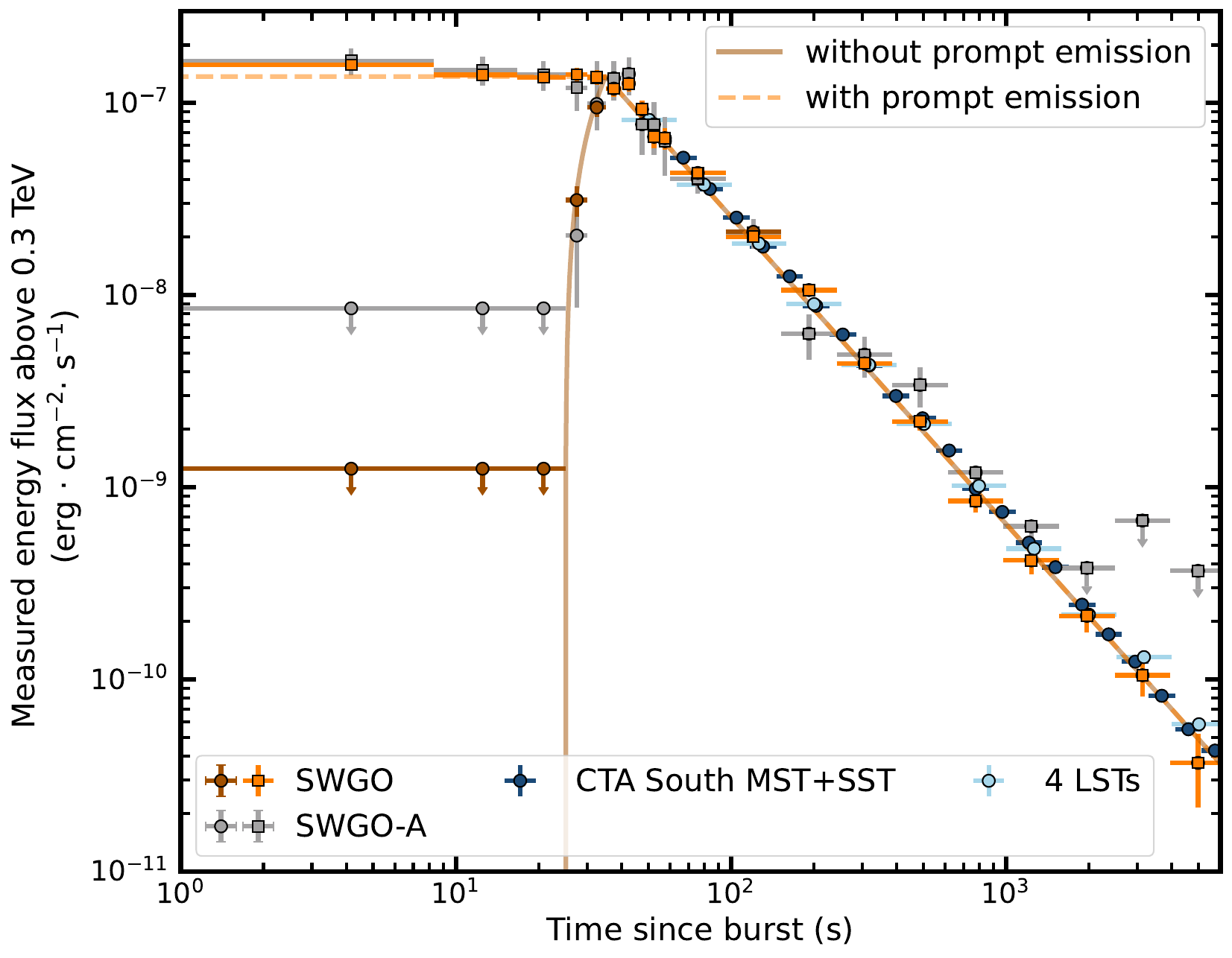}
    \caption{Simulated light curve of a GRB at $z=0.1$ and zenith angle of $20^\circ$, located above both SWGO and CTAO. The spectral and temporal evolution are modeled after GRB~190114C. Two prompt emission scenarios are shown: a sudden rise (without prompt emission) and a plateau (with prompt emission), illustrating how both SWGO and SWGO-A can probe the presence of prompt VHE emission.}
    \label{fig:grb_lc}
\end{figure}

To evaluate SWGO's routine detection capabilities, a representative baseline is required. While the recently observed GRB 221009A provided extraordinary VHE coverage, its extreme energetics make it an atypical outlier. Therefore, we focused instead on simulating a less energetic, more statistically probable event like GRB 190114C.
Taking the measured light curve of GRB190114C~\citep{2019Natur.575..455M}, two scenarios are considered for the early development of the GRB light curve at VHEs: with and without prompt-phase emission. This flux evolution is simulated as would be observed with SWGO-A at a zenith angle of 20$^\circ$. The spectral and temporal properties are selected to match those of GRB190114C, with a spectral index of 2.2 and temporal evolution $\propto t^{-1.6}$. To explore a domain more relevant to GW-detected mergers, a smaller redshift of $z=0.1$ is adopted for this simulation. Note that, in reducing the GRB's redshift, the event becomes less luminous as the (unabsorbed) flux normalization was kept to be the same. The effect of EBL absorption is accounted for with the model proposed in~\citet{2021MNRAS.507.5144S}. The prompt phase lasts 25~s (analogus with GRB190114C) and is simulated as either no emission followed by a rapid rise or a flat plateau. To illustrate the complementarity of the (almost) co-located CTAO-South, the GRB is considered to be observed with the CTAO-South configuration with a delay of 60~s. Observations with four Large-Sized Telescopes (LSTs; the largest component telescopes of CTAO) are also considered, which have a more rapid follow-up response, reducing the delay to 40~s ~\citep{2019scta.book.....C}. The results are shown in Figure~\ref{fig:grb_lc} and highlight how even the rapid response of the LSTs misses the prompt phase, such that only SWGO (and SWGO-A) can distinguish between the two scenarios.
While the precision and sensitivity of pointed IACTs like CTAO-South are unmatched once on-target, SWGO's (and SWGO-A's) complementary assets are its zero-delay, pre-trigger coverage of the prompt phase and its high duty cycle, which guaranties observations during daytime or prohibitive weather conditions when IACTs cannot operate.

\subsubsection{SWGO Gamma-Ray Burst Detection Rates}
\label{subsec:GRB_rates}

To perform the detectability study for SWGO, presented here, a synthetic population of GRBs was simulated with the Conditional Tabular Generative Adversarial Networks~\citep[CTGANs,][]{2019arXiv190700503X}. A CTGAN is a specific type of GAN, or machine-learning method, in which a generator and a discriminator network compete to create increasingly realistic data. CTGANS are designed explicitly for synthesizing realistic tabular data by learning conditional relationships between features.
To apply CTGANS to GRBs, the GRB data from \citet{Zhu_2023} and \citet{Tang_2019} that contains a clean sample of short and long GRBs was used, respectively, with parameters such as redshift, duration (T$_{90}$), fluence, and isotropic energy (E$_{\rm iso}$) in a tabulated form. This GRB list was cross-matched with the \emph{Swift} GRB table~\citep{swift_grb_table} to include the BAT photon index and XRT 11 hrs flux. The feature engineering process before training the CTGAN applies a logarithmic transformation to these variables. For each column, the standardized (zero-mean, unit-variance) version of that column (z-score) was added. To improve generalization and reduce overfitting, a multiplicative Gaussian noise function was used for both the logarithmic and linear values. An isolation forest with contamination set to 0.05 was used, to ensure that extreme outliers do not bias the generative model. 
The model was trained with the \texttt{ctgan.CTGAN} library of the Synthetic Data Vault Project and the tuned parameters include epochs, batch size, generator and discriminator learning rate, and pac. 

After cleaning, 80\% of the data were used to train the model with the remaining 20\% held for validation. 
The evaluation metric employed to select the best model was the {\it LogisticDetection}, which picks the model with the lowest value. With this best model, a sample of thousands of GRBs was created to compare features like E$_{\rm iso}$, T$_{90}$, fluence, and the relations between the input/real data and the synthetic one. This pipeline ensures the synthetic GRB population is physically meaningful and statistically consistent. The final GRB population sampled 10,000 GRBs with a ratio of 1/4 between short and long GRBs. This number of GRBs is equivalent to approximately 100 years of \emph{Swift} observations, since it detects roughly 100 GRBs per year~\citep{swift_overview}. The distribution of GRBs in the sky was randomly sampled within the \emph{Swift} range of ($-20^{\circ}$, $70^{\circ}$) in declination, with random values for the right ascension coordinates and onset times. Although \emph{Swift} has detected GRBs across the whole sky, the declination region considered encompasses the majority of its cumulative sky exposure and GRB detections ($\sim$90\%), while excluding zones with limited visibility due to Sun/Earth avoidance constraints. This conservative cut ensures uniformity of detection efficiency without a requirement to model detailed exposure maps.

\begin{figure*}
    \centering
    \includegraphics[width=0.45\linewidth]{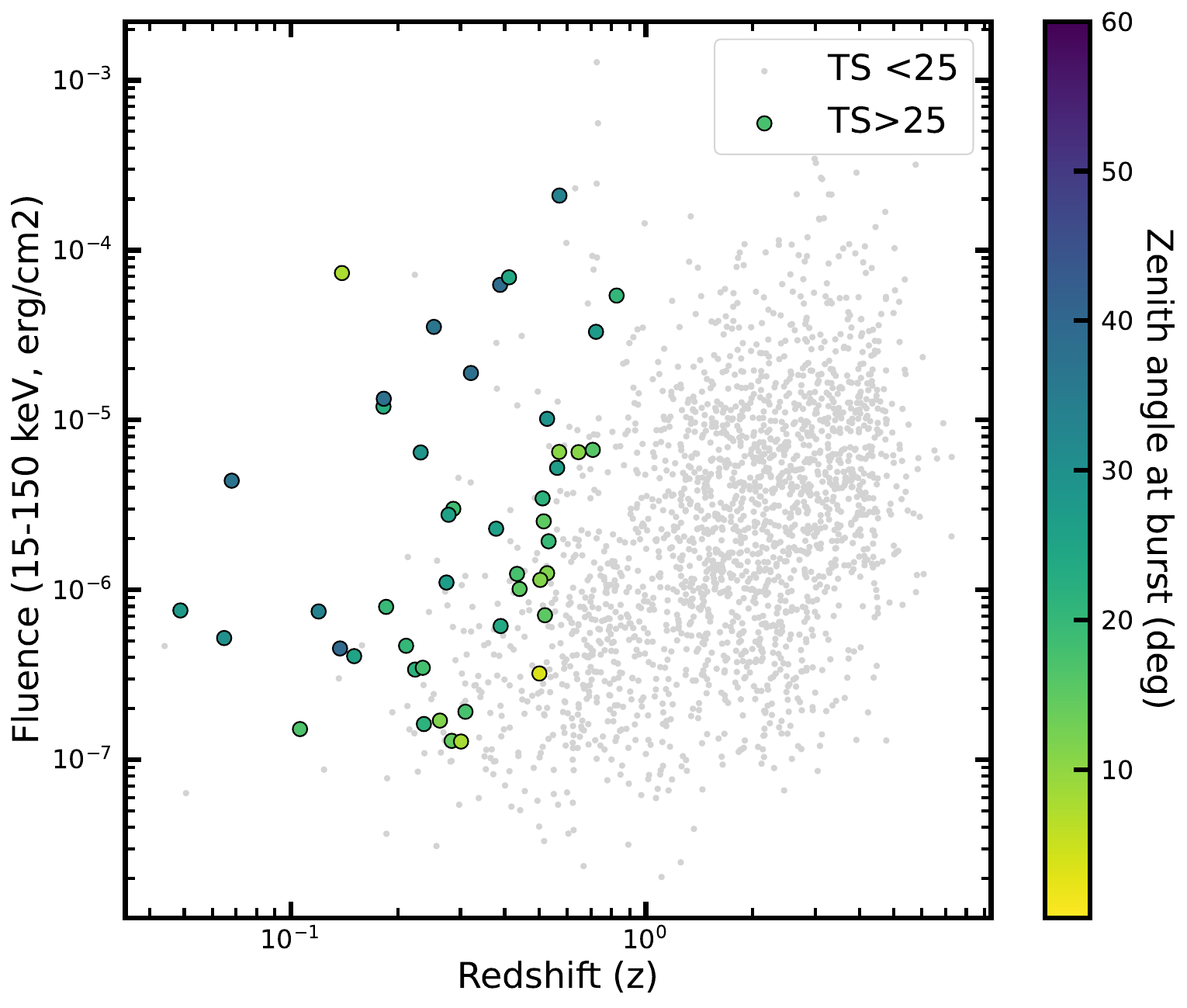}
    \includegraphics[width=0.45\linewidth]{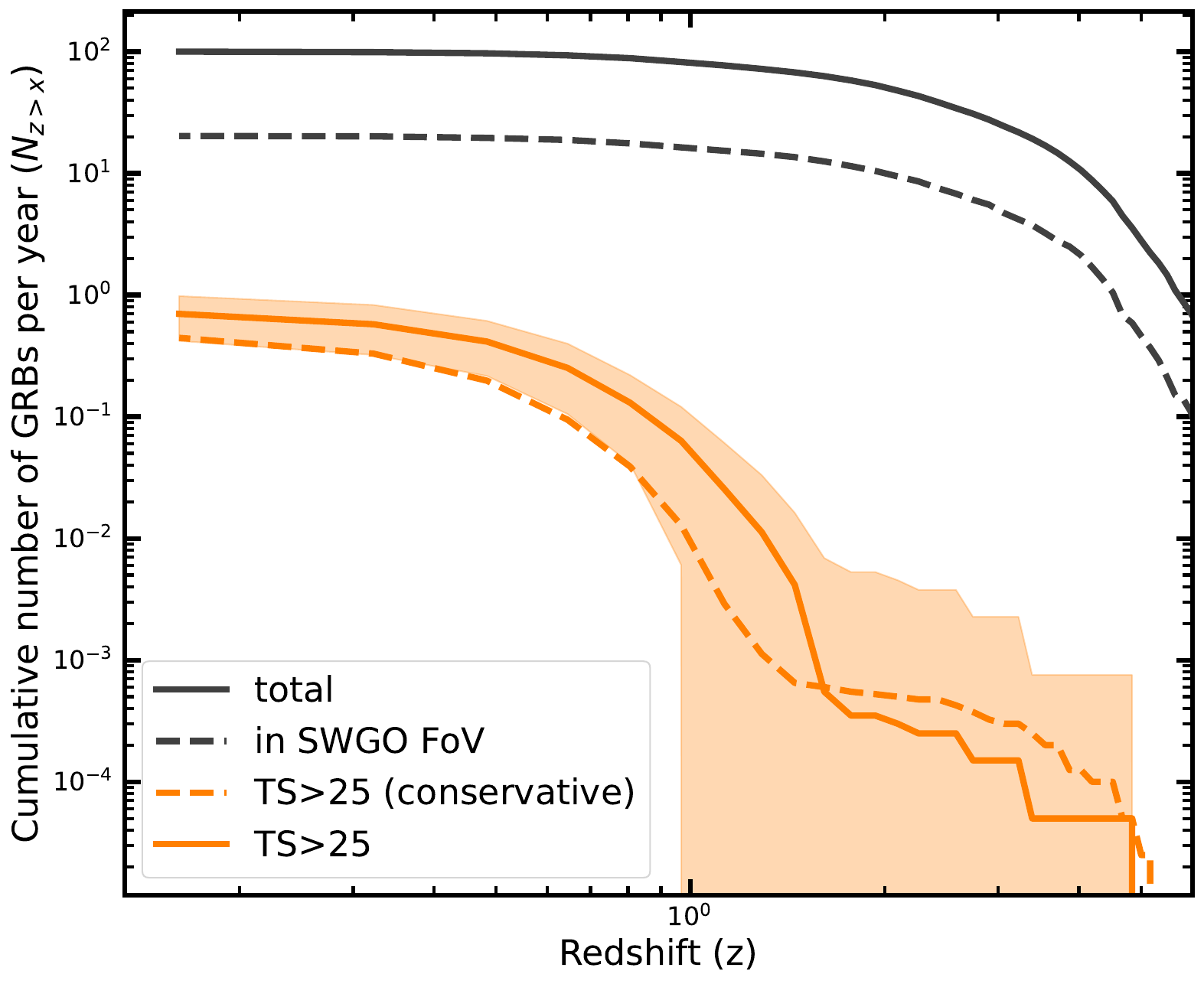}
    \caption{\textbf{Left:} Redshift and fluence distributions of the GRBs detected (colored circles) and undetected (gray dots) by SWGO from one random sample of the 100~yr of \emph{Swift} GRBs simulation. Only detectable GRBs (those that fall within the  SWGO FoV of 0--52\degree zenith angle) are shown here. The color of the dots represents the zenith angle at which the burst occurred. 
    \textbf{Right:} Cumulative yearly rate of detected (orange), total (black), and total detectable (black dashed) GRBs as a function of redshift in the synthetic GRB population. {\bf The orange dashed line shows the cumulative detection rate for the high-threshold scenario as described in Section~\ref{sec:EventReco}.} }
    \label{fig:grb_population}
\end{figure*}

The synthetic GRB population was injected into the SWGO simulation pipeline, randomly distributing the bursts in time. A PL spectral model was assumed, with an index of 2, and normalized under the assumption that the fluence per decade remains constant. The effect of EBL absorption was taken into account for the redshift of each GRB. For the temporal evolution, a linear rise up to T$_{90}$ was assumed, and followed by a power law $\propto t^{-1.2}$. For each event, the detection significance was derived by integrating within a window of length 3~$\times$~T$_{90}$. Note that this integration window is chosen for simplicity and could be further optimized depending on the brightness of the GRB.

The random distribution of bursts in time translates to a random distribution of sky coordinates. The position of the GRB is tracked within the integration window in steps of 1~min to account for the changing response as a function of zenith angle. The instrument response is simulated for the zenith range $0^{\circ}$--$52^{\circ}$, so emission from the GRBs is only simulated while they fall within this range. The simulation was run 200 times, each with a different random seed for the simulated counts and burst time assignment. An example realization of the simulation and the cumulative yearly rate of detected GRBs are shown in Figure~\ref{fig:grb_population}. From the right panel in the figure it can be seen that the total rate of GRBs detected by SWGO is estimated as 0.7~events per year from externally triggered GRBs or those with X-ray counterparts.
Looking ahead, the number of GRB triggers is expected to increase with the operation of recent missions such as the Space Variable Objects Monitor (SVOM), launched in mid-2024, and the Einstein Probe, launched in early 2024, both of which have been successfully commissioned. The inclusion of these additional triggers could potentially increase the SWGO detection rate to 1.4 GRBs per year. We note that the GRB detection capabilities of SWGO do not rest solely on events close to the trigger threshold.
Figure~\ref{fig:grb_population} provides a second, dashed curve to show the relatively modest impact of using the high-threshold trigger as described in Section~\ref{sec:EventReco}.

The GRB detection prospects study for SWGO assumes that only GRBs that can be associated with an X-ray/soft-gamma-ray trigger. Blind searches are of considerable interest, but will contribute at a lower rate to the detection of SWGO GRBs.
 
\subsection{Exploratory Searches for New Transient Phenomena}\label{sec:exploratory}



The Universe abounds with transient phenomena that reveal extreme and energetic processes. From the explosive deaths of massive stars to the violent interactions in binary systems, these events provide a unique opportunity to understand the physics of compact objects, particle acceleration, and the behavior of matter under extreme conditions. While phenomena such as AGN (Section~\ref{sec:AGN}) and GRBs (Section~\ref{sec:GRB}) have established VHE detection records and well-characterised SWGO detection prospects, there remains a vast landscape of poorly explored transient events with potential VHE counterparts. Among them are magnetar outbursts, gamma-ray novae, jetted tidal disruption events (TDEs) near the Galactic Center,  flares from young stellar objects, and GW counterparts from compact binary mergers. 

SWGO's wide FoV and high duty cycle make it uniquely suited to probe these source classes; however, the VHE properties of most are poorly constrained and the rate estimates presented here should be understood as order-of-magnitude projections.

\subsubsection{\textbf{Methodology and Validation}}
 
For each source class, a spectral template is adopted from the best available VHE observation or physical model, scaled to the distance or redshift of a hypothetical event, attenuated by EBL absorption~\citep{2021MNRAS.507.5144S} for extragalactic sources, and convolved with the SWGO instrument response to compute expected signal and background counts per energy bin. Detection significance is evaluated via the Li \& Ma~(\citeyear{LiMa1983}) test statistic: a $5\sigma$ threshold is adopted for externally triggered events and $\sim$$8\sigma$ pre-trial (corresponding to $5\sigma$ post-trial for $\sim$$10^9$ independent spatial-temporal trials) for untriggered blind searches. For Galactic sources, the rate is the product of the occurrence rate, the fraction within the detection horizon, the visible sky fraction, and, for blind searches, the instantaneous FoV fraction\footnote{Rates were evaluated with the high-threshold IRFs (50-hit threshold for SWGO-A, 100-hit for SWGO)}.

Among the 166 confirmed Galactic transient sources in current catalogs, 140 fall within SWGO's optimal viewing window ($\theta_z < 52^\circ$), including six high-energy binary systems, four black holes, three pulsars, and six SGRs (one confirmed high-energy emitter). A broader analysis of all transient alerts shows that 22.2\% of reported events fall within SWGO's visible region, with an additional 4.7\% in challenging, but potentially observable, zones.

\subsubsection{\textbf{Magnetar Giant Flares}}
 
Magnetars are highly magnetized neutron stars ($B \sim 10^{14}$--$10^{15}$\,G), classified as Soft Gamma Repeaters (SGRs) or Anomalous X-ray Pulsars (AXPs), that produce bursts via magnetic reconnection or crustal instabilities~\citep[e.g.,][]{1Thompson_1995, 2Kaspi_2017, Harding_2006}. Approximately 30 are known in our Galaxy~\citep{Esposito_2021, Palmer_2005}.
 
Magnetar giant flares (MGFs) release up to $E_{\mathrm{iso}} \sim 2 \times 10^{46}$\,erg~\citep{Palmer_2005}, with a pulsating tail lasting $\sim$400\,s that represents the most promising VHE window. The \textit{Fermi}-LAT detected three photons above 100\,MeV from the extragalactic MGF candidate GRB\,200415A in NGC\,253~\citep{Ajello2021}, measuring $\Gamma = 1.7 \pm 0.3$. The VHE fraction $f_{\mathrm{VHE}}$ remains unconstrained, making any rate estimate inherently speculative.
 
A Galactic MGF is detectable far above the blind-search threshold at any distance within the Galaxy, provided $f_{\mathrm{VHE}} \geq 10^{-4}$. The detection rate is therefore set entirely by the Galactic occurrence rate ($\sim$0.03--0.06\,yr$^{-1}$; \citealt{Palmer_2005, Burns2021}; see also \citealt{Beniamini2025}), the visible sky fraction ($\sim$$70\%$), and the instantaneous FoV fraction ($\Omega_{\mathrm{FoV}}/4\pi \approx 16\%$), yielding $R_{\mathrm{MGF}} \approx 0.005$\,yr$^{-1}$. We stress that while the \emph{premise} of detectable VHE emission is speculative ($f_{\mathrm{VHE}}$ unconstrained), the \emph{rate} itself is robust: detection is essentially guaranteed for any $f_{\mathrm{VHE}} \geq 10^{-4}$, so the value is set by geometry and occurrence rather than by the emission model or the instrument response. No instrumental improvement other than a wider FoV can increase \textbf{it}.
 
Intermediate flares ($E_{\mathrm{iso}} \sim 10^{41}$--$10^{43}$\,erg; \citealp[e.g.][]{WoodsThompson2006, TurollaZaneWatts2015}) occur more frequently by a factor $10^{3(\gamma-1)}$, where $\gamma \sim 1.4$--$2.0$ is the burst energy distribution index~\citep{Cheng1996, Gogus1999, Gogus2000}. Adopting $\gamma = 1.5$ yields $\sim$0.2\,yr$^{-1}$, though VHE extrapolation to these lower energetics is highly uncertain.

 
 

\subsubsection{\textbf{Binary Neutron Star Merger Afterglows}}
 
GW170817 ($\sim$41\,Mpc) was not detected at VHE despite deep H.E.S.S. follow-up~\citep{HESSGW170817}, consistent with the structured-jet geometry ($\theta_{\mathrm{obs}} \approx 15$--$25^\circ$; \citealt{Ghirlanda2019, Mooley2018}), where SSC flux is suppressed by de-beaming and Klein-Nishina corrections~\citep{Pellouin2024, Mondal2025}. Using the LVK O5 BNS detection volume ($\sim$200\,Mpc) and merger rate ($49^{+121}_{-42}$\,Gpc$^{-3}$\,yr$^{-1}$; \citealt{GWTC4pop}), the detection rate for Binary Neutron Star merger afterglows is $R_{\mathrm{BNS}} \lesssim 0.001$--$0.025$\,yr$^{-1}$, the upper bound corresponding to the full SWGO array's enhanced low-energy effective area. SWGO's role will be primarily through rapid upper limits and wide-field coverage complementary to CTAO.

\subsubsection{\textbf{Jetted Tidal Disruption Events}}
 
No TDE has been detected at VHE; MAGIC observed Swift\,J1644+57 ($z = 0.353$) without detection~\citep{Aleksic2013_TDE}, consistent with theoretical expectations that both $\gamma\gamma$ pair-production in the intense accretion-powered photon fields~\citep{Murase2020} and Compton scattering in the dense circum-nuclear debris~\citep{Senno2017_TDE} renders the TDE environment opaque at GeV--TeV energies. For assumed escaping luminosities $L_{\mathrm{VHE}} = 10^{43}$--$10^{45}$\,erg\,s$^{-1}$ ($\Gamma = 2.0$, 30-day integration), SWGO's detection horizon reaches $z_{\mathrm{hor}} \approx 0.1$--$0.2$, already beyond the volume within which the jetted TDE rate would yield a detection, demonstrating that sensitivity is not the \emph{primary} limiting factor. The bottleneck is internal absorption: no published model predicts escaping VHE flux above threshold, and the jetted TDE volumetric rate is only $\sim$0.03\,Gpc$^{-3}$\,yr$^{-1}$~\citep{Sun2015}. Detection remains speculative; SWGO would contribute through serendipitous discovery or constraining upper limits.

%% file: ParticleBSM.tex
    \section{Particle Physics and Beyond the Standard Model}

This section explores how SWGO's unique combination of wide-field, wide-energy range performance and Southern Hemisphere location will contribute to our knowledge of particle and fundamental physics. 

This includes probing Lorenz Invariance, testing various dark matter (DM) particle candidates including WIMPs and primordial black holes (PBHs), and in general probing subtle signatures predicted by beyond-the-standard-model (BSM) theories.

\subsection[Dark Matter]{Dark Matter}
\label{sec:DM}

Among various BSM scenarios, WIMPs in the GeV--TeV mass range remain a primary target for indirect detection. The annihilation of WIMPs into secondary particles is predicted to produce VHE gamma-ray signals peaking in high-density regions such as the GC and satellite galaxies. Given its continuous, wide-FoV monitoring of the Southern sky, SWGO provides key advantages over Southern pointing instruments like H.E.S.S. and CTAO.

In this work, we restrict our analysis to the case of self-annihilating DM; the scenario of decaying DM will be explored in a future publication \citep[for details on the statistical and modeling techniques, see, e.g.,][]{2019JCAP...12..061V}. For $m_{\rm DM}$ up to 100~TeV, we employ gamma-ray spectra from the PPPC4DMID package \citep{2011JCAP...03..051C}. For masses extending above 100~TeV into the multi-PeV range, spectra are computed using the HDMSpectra package \citep{2021JHEP...06..121B}, noting that such ultra-heavy ranges generally face constraints from unitarity bounds \citep{1990PhRvL..64..615G, 2019PhRvD.100d3029S}. To calculate expected sensitivities against the cosmic-ray background, we implement a 2D spatial and energy binned joint-likelihood method assuming a Poissonian distribution for the detected events \citep{2019JCAP...12..061V}.

\subsubsection{Dark Matter Annihilation Searches towards the Galactic Halo}
\label{subsubsec:dm_ann_searches_gal_halo}

Due to its high DM density, the GC is a primary target for setting upper limits on the annihilation cross-section. To avoid the heavy astrophysical background contamination concentrated in the inner Galactic plane, we exclude the plane ($|b| \leq 0.3^\circ$) from our analysis and focus on the surrounding Galactic halo. The spatial binning employs $0.3^\circ \times 0.3^\circ$ bins over a $6^\circ \times 6^\circ$ square region centered on the GC, while the energy range is divided into 25 logarithmically-spaced bins between 31.6~GeV and 2~PeV. For comparison with other observatories, a standard Einasto profile \citep{CTA:2020qlo, 2022PhRvL.129k1101A} is adopted for the DM density distribution.

\begin{figure*}
    \centering
    \includegraphics[width=0.47\linewidth]{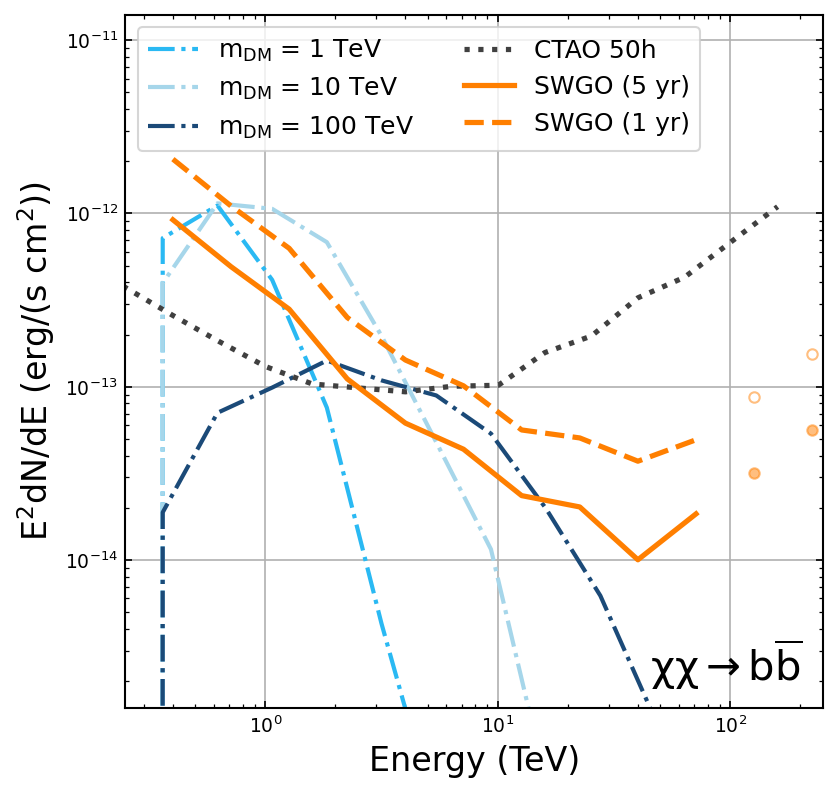}
    \hspace{0.5 cm}
    \includegraphics[width=0.47\linewidth]{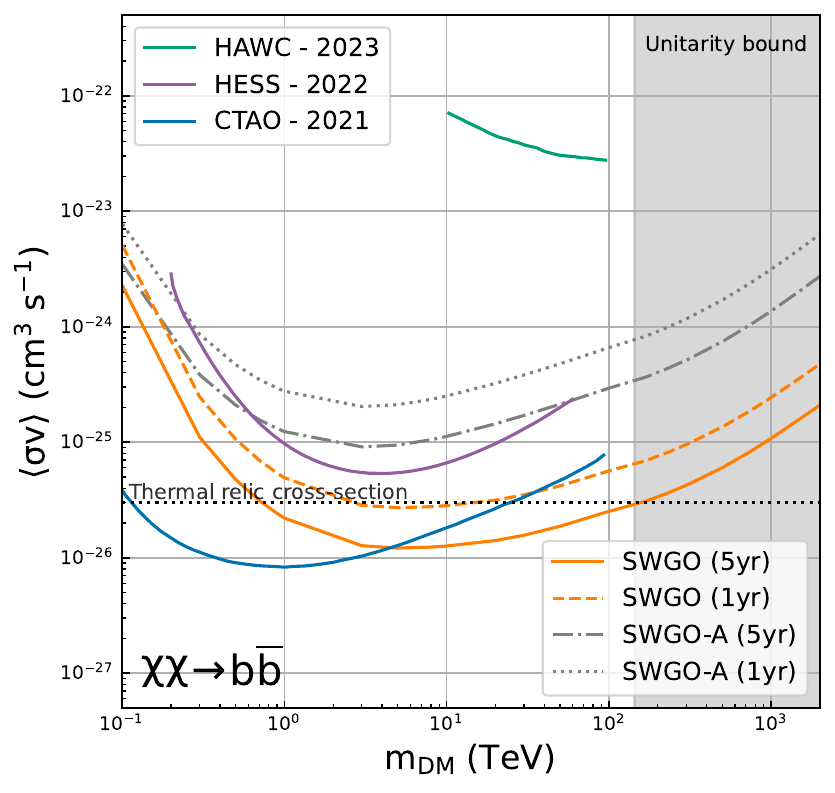}
    \caption{\textbf{Left:} SWGO flux sensitivity curves for point sources as function of reconstructed gamma-ray energy, for one and five years of observation of the Galactic Halo are shown as dashed and solid lines, respectively. The dotted regions indicate extrapolated vales. Also plotted is the DM annihilation flux into $b \bar{b}$ per reconstructed energy bin for different DM particle masses and an arbitrary normalization (dash-dotted lines). \textbf{Right:} 95\% C.L. sensitivity on the velocity-weighted cross-section for DM self-annihilation into $b\bar{b}$ as a function of $m_{\rm DM}$ for an Einasto profile for the Galactic Halo. SWGO and SWGO-A sensitivities are calculated in the inner $6^{\circ}$, excluding a $\pm0.3^{\circ}$ band in Galactic latitude. Additionally, we compare these results with the upper limits derived from HAWC~\citep{2023JCAP...12..038A}, H.E.S.S.\citep{2022PhRvL.129k1101A}, and the projected sensitivity of CTAO in the case with misidentified CRs and a Galactic Plane mask on $|b| \leq 0.3^{\circ}$\citep{CTA:2020qlo}. The green line represents HAWC, the purple line H.E.S.S., and the blue line CTAO. The solid and dashed orange lines depict the SWGO sensitivity for five years and one year, respectively, while the grey dash-dotted and dotted lines indicate the corresponding sensitivities for SWGO-A.}
    \label{fig:gc_dm_limits} 
\end{figure*}

Figure~\ref{fig:gc_dm_limits}, left panel, shows the 95\% C.L. sensitivity on $\langle \sigma v \rangle$ versus $M_{\text{DM}}$ for 1 and 5 years of observation by SWGO-A and SWGO in the case of DM particles annihilating and producing a $b\bar{b}$ pair. This choice of annihilation channel is conservative, as other channels (e.g., $\tau^+\tau^-$ or $W^+W^-$) tend to yield harder gamma-ray spectra and thus stronger signals at the high-energy end of the spectrum~\citep{2019JCAP...12..061V}. The sensitivity of CTAO (for the case of only misidentified CRs and with a Galactic Plane mask on $|b| \leq 0.3^\circ$)~\citep{CTA:2020qlo}, HAWC~\citep{2023JCAP...12..038A}, and H.E.S.S.~\citep{2022PhRvL.129k1101A} are plotted for comparison. A sensitivity smaller than the thermal relic cross-section $\langle \sigma v \rangle \lesssim 3 \times 10^{-26}$ cm$^3$ s$^{-1}$ is reachable for SWGO in the mass range of $\sim$800~GeV to $\sim$140~TeV. The sensitivity is better than that of CTAO for all masses $\gtrsim$5~TeV and several orders of magnitude better than that of HAWC. Within five years of operation, SWGO-A will reach a sensitivity comparable to that of H.E.S.S.~\citep{Cirelli:2024ssz}, obtained after two decades of operation and currently one of the most sensitive instruments to annihilating dark matter in the mass range from a few TeV to a hundred TeV.

SWGO’s DM searches are highly complementary to those of the \emph{Fermi}-LAT and CTAO, collectively placing stringent constraints on the thermal relic annihilation cross-section for masses up to 100~TeV (see Figure~\ref{fig:gc_dm_limits}, left panel, and \cite{McDaniel:2023bju}). In the mass range from a few hundred GeV to a few tens of TeV, joint observations of the GC by CTAO and SWGO can significantly boost statistical significance of a detection, where SWGO probes the high-energy spectral cut-off and CTAO resolves the spatial morphology of a potential signal (see Figure~\ref{fig:gc_dm_limits}, right panel).

\subsubsection{Dark Matter Searches in Satellite Galaxies} 

Dwarf spheroidal galaxies (dSphs) represent exceptionally clean targets for DM searches due to their high mass-to-light ratios and low intrinsic astrophysical backgrounds. For this analysis, data are combined from 32 distinct dSphs that fall within the FoV of SWGO out of a reference list of 50 recently evaluated by the \emph{Fermi}-LAT \citep{McDaniel:2023bju}. They include dSphs with high measured J-factors, such as Coma Berenices ($\log_{10} J = {19.00 \pm 0.35} \;\mathrm{GeV}^2\,\mathrm{cm}^{-5}$ ), Reticulum-II ($\log_{10} J = {18.9 \pm 0.38}$), and Horologium-I ($\log_{10} J = {19.00 \pm 0.81}$), as well as dSphs with high estimated J-factors, such as Cetus-II ($\log_{10} J = {19.10 \pm 0.60}$) and Carina-III ($\log_{10} J = {19.70 \pm 0.60} $). The galaxies that dominate the combined sensitivity are determined by a combination between high J-factor and large yearly exposure to SWGO.

\begin{figure*} 
\centering 
\includegraphics[width=0.47\linewidth]{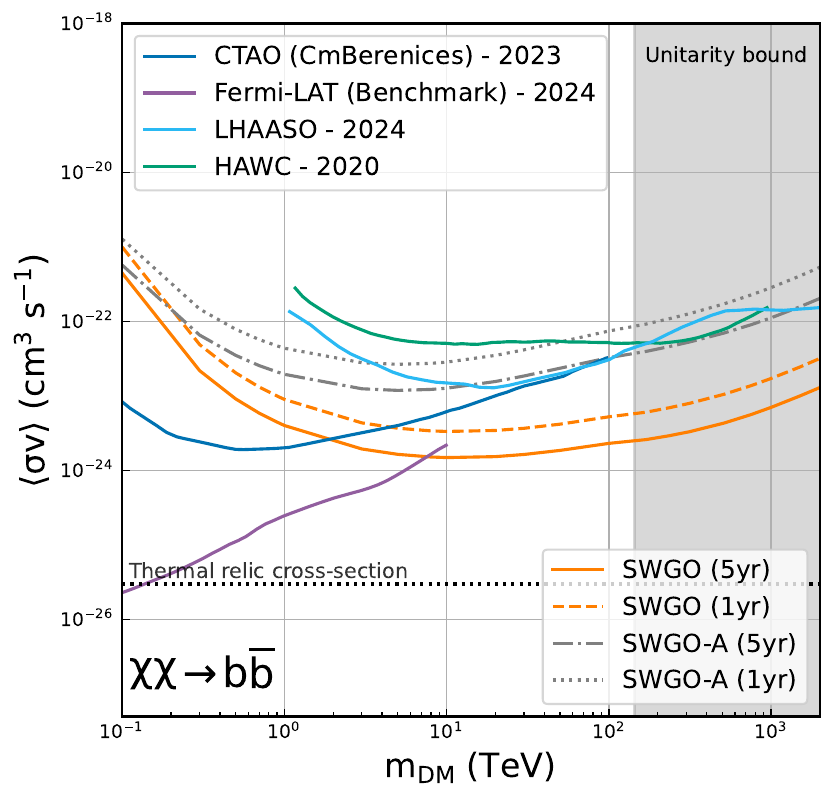} 
\hspace{0.5 cm}
\includegraphics[width=0.47\linewidth]{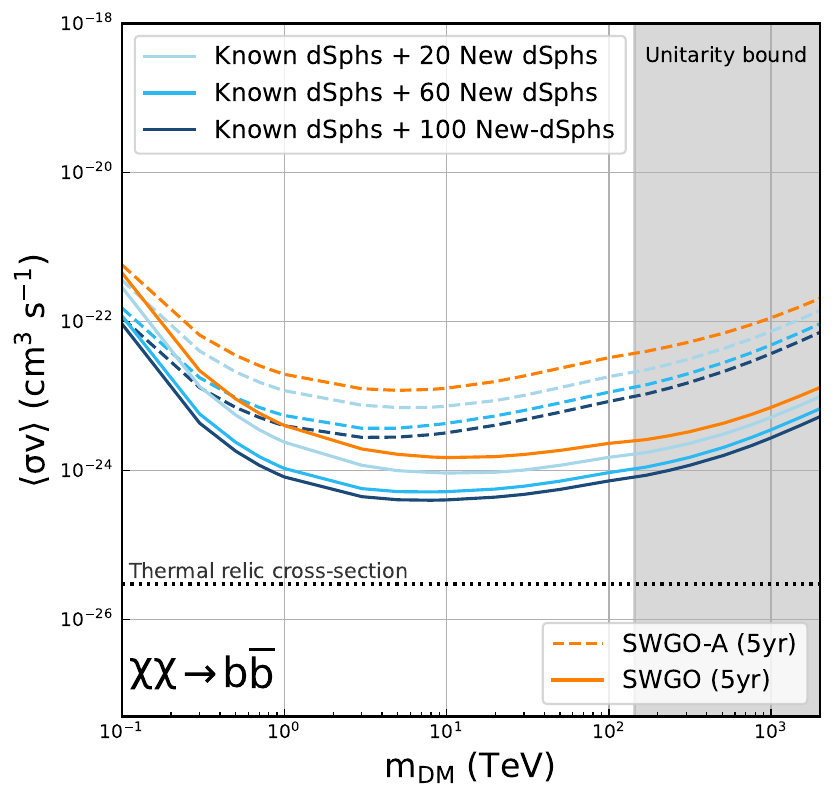} 
\caption{\textbf{Left:} 95\% C.L. sensitivity on the velocity weighted cross-section for DM self-annihilation into $b\bar{b}$ as a function of $m_{\rm DM}$ for 32 known dwarf galaxies observable by SWGO~\citep{McDaniel:2023bju}. Additionally, we compare these results with the upper limits derived from the HAWC~\citep{2020PhRvD.101j3001A}, \emph{Fermi}-LAT~\citep{McDaniel:2023bju}, and LHAASO~\citep{2024PhRvL.133f1001C} experiments, as well as the future sensitivity of CTAO~\citep{2023arXiv230909607S}, assuming $\sim$500~hours of observation time on Coma Berenices alone. CTAO limits are denoted in blue, the \emph{Fermi}-LAT limits  in purple, LHAASO limits in cyan, and HAWC limits in green. The solid and dashed orange lines depict SWGO sensitivities for five years and one year, respectively, while the grey dash-dotted and dotted lines indicate the corresponding sensitivities for SWGO-A. \textbf{Right:} Extension of the sensitivity curves, accounting for potential new dSphs discovered in the coming years. The three curves illustrate the impact of future discoveries on the sensitivity: the light blue line includes 20 new dSphs, the cyan line includes 60 new dSphs, and the blue line includes 100 new dSphs, all assumed to follow the same J-factor distribution as that of the currently known population and to be isotropically distributed across the sky.} 
\label{fig:dwarf_dm_limits} 
\end{figure*}

Figure~\ref{fig:dwarf_dm_limits} (left panel) displays the 95\% C.L. sensitivity on $\langle \sigma v \rangle$ for the $b\bar{b}$ channel. The results are compared against projected CTAO sensitivity for Coma Berenices alone, assuming $\sim$500~hours of observation time~\citep{2023arXiv230909607S}, as well as against the combined limit from HAWC~\citep{2020PhRvD.101j3001A}, the combined limit from the \emph{Fermi}-LAT~\citep{McDaniel:2023bju}, and the combined limit from LHAASO~\citep{2024PhRvL.133f1001C}. Due to its higher exposure and simultaneous coverage of multiple targets, SWGO is projected to exceed the sensitivity of CTAO in the multi-TeV DM mass range, while notably improving upon the limits set by existing wide-field instruments like HAWC.

Due to its continuous monitoring capabilities, SWGO can accumulate data across all regions of the Southern gamma-ray sky, including those where new dSphs may later be discovered, unlike IACTs, which require targeted observations. For instance, the Vera C. Rubin Observatory's Legacy Survey of Space and Time (LSST)~\citep{2019ApJ...873..111I} is expected to discover hundreds of new dSphs in the Southern Hemisphere with unprecedented sensitivity~\citep{2022MNRAS.516.3944M}. Once identified, these new targets can be promptly analyzed using existing SWGO data, offering a major advantage for indirect DM searches. 

To quantify this potential, the right panel of Figure~\ref{fig:dwarf_dm_limits} illustrates the projected improvement in the sensitivities of SWGO and SWGO-A by extending the baseline joint-likelihood of 32 known dSphs with 20, 60, and 100 hypothetical new systems. These new targets are assumed to follow the same J-factor distribution as the current sample and to be distributed isotropically across the sky. The projected improvement comes from the increased statistical power in the joint-likelihood analysis simulations, not from a change in the underlying DM physics. Under these assumptions, the sensitivity can improve by a factor $\geq$5 relative to the 32-target baseline. The scaling is sub-linear because the combined likelihood is dominated by the few targets with the largest J-factors and most favorable yearly exposure to SWGO, meaning the actual gain will depend on whether the newly discovered dSphs possess J-factors comparable to the most prominent systems currently known.

\subsection[Primordial Black Holes]{Primordial Black Holes}

Beyond particle DM candidates, SWGO may also probe early-Universe relics such as PBHs, which could contribute to the DM abundance. Black holes (BHs) are solutions of Einstein’s field equations for a point-like source of mass $M$. The simplest solution is for the Schwarzschild metric (spherical symmetry), which defines the event horizon at a radius, $r_{\text{S}} = 2GM/c^{2}$, where $G$ and $c$ denote the gravitational constant and the speed of light, respectively. BHs resulting from stellar gravitational collapse span a mass range from a few to millions of solar masses. However, in early stages of the Universe during the radiation-dominated era, the collapse of over-densities resulting from strong inhomogeneities may have led to the formation of low-mass BHs, known as PBHs. Furthermore, this hypothetical mechanism may extend the BH mass spectrum up to the Planck scale, $M_{\text{Pl}}\sim10^{-5}$g \citep{PBH_hawking1}. Since PBHs are non-baryonic, they are considered viable candidates for cold DM, attracting increasing interest as partial contributors to (or even the main constituents of) DM in the Universe \citep{PBH_DM}.

In the 1970s, S. Hawking applied quantum field theory to spacetime in the vicinity of the event horizon, concluding that BHs can radiate particles of any spin that follow a thermal spectrum with temperature $kT=\hslash c^{3}/8\pi GM\sim (10^{13}$ g$/M)$ GeV, where $\hslash$ is the reduced Planck constant and $k$ the Boltzmann constant \citep{PBH_hawking2}. As the BH increases in temperature, heavier particles start to be produced, which decreases the BH mass at a rate $dM/dt=-\alpha(M)/M^{2}$, where $\alpha(M)$ accounts for the degrees of freedom of the emitted particles. Integrating the mass-loss rate gives the initial mass, $M_{0}$, required to be reduced to zero (evaporation) in time, $\tau$, according to the relation $M_{0} \approx [3\alpha(M_{0})\,\tau]^{1/3} \approx 1.3\times10^{9}\,[\tau/1\text{s}]^{1/3}$~g, where the value of $\alpha$ is from \cite{PBH_ukwatta}, which considers only Standard Model (SM) particles. PBHs with initial masses $M_{0}\sim10^{15}$~g, should have $\tau$ comparable to the age of the Universe and, therefore, be evaporating today. 

Primordial Black Hole evaporation ends with an energetic short-lived particle burst. The energy of the emitted gamma rays depends on the physics that govern the last stages, i.e., the number of degrees of freedom of the particles that are being emitted. In the SM evaporation scenario, the energy of gamma rays in the final explosion covers the range from MeV to TeV. In particular, photons in the 1--100~TeV band are emitted for evaporation times $\mathcal{O}(10^{-1})$ to a few seconds \citep{PBH_Halzen}. Gamma-ray emission primarily occurs through two channels: (a) photons from Hawking radiation and (b) photons from hadronic decays during quark and gluon fragmentation (when the temperature of the PBH reaches $kT\sim \Lambda_{QCD}\sim 200$--$300$~MeV). In the latter, the neutral-pion decay into two photons is the major contribution to the spectrum. For a PBH with evaporation time $\tau$ and temperature at the beginning of the evaporation time $kT_{\tau}$, the number of photons per unit energy $dN/dE$ can be parametrized as \citep{PBH_spectrum}:
\begin{equation}
    \frac{dN}{dE} \approx 9 \times 10^{35} 
    \left\{
    \begin{array}{ll}
        \left( \frac{E}{\mathrm{GeV}} \right)^{-3} \ \mathrm{GeV}^{-1} \ , & \text{for } E \geq kT_{\tau}\\
        \left( \frac{E}{\mathrm{GeV}} \frac{kT_{\tau}}{\mathrm{GeV}} \right)^{-3/2} \ \mathrm{GeV}^{-1} \ , & \text{for} \ E < kT_{\tau} 
    \end{array} \right. \, ,
\end{equation}
for photons with $E\ge1$ GeV. Although the final stages of PBH evaporation may resemble short GRBs, their light-curve exhibits some distinctive features \citep{PBH_profumo}: 
\begin{itemize}
    \item Short duration: The burst typically lasts from a few seconds to several tens of seconds.
    \item Non-decreasing luminosity: Under the assumption that only SM particles are emitted and no additional degrees of freedom are involved, the luminosity increases steadily until the final moment of evaporation.
    \item Locally-constrained signal: Given the sensitivities of current instruments, PBH bursts are expected to be detected within a few parsecs of Earth.
    \item Universal intrinsic spectrum: The intrinsic emission spectrum is characteristic of the evaporation process, implying that different PBH bursts should appear spectrally identical.
\end{itemize}

Several instruments have set upper limits for the PBH burst-rate density, $\dot{n}$. In the case of Cherenkov telescopes such as H.E.S.S. and VERITAS, the upper limits are $\dot{n}<2\times 10^{3}$~pc$^{-3}$yr$^{-1}$ with $95\%$ confidence level, and $\dot{n}<2.22\times 10^{4}$~pc$^{-3}$yr$^{-1}$ with $99\%$ confidence level, respectively \citep{PBH_hess,PBH_veritas}. In the case of wide-FoV WCDs, HAWC has established an upper limit $\dot{n}<3.4\times10^{3}$~pc$^{-3}$yr$^{-1}$ with $95\%$ confidence level using three years of data with events located out to 0.5~pc \citep{PBH_HAWC}. Most recently, prospective observations of LHAASO have set more constraining upper limits with $\dot{n}<7\times10^{2}$~pc$^{-3}$yr$^{-1}$ with five years of data with $99\%$ confidence level, considering bursts at distances up to $\sim$0.1~pc from the Earth \citep{PBH_lhasso}.

SWGO's approach to PBH burst detection (or the setting of upper limits) will be based on previous gamma-ray detector strategies. This will consist of counting the number of photons, $\mu(r,\tau)$, from a PBH burst with an evaporation time $\tau$ inside a volume of radius $r$, which can be expressed as:
\begin{equation}
    \mu(r, \tau) = \frac{1 - f}{4 \pi r^2} \int_{E_1}^{E_2} \mathrm{d}E' \int_{0}^{\infty} \mathrm{d}E \frac{\mathrm{d}N(\tau)}{\mathrm{d}E} A(E)G(E, E')\enspace ,
    \label{eq:placeholder}
\end{equation}
where $E$ $(E')$ is the true (reconstructed) energy, $A$ is the effective area, $G$ is the resolution, and $f$ is the dead time of the detector (given SWGO's high duty cycle, $f$ will be negligible). There will be a particular focus on detecting very short bursts in order to capture emission from the TeV energy band and to minimize dependence on the zenith angle of the source. Access to the Southern sky plus the improved sensitivity of SWGO compared to other WCD facilities present a major opportunity to search for signatures of PBH evaporation and to further constrain existing upper limits \citep{PBH_swgo}. Although this is a serendipitous phenomenon, its detection would have a profound implication for fundamental physics through its interplay in cosmology (evidence of non-stellar BH formation), high-energy physics (probing BSM physics), and gravitation (testing Hawking radiation as a key quantum gravitational phenomenon).

\subsection[Axion-Like Particles]{Axion-Like Particles}
\label{sec:ALPs}

Axion-Like Particles (ALPs) are hypothetical particles predicted in many extensions of the Standard Model \citep{alp_theory}, related to the axion \citep{axions}, and considered viable DM candidates \citep{alp_dm}. Each ALP candidate is uniquely characterized by two parameters: its mass, $m_a$, and its photon coupling factor, $g_{a\gamma}$.

If ALPs exist, their conversion into photons could allow for the detection of gamma rays that would otherwise be absorbed. A potential observational signature of ALPs may involve the detection of photons with energy $\gtrsim$30~TeV coming from a hard-spectrum extragalactic source or $\gtrsim$10~TeV from soft-spectrum Galactic sources. These high-energy regimes are particularly suited for indirect searches with SWGO, the energy range of which is expected to go beyond 100~TeV.

The existence of ALPs is still an open question in current physics; their indirect detection from astrophysical sources would allow us to test various theoretical models and improve our understanding of the systematic uncertainties associated with the EBL models \citep{EBLFran,EBLDomin,EBLFinke} and, by extension, the intrinsic energy spectrum. 

\begin{figure*}
    \centering
    \includegraphics[width=0.47\linewidth]{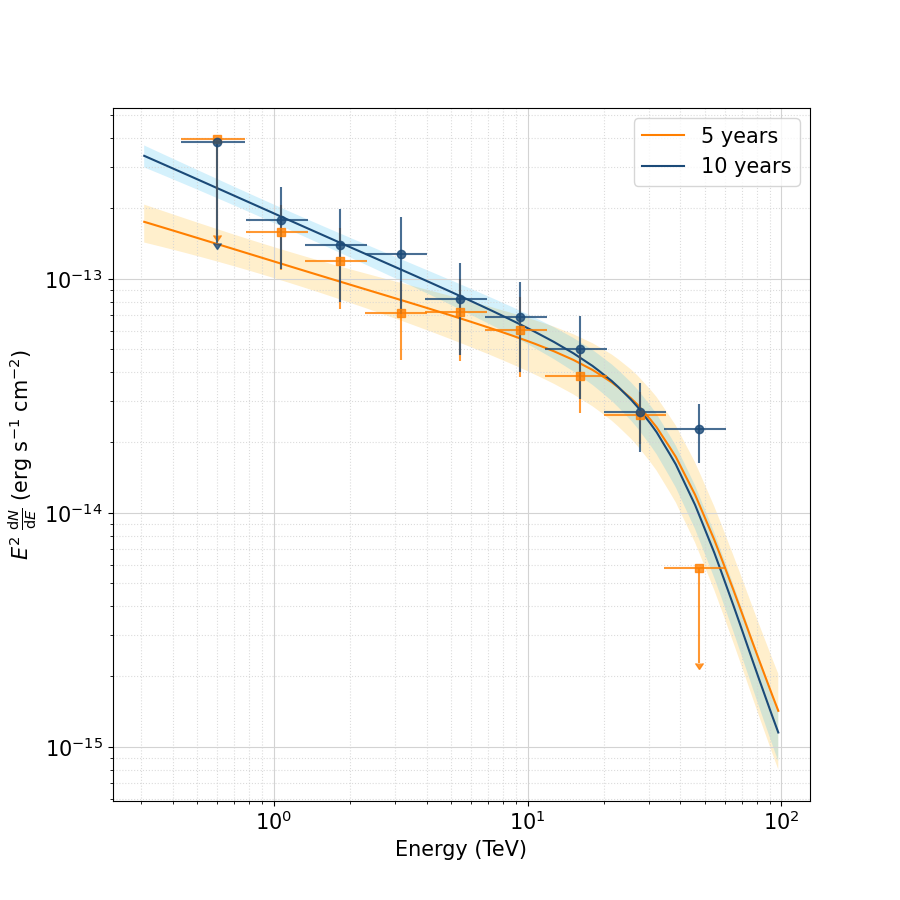}
    \includegraphics[width=0.47\linewidth]{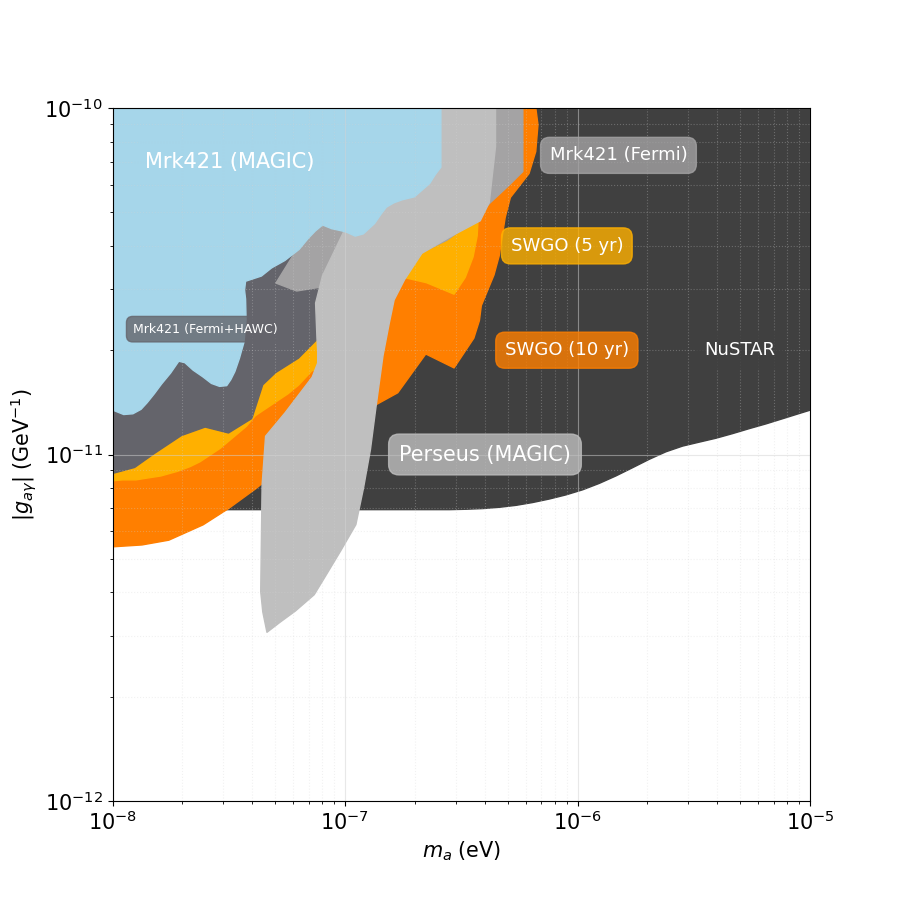}
    \caption{\textbf{Left:} Simulated emission of Cen~A detected with the full SWGO array for five (orange) and ten (blue) years of observation, assuming a power-law emission spectrum. \textbf{Right:} Exclusion regions in the $(m_a,g_{a\gamma})$ phase-space associated with the observations of Cen~A with the full SWGO array for five and ten years, in comparison to exclusion regions from other experiments.}
    \label{fig:ALPsPlots}
\end{figure*}

A particularly relevant source for study is the AGN Centaurus A (Cen A) due to its intense emission of very-high-energy gamma rays, its proximity to Earth ($z=0.0018$), and the astrophysical conditions favorable for photon-to-ALP conversion due to the source's magnetic field, $B=6.2\;{\rm G}$ \citep{CenARef1, CenARef2}. The emission of this point-like extragalactic source has been modeled as a PL spectrum with index $\alpha = 2.52 \pm 0.19$ and normalization $\phi_0 = (1.49 \pm 0.43) \times 10^{-13} \ \text{TeV}^{-1}\text{cm}^{-2}\text{s}^{-1}$ at an energy of 1~TeV \citep{CenARef3}. Photons up to $\sim$50~TeV are expected, even accounting for EBL absorption.  With its wide energy coverage extending beyond 100 TeV, SWGO offers the opportunity to probe emissions from Cen~A, potentially providing indirect evidence for photon-ALP conversion. 
Figure~\ref{fig:ALPsPlots}, left panel, shows the
expected spectrum, assuming an emitted power law, of Cen~A with the full SWGO array for five and ten years of observation. Figure~\ref{fig:ALPsPlots}, right panel, shows
the exclusion regions in the $(m_a,g_{a\gamma})$ phase-space with a 95\% C.L., computed using these two observation times, and in comparison with exclusion regions from different experiments. For these simulations, the EBL model from \cite{EBLFran} was assumed. As the Figure shows, SWGO will probe new ALP parameter space with such a Cen~A observation. In addition, its wide FoV provides the potential for finding new targets for such ALP searches.

\subsection{Testing Lorentz Invariance}
\label{tLIV}

Very-high-energy photons serve as pivotal probes in testing fundamental physics principles, such as Lorentz symmetry. Examining the limits of any symmetry's validity is essential for theoretical advancements and experimental physics. These investigations are motivated by the hypothesis that Lorentz invariance violation (LIV) may arise from BSM theories, including quantum gravity and string theories \citep[][and references therein]{Addazi:2021xuf, Martinez-Huerta:2020cut, Kostelecky:2008ts}. 

Previous studies indicate that LIV may introduce modifications in the propagation of high-energy photons, including changes in their velocity, interaction thresholds, and arrival times. These effects can be modeled by modifying the photon dispersion relation, which takes the form:
\begin{equation}
\label{LIV:eq1}
E_{\gamma}^2 - p_{\gamma}^2 = \pm \frac{E_{\gamma}^{n+2}}{\left(E_{\text{LIV}}^{(n)}\right)^n}\enspace , 
\end{equation}
where $E_{\gamma}$ and  $p_{\gamma}$ are the energy and momentum of the photon, respectively, and $E_{\text{LIV}}^{(n)}$ denotes the scale of LIV at order $n$. These scales indicate the energies at which corrections of the corresponding order become relevant. 

The sign ambiguity in Eq.~\ref{LIV:eq1} gives rise to two distinct scenarios: superluminal ($+$) and subluminal ($-$). Phenomenological consequences of such modifications are typically studied on a term-by-term basis for specific orders, with a focus on the linear ($n=1$) and quadratic ($n=2$) cases, as these yield the most significant effects and are more readily constrained by observations. The observable consequences of such modifications include, among others, anomalies in particle interaction thresholds, photon decay, photon splitting, birefringence effects, and time delays \citep{Addazi:2021xuf}. For instance, in the linear case, birefringence (where photons of opposite polarization states propagate at different speeds) can place a lower bound on the energy scale at which new physics may emerge, $E_{\text{LIV}}^{(1)}$, many orders of magnitude above the Planck scale, $E_{\text{Pl}}\approx 1.22 \times 10^{28}$eV \citep{Wei:2019nhm, Gotz:2014vza}.

In the sub-luminal scenario of Eq.~\ref{LIV:eq1}, attention is given to anomalies in particle interaction thresholds ~\citep{Scully:2008jp, Stecker:2009hj, Galaverni:2007tq}. 
These effects alter the kinematics of the pair-production process $\gamma\gamma \to e^-e^+$,  which governs the attenuation of VHE gamma rays through interactions with low-energy background photons.
Specifically, the threshold energy for this pair-production process is modified. For head-on collisions, the threshold energy, $\epsilon_{\text{thr}}$, of the low-energy photon is given by:
\begin{equation}
\label{LIV:eqthr}
\epsilon_{\text{thr}} = \frac{m_{e^-}^2}{E_\gamma} + \frac{E_\gamma^{n+1}}{4E_{\text{LIV}}^{(n)}} \enspace,
\end{equation}
where the second term on the right-hand side, introduced by LIV, results in the appearance of a minimum in $\epsilon_{\text{thr}}$. In contrast to the Special Relativity (SR) scenario, where the energy threshold decreases monotonically with increasing gamma-ray energy, the presence of a minimum in $\epsilon_{\text{thr}}$ leads to greater transparency of the background medium beyond a certain critical gamma-ray energy~\citep{Jacob:2008gj}. The critical energy $E_\gamma^*$ of gamma rays corresponding to this minimum is determined by differentiating Eq.~\ref{LIV:eqthr}. For the linear case, the critical energy is:
\begin{equation}
E_\gamma^* \sim 18.5\ \text{TeV} \left(\frac{E_{\text{LIV}}^{(1)}}{E_{\text{Pl}}}\right)^{1/3} \,.
\end{equation}

Given the strong constraints arising from birefringence effects, the quadratic case provides a particularly relevant phenomenological window. In this case, the critical energy can be expressed as:
\begin{equation}
E_\gamma^* \sim 8.5 \times 100 \ \text{TeV} \left(\frac{E_{\text{LIV}}^{(2)}}{10^{-4} E_{\text{Pl}}}\right)^{1/2}.
\end{equation}
Current bounds on $E_{\text{LIV}}^{(2)}$ from anomalous transparency studies, derived from single-source analyses, reach approximately $7.8 \times 10^{20}$~eV ($6.4 \times 10^{-8}E_{\text{Pl}}$), based on observations of gamma rays up to 20~TeV from the Mrk~501 flare~\citep{HESS:2019rhe}. Observations at energies of order $\mathcal{O}(1)$~PeV with SWGO are expected to extend the sensitivity to significantly higher LIV scales.

Recent work~\citep{Carmona:2024thn} has provided a first-principles calculation of the pair-production cross section within this same quadratic subluminal LIV framework. Earlier studies typically relied on approximations based on the SR cross section with modified thresholds, which may lead to an overestimation of the LIV-induced transparency and consequently to an underestimation of the corresponding LIV scale. The improved treatment of the interaction process therefore leads to more reliable predictions for photon survival probabilities and the associated sensitivity reach.

In super-luminal scenarios, Eq.~\ref{LIV:eq1} permits photon decay into electron-positron pairs $(\gamma\rightarrow e^+ + e^-)$ above a certain energy threshold, leading to rapid decay of high-energy photons~\citep{Martinez-Huerta:2016azo}. This effect would result in a suppression of photon emission from sources exceeding this energy threshold, which would manifest as a hard cutoff on the high-energy tail of spectral sources~\citep{HAWC:2019gui}. As a result, the non-observation of such a cutoff in measurements of astrophysical VHE photons constrains the energy scale at which photon decay becomes kinematically allowed, based on the observation of photons with energies up to $E_{\text{obs}}$. This allows us to refine the bounds on the LIV energy scale, given by:
\begin{equation}
E_{\text{LIV}}^{(n)} > E_{\gamma} \left[ \frac{E_{\gamma}^2 - 4m_{e}^2}{4m_{e}^2} \right]^{1/n}.
\end{equation}

In this context, observing photons with energies ranging from 1 to 5~PeV  by SWGO may substantially enhance constraints on LIV derived from Southern Hemisphere observations. The absence of decay signatures can be used to establish limits on LIV, potentially to or beyond the Planck scale, $E_{\text{Pl}}$, depending on the decay model (linear or quadratic). For a linear modification, the resulting super-luminal LIV bounds may range from  $1.93 \times 10^{33}$~eV to $1.2 \times 10^{35}$~eV. For a quadratic modification, using similar calculations, they range from $9.78 \times 10^{23}$~eV to $2.45 \times 10^{25}$~eV.  Current stringent super-luminal limits for $n=1$ are on the order of $10^{33}$~eV~\citep{Li:2022ugz}. 

Additionally, photon splitting processes, a phenomenon where high-energy photons split into multiple lower-energy photons under LIV scenarios with $n=2$, may also offer additional constraints on LIV~\citep{Gelmini:2005gy,Satunin:2019gsl}. Observations of sources like RX J1713.7-3946 without any signatures of photon splitting at energies $10^{14}-10^{15}$ eV can further limit the LIV scale, as predicted by:
\begin{equation}
E_{\text{LIV}}^{(2)} > 3.33 \times 10^{19} \, \text{eV} \left(\frac{L}{\text{kpc}}\right)^{0.1} \left(\frac{E_{\gamma}}{\text{TeV}}\right)^{1.9}.
\end{equation}
The calculated LIV limits for photon splitting at $10^{14}$ eV and $10^{15}$ eV are $2.10 \times 10^{23}$ eV and $1.67 \times 10^{25}$ eV, respectively, providing further empirical support for refining the bounds on LIV.

A comprehensive exploration of both sub-luminal and super-luminal regimes is essential to systematically test the full range of predictions arising from Lorentz-violating extensions. Addressing both cases independently enhances the interpretative power of observational constraints and ensures sensitivity to the diverse phenomenology predicted by different theoretical frameworks. Furthermore, as explored in Section~\ref{sec:transients}, gamma-ray observations of transients may also serve as critical probes for LIV scenarios and ALP-induced modulations.


%% file: CosmicRay.tex
\section[Cosmic-Ray Measurements]{Cosmic-Ray Measurements}
\label{sec:CR}

In addition to a vast range of gamma-ray topics, SWGO will act as a unique tool for studying charged CRs in the TeV--PeV regime. Its large effective area and advanced muon counting capability, combined with its privileged location in the Southern Hemisphere, will allow the pursuit of relevant open questions in TeV--PeV CR astrophysics. The main topics of interest are described in detail below.

\subsection{Spectrum and Composition of Cosmic Rays around the Knee}
\label{subsec:spec_and_comp_measurements_knee}

The mechanism behind the acceleration of CRs and the physical processes that affect their propagation through space are subjects of intense debate. An indirect approach to investigating these is to study the CR energy spectrum and mass composition. In the energy region around and below the knee (1~TeV~$\lesssim E \lesssim$~10~PeV), in which SWGO will be sensitive, there have recently been several experimental efforts to study the spectrum and the relative abundances of elemental mass groups of CRs. These aim to fill the gap between direct and indirect CR data and to refine the measurements around the knee to investigate the properties of Galactic CRs. Between 10~TeV and 1~PeV, measurements on the all-particle spectrum of CRs from the NUCLEON satellite \citep{Atkin17, GREBENYUK20192546} and the HAWC Observatory \citep{Alfaro17a, Alfaro25} have established a first bridge with high statistics and large precision between the TeV CR balloon- and space-borne experiments and the PeV air-shower observations. In addition, they have shown the existence of a softening at tens of TeV in the total spectrum of CRs.  Besides this new feature, in the last decade, similar features have also been discovered in the spectra of elemental mass groups of TeV CRs by both direct (e.g., NUCLEON  \citep{GREBENYUK20192546}, DAMPE \citep{An19, Alemanno21, Alemanno24, Alemanno26}, and CALET \citep{Adriani19, Adriani23}) and indirect experiments \citep[HAWC,][]{Albert22}. DAMPE has shown that the softenings in the elemental spectra of CRs are located at the same rigidity value, around 15~TeV \citep{Alemanno26}, which could point out a Peters cycle for the acceleration of TeV CRs \citep{Peter1961}. These results could also indicate the presence of an individual nearby TeV CR source inside our Galaxy \citep{Liu19} or a complete population of such a kind of CR sources in the Milky Way \citep{Yue20}. It could also point out an unknown mechanism affecting the propagation of TeV CRs in our Galaxy \citep{Malkov22}. 

 Hardenings at around 100~TeV in the spectra of H and He nuclei have also been reported by HAWC \citep{Arteaga23}. GRAPES-3 has confirmed a hardening in the spectrum of protons at $\sim$166~TeV \cite{Varsi24}. LHAASO has also recently detected two hardenings in the CR spectra of protons and helium close to 342~TeV \citep{LHAASO_protons} and  1.1~PeV \citep{LHAASO_helium}, respectively. There seems to be a discrepancy in the location of these hardenings when comparing LHAASO data with HAWC and GRAPES-3 results, which calls for additional CR measurements at $\sim$100~TeV. In any case, the hardenings detected at TeV energies in the spectra of the light CR nuclei may indicate the point where the contribution from possible TeV CR sources is surpassed by that from the background of PeV CR accelerators, which are expected to be responsible for the knee. 
 
At PeV energies, measurements of the air-shower experiments KASCADE \citep{kascade1, kascade2, APEL201354}, IceTop / IceCube \citep{IceCube_all}, and MAKET-ANI \citep{MaketANI, MaketANI1} indicate that the knee is produced by cutoffs in the spectra of light primaries (H and He nuclei). Also, the latest results of LHAASO on the mean logarithmic mass support the picture that the knee is dominated by light primaries \citep{LHAASO_knee}. KASCADE-Grande data have also shown individual cutoffs in the spectra of the CNO and Fe components. Interestingly, within systematic and statistical errors, the positions of the breaks detected by KASCADE and KASCADE-Grande in the elemental groups of CRs seem to depend on the charge of the primary particle \citep{kascade1, kascade2, APEL201354}, an observation that may be hinted at by LHAASO, at least, for the breaks at the spectra of protons \citep{LHAASO_protons} and helium nuclei \citep{LHAASO_helium}. If confirmed, these results may imply a rigidity-dependent scenario consistent with a Peters cycle \citep{Peter1961}. However, propagation effects on the CR spectra are expected to explain these results. If so, the knee would indicate the point where Galactic PeV CRs start to escape from their magnetic confinement in the Milky Way \citep{Candia02, horandel04, Giacinti23}.

In order to test different hypotheses about the origin of CRs at TeV and PeV energies, the shape of the energy spectra and the location of the knee-like features for the elemental mass groups of CRs must be measured with enough precision. Additionally, the  TeV softenings observed by DAMPE in the elemental spectra of CRs need confirmation with large statistics and high precision. In HAWC, the energy resolution for CRs is found between $21\%$ and $53\%$ from 10~TeV to 1~PeV \citep{Alfaro25} and for LHAASO, between $15\%$ and $40 \%$ in the 10~TeV--4~PeV interval \citep{LHAASO:2021Detector}. DAMPE has a resolution in the range of 25--35\% for energies between 100~GeV and 10~TeV \citep{An19}. The resolution of SWGO is expected to be better than that of HAWC (see, for example, Section~\ref{sec:EventReco}), hence, SWGO can contribute to the above scientific tasks by measuring the elemental spectra in the TeV energy regime for H, He, CNO, and Fe primary nuclei.

SWGO data would also be important in investigating the transition from the TeV to the PeV energy regime in the spectra of CRs by measuring the TeV hardenings in the spectra of CRs with high precision. In addition, SWGO can contribute to measuring the precise location of the p, He, and CNO knee-like features at PeV energies to test models about the production and propagation of Galactic CRs. LHAASO results on the energy spectrum of protons and helium, and on the abundance of light primaries around the knee, need to be confirmed. The question of whether the knee is dominated by light primaries is of importance to constrain the limit of maximum acceleration of CRs by Galactic sources \citep{horandel04, Vink22} and the effect of their propagation through the intergalactic medium \citep{horandel04, Giacinti23}. It is also relevant to establish the relation between the knee and the recent observations of hadronic PeVatrons in our Galaxy (see Section~\ref{sec:PeVatron}). The TeV cutoff in the total spectrum of CRs must also be investigated because of its implications on the type of CR sources and the proximity of the CR accelerators to Earth.  It must be mentioned that SWGO measurements in the TeV energy range will also be important to cross-check the energy scales of current direct and indirect experiments.
  
\begin{figure}[ht]
    \centering
    \includegraphics[width=\linewidth]{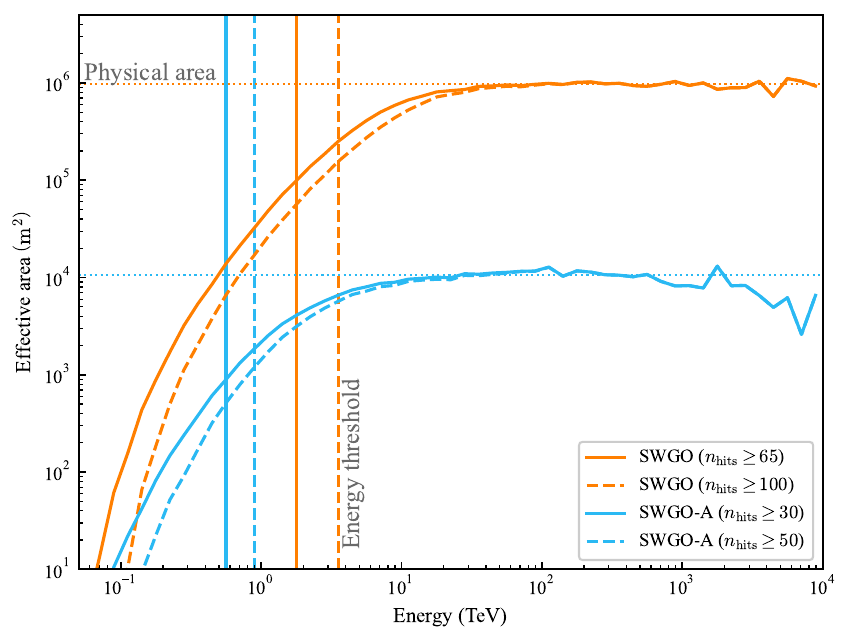}
    \caption{Effective area for protons for the SWGO (orange) and SWGO-A (blue) configurations. Only events with their core within the array and zenith angles between $0^{\circ}$ and $65^{\circ}$ are considered. An optimistic and a high-threshold scenario were considered, each with different selection criteria based on the number of hits and represented by full and dashed lines, respectively. Horizontal dotted lines denote the physical area of the array for each configuration, $A_{\rm{SWGO}}= \pi (560 \, \rm{m})^2 \approx 9.8 \times 10^5 \, \rm{m^2}$ and $A_{\rm{SWGO-A}}= \pi (58 \, \rm{m})^2 \approx 1.05 \times 10^4 \, \rm{m^2}$. Vertical lines denote the energy threshold (defined as the energy at which the effective area reaches $10\%$ of its saturated value), $E_{\rm{thr}}^{\rm{SWGO (\geq 65)}} \approx 1.75 \, \rm{TeV}$, $E_{\rm{thr}}^{\rm{SWGO (\geq 100)}} \approx 3.5 \, \rm{TeV}$, $E_{\rm{thr}}^{\rm{SWGO-A (\geq 30)}} \approx 550 \, \rm{GeV}$, and $E_{\rm{thr}}^{\rm{SWGO-A (\geq 50)}} \approx 900 \, \rm{GeV}$.}
    \label{fig:CREffectiveArea}
\end{figure}

SWGO will offer insight into the above questions by: (a) providing a large effective area and high duty cycle for the detection of CRs in the TeV--PeV energy regime, and (b) taking advantage of improved muon-counting capabilities compared to current instruments that are sensitive to the same energy range, e.g., HAWC. 
An estimate of the proton effective area achievable with SWGO is shown in Figure~\ref{fig:CREffectiveArea}. For this estimation, we apply conservative selection criteria regarding the core location, only considering events for which the shower core falls within the array boundaries and with a zenith angle between $0^{\circ}$ and $65^{\circ}$. As described in Section~\ref{sec:performance}, an optimistic and a high-threshold scenario are considered for the minimum number of hits, 65 and 100 for SWGO and 30 and 50 for SWGO-A, respectively. We define the energy threshold as the energy for which the effective area reaches 10\% of its saturated value, leading to $E_{\rm{thr}}^{\rm{SWGO (\geq 65)}} \approx 1.75 \, \rm{TeV}$, $E_{\rm{thr}}^{\rm{SWGO (\geq 100)}} \approx 3.5 \, \rm{TeV}$, $E_{\rm{thr}}^{\rm{SWGO-A (\geq 30)}} \approx 550 \, \rm{GeV}$, and $E_{\rm{thr}}^{\rm{SWGO-A (\geq 50)}} \approx 900 \, \rm{GeV}$. This enables SWGO to achieve the high event statistics necessary for precise studies in the TeV--PeV energy range.

In addition, enhanced muon-counting capabilities will improve the determination of the primary CR composition. Generally, identifying the species of a CR primary using WCDs is challenging, as the longitudinal profile of the shower is not measured directly. However, the dual-layer tank design chosen for the inner array makes the larger top layer act as a calorimeter for the electromagnetic component of the shower while allowing the muons to reach the bottom layer, making it a crucial tool for identifying and, ideally, counting them. The relation between the total electromagnetic energy, $E_{\rm{EM}}$, and the total number of muons, $N_{\mu}$, of an air shower may be used as a strong separator between different CR species. See Section~\ref{sec:muon} for more details regarding SWGO's study of the muon lateral distributions. 

The possibility of separating the composition into four main groups, H-like, He-like, N-like, and Fe-like, is desirable. To that end, different methods for estimating $N_{\mu}$ and $E_{\rm{EM}}$ have been explored in the context of SWGO, with the most promising one being a template-based search \citep{Lang:2023oon}. Template-based reconstructions have already proven to be accurate for gamma-ray events detected with WCDs \citep{Joshi:2018qbo} and are one of the methods considered for core, energy, and direction reconstruction of gamma rays in SWGO. 
For CR studies, preliminary implementations of the method have shown to reach a resolution in the number of muons of the order of 10--20\% for He-like, N-like, and Fe-like, and of the order of $\sim$30\% for protons~\citep{Lang:2023oon}. 
While more realistic conditions, such as uncertainties in core reconstruction, still need to be accounted for, current estimates indicate that using a dual-layer tank significantly improves primary CR separation when compared to single-layer tanks, with optimistic estimates achieving a purity of nearly 90\% for protons and at least 95\% for heavier primaries (an improvement of $\sim$50--150\% with respect to single-layer estimates)~\citep{Lang:2023oon}. This can help SWGO measure the energy spectra for different mass groups, contributing to resolving existing tensions in CR composition measurements around the knee and providing crucial insights into the origin of PeV CRs.

\subsection{Anisotropy}
\label{sec:CR-Anisotropy}

Due to their electric charge, CRs are deflected by magnetic fields, resulting in strongly diffusive propagation within our Galaxy. This process masks out information about their original acceleration sites. Therefore, the distribution of arrival directions of CRs (on Earth) is highly isotropic. Nevertheless, ground-based observatories covering large areas have recently reached sufficiently large exposures to resolve small but significant anisotropic structures. These anisotropic structures are typically characterized using the angular power spectrum derived from a spherical harmonic decomposition of the arrival direction distribution. For the TeV--PeV range, the most prominent deviation from isotropy comes in the form of a significant, but small, dipole with an amplitude of the order $\sim$$10^{-3}$ \citep{Abbasi_2025}.

The study of large-scale anisotropies can lead to strong biases if only partial FoVs are considered. Ground-based instruments have access to only a fraction of the sky and complementary measurements from different experiments are desirable for achieving full-sky coverage. This was first done in the TeV region by HAWC and IceCube \citep{HAWC:2018wju}. However, combining data remains challenging due to differences in detection techniques as well as sensitivities to energy and composition across experiments. Any attempt to align the systematic uncertainties is further complicated by the small region of overlap ($\sim$$10^{\circ}$ for the HAWC and IceCube measurements), in which the direction reconstruction quality is the poorest for both experiments. SWGO will contribute significantly to studying the full-sky dipole anisotropy by providing large statistics and high-quality data, employing a detection mechanism similar to that of HAWC and a larger overlap with its FoV. For all considered primaries, even a simple plane-fit reconstruction for the arrival directions can help achieve angular resolutions <$0.5^{\circ}$, enough to explore mid-scale anisotropies, for which the size of structures is of the order of tens of degrees.

The amplitude and phase of the measured dipole component also present an interesting evolution with energy, as already observed by IceCube \citep{Abbasi_2025}. A ``swing'' in the dipole component direction is observed between $\sim$10~TeV and $\sim$1~PeV--- 
its phase in Right Ascension gradually changing from pointing $\sim$$150^{\circ}$ away from the GC for $E \lesssim 80$~TeV to pointing roughly close to the GC for $E \gtrsim 200$~TeV. Correspondingly, a drop in amplitude is seen between $\sim$25~TeV and $\sim$100~TeV, followed by a steep increase up to a few tens of PeV. Moreover, increasing multipole contributions become evident at higher energies \citep{Abbasi_2025}.

Although the origin of the observed dipolar anisotropy remains an open question, its energy-dependent evolution and its decomposition by mass group may hold the key to understanding its nature. If the anisotropy arises from propagation effects, one would expect this evolution (or ``swing'') to appear at different energies for different CR nuclei. Furthermore, magnetic field effects are modulated by particle rigidities, implying that a rigidity-dependent behavior (instead of an energy-dependent one) is foreseen. Therefore, a reasonable event-by-event species classification is desired. As described in Section~\ref{subsec:spec_and_comp_measurements_knee}, a template-based method has demonstrated effective separation of CR primaries into four mass groups. As a result, SWGO's enhanced muon-counting capabilities can contribute to the study of the composition-dependent anisotropy in the TeV--PeV energy range, offering new insights into the origin of the Galactic CR dipole.

In the literature, large-scale anisotropy (e.g., the dipole component) is usually associated with diffusive propagation of CRs through interstellar magnetic turbulence \citep{Pasquale_Blasi_2012, PhysRevLett.117.151103, Qiao_2023, Li_2024} and may, therefore, provide crucial insights into the properties of this turbulence \citep{giacinti_sigl_2012, giacinti_kirk_2017, Giacinti_2021, bian2024angularpowerspectrumtevpev}. However, the observed CR anisotropy exhibits a complex angular-scale structure, typically described using spherical harmonic functions. This mathematical description enables investigation into the origin of these observations, including non-diffusive propagation effects occurring within the mean scattering length, correlations among particles traversing the same turbulent magnetic fields, and the effects of turbulent convection \citep{Ahlers:2013ima, mertsch_ahlers_2015, ahlers_mertsch_2016, mertsch_2019, genolini_2021, Kuhlen_2022, zhang2024smallscale}. By offering an unbiased perspective on the distribution of power across various angular scales, this approach is essential for testing and distinguishing between various models that explain the observed medium- and small-scale anisotropies. SWGO will further enhance studies on the properties of interstellar turbulence by leveraging its detection redundancy and providing high-quality CR data, thereby advancing our understanding of CR propagation.

\subsection{Muon Puzzle and Hadronic Interaction Models}
\label{sec:muon}
The inference of the mass composition of air showers is deeply intertwined with our understanding of the hadronic interactions that govern their development. To accurately model the shower evolution, hadronic interaction models must describe the full phase space of multi-particle production. This relies on phenomenological parameterizations that are calibrated to available accelerator data and extrapolated to higher energies and kinematic regions that lie beyond current experimental reach.

Despite extensive efforts, state-of-the-art hadronic interaction models, tuned to data from the Large Hadron Collider, fail to accurately predict the muon content of air showers at UHEs \citep{PierreAuger:2014ucz,PierreAuger:2016nfk,PierreAuger:2024neu}. This discrepancy, commonly referred to as the Muon Puzzle, has been observed across multiple experiments, suggesting that the divergence between data and simulations may emerge as early as at PeV energies \citep{ArteagaVelazquez:2023fda}.  

Measurements of the relative fluctuations in the muon number from the Pierre Auger Observatory \citep{PierreAuger:2021qsd} align with mass-composition expectations derived from the analysis of the depth of the shower maximum in the longitudinal profile. Combined with the findings of \cite{Cazon:2018gww}, this evidence strongly suggests that the origin of the Muon Puzzle may lie in subtle deviations across numerous low-energy hadronic interactions, rather than in a substantial modification at the highest energies.
Additionally, muon distributions at low energies have been found to depend on the choice of both low- and high-energy hadronic interaction models \citep{Pastor-Gutierrez:2021lxs}. 

For all these reasons, SWGO must validate the modeling of shower physics and accurately measure the muon component of Extensive Air Showers (EASs) at ground level. In fact, with its capability to identify EAS muons --- necessary to perform composition studies and gamma/hadron separation --- and high-fill-factor of detector units, SWGO is particularly well-suited for this task in the energy range from a few hundred GeV to several tens of PeV. It is also worth noting that the transition between low- and high-energy hadronic interaction models occurs at $\mathcal{O}$(100~GeV), further enhancing the relevance of this energy regime.

In the following, we describe one possible strategy that SWGO can employ to discriminate between high-energy hadronic interaction models. 
As described in Section~\ref{subsec:spec_and_comp_measurements_knee}, the template-based method used for the energy reconstruction and composition separation relies on the muon lateral distribution functions (LDFs) derived from simulations and is, therefore, highly sensitive to variations in hadronic interaction models. However, the correlation between measurements from the bottom and top layers of the detector offers a nearly independent estimate of the number of muons in each individual tank. Since this method does not depend on the overall shower distribution, it is less sensitive to uncertainties in hadronic interaction models. Differences in the muon energy spectrum and the punch-through effect are expected across models, however, they represent second-order effects. 

Preliminary studies \citep{Kunwar:2022iav} have demonstrated the feasibility of using the bottom layer as a muon tagger, allowing for the identification of tanks where one or more muons contribute to the signal. 
This capability is important for gamma/hadron separation, as the expected number of muons in gamma-ray air showers is very low. In the case of CRs, however, the number of muons increases substantially with energy, reaching $10^{3}$--$10^{4}$ muons at the detector level for Fe nuclei at TeV energies. In this regime, precise muon counting, rather than mere tagging, becomes essential.  

\begin{figure}[ht]
    \centering
    \includegraphics[width=\linewidth]{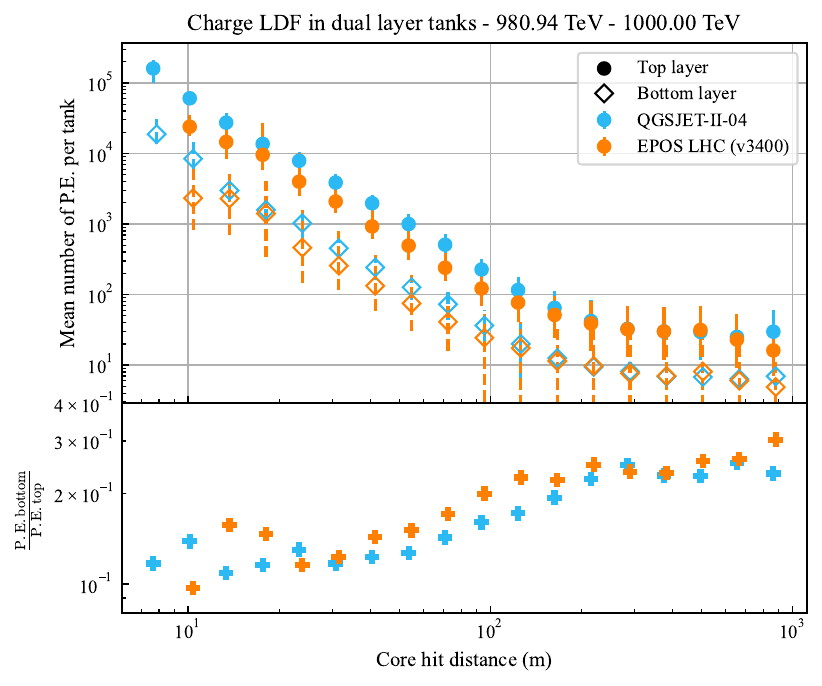}
    \caption{Lateral distribution functions of the integral charge in different tank layers as a function of the distance to the shower core in the shower plane. The mean of events and individual tanks is used for each distance bin. LDFs for the top and bottom layers are shown as closed and open markers, respectively. Two different high-energy hadronic interaction models are used: \texttt{QGSJET-II.04} (blue) and \texttt{EPOS-LHC} (orange). The \texttt{UrQMD-1.3.1} low-energy hadronic interaction model~\citep{urqmd1, urqmd2} is used in all the cases.}
    \label{fig:muonLDF}
\end{figure}

An initial assessment of SWGO’s ability to estimate the muon LDF from individual tank measurements, largely independent of hadronic interaction models, can be obtained by analyzing the integrated charge in each layer of every tank. Figure~\ref{fig:muonLDF} presents the median integrated charge from the top and bottom layers as a function of the distance to the shower core in the shower plane for several proton-induced events within an example energy range of $\sim$980~TeV up to 1,000~TeV, simulated using two different hadronic interaction models: \texttt{QGSJET-II.04}~\citep{Ostapchenko:2010vb} and \texttt{EPOS-LHC}~\citep{Pierog:2013ria}. The predicted LDFs for the measured charges exhibit notable differences between the two models, highlighting SWGO's potential to place strong constraints on hadronic interaction models and contribute to a more model-independent investigation of the muon LDF and the Muon Puzzle. It's worth noting that the ratio of the charges is a largely model-independent, first-order estimate of the muon content and already shows significant differences between different hadronic interaction models. More complex methods, e.g., machine learning or templates, could be even more sensitive to such differences.

The impact of uncertainties in energy reconstruction, arising from assumptions about hadronic interaction models in such a simplified analysis, remains to be evaluated. Nevertheless, more sophisticated and robust approaches to individual tank muon counting can be developed using machine-learning techniques. For instance, deep neural networks (DNNs) have demonstrated improvements in gamma-ray reconstruction for detectors like SWGO~\cite{Assis:2022vfk,SWGO:2023ndi,Conceicao:2024vos,Glombitza_wcd_2025}, suggesting their potential for enhancing muon detection as well.

\subsection{Heliosphere and Solar Physics}

\begin{figure}[ht]
    \centering
    \includegraphics[width=\linewidth, trim={0.4cm 0.8cm 0.1cm 0.1cm},clip]{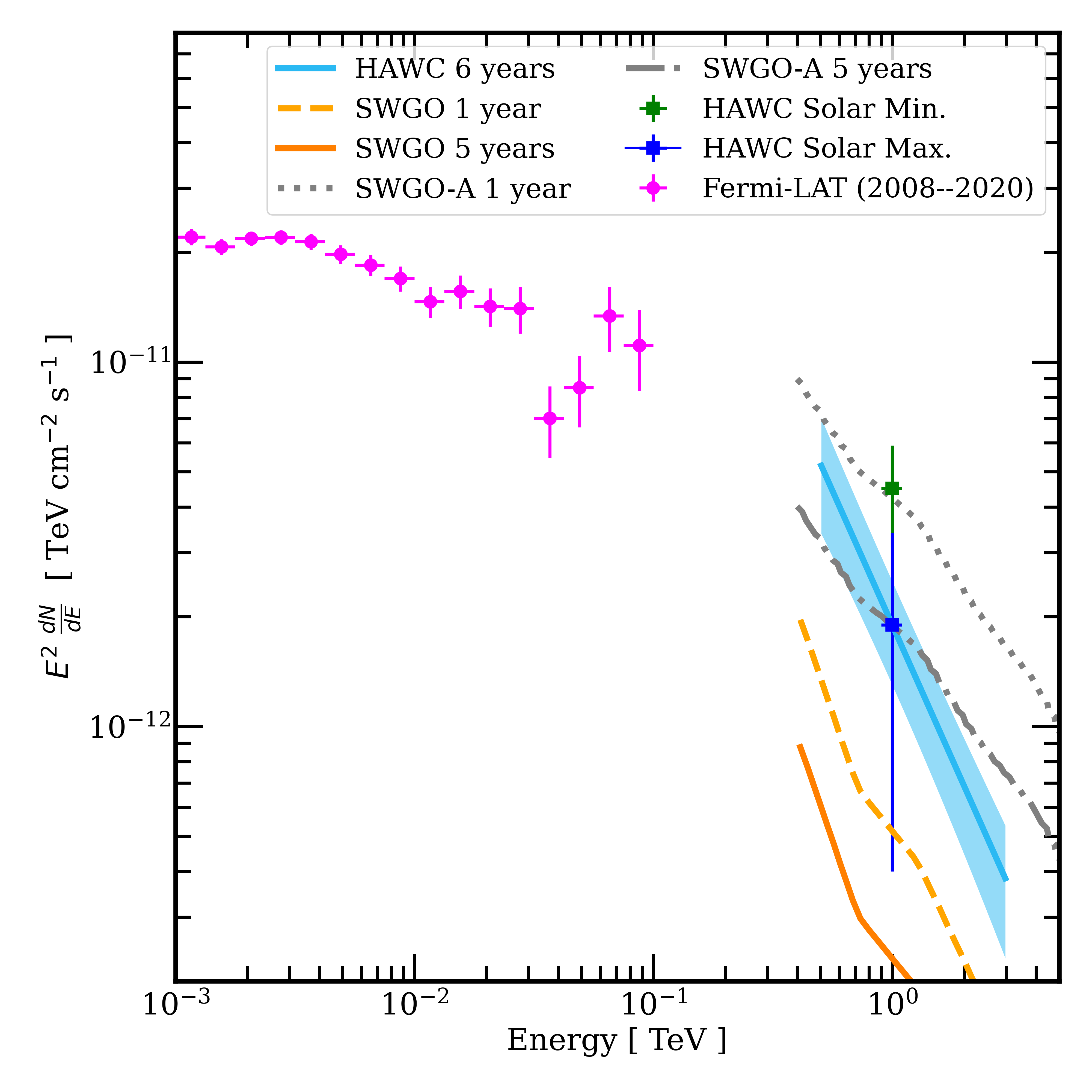}
    \caption{Gamma-ray emission from the solar disk at GeV--TeV energies during the solar cycles 24--25 with measurements from the \emph{Fermi}-LAT (magenta) and HAWC (green for solar minimum and blue for solar maximum). The time-integrated spectrum measured by HAWC over six years is shown in blue. The sensitivity curves for SWGO (orange lines) are also shown (dashed for one year of operation and solid for five years). The one-year and five-year sensitivity of SWGO-A is shown as the gray dotted and dashed-dotted lines, respectively.}
    \label{fig:sun}
\end{figure}

The Sun has emerged in recent years as a new source of interest in gamma-ray astronomy, with a puzzling VHE signal observed by the \emph{Fermi}-LAT and HAWC in the GeV--TeV range \citep{2011ApJ...734..116A,2023PhRvL.131e1201A}. Nominal models of CR interactions with the Sun’s atmosphere are unable to explain the hard spectrum and morphology of these solar gamma rays. An improved characterization of the systematics on the spectral measurements of the Sun in different activity states, including a potential resolution of the solar gamma-ray puzzle, would be a significant step forward in our understanding of local CR propagation in the Sun’s dynamic environment and its magnetic fields, as well as in searches for dark matter and new physics \citep{2021PhRvD.104b3024B}. The Sun, being a bright, moving source, can only be efficiently probed using an all-sky survey instrument that is capable of daytime operations, effectively ruling out observations from even the most sensitive IACTs. 
SWGO, with its improved sensitivity over HAWC, is perfectly suited for a high-statistics measurement of the gamma-ray spectrum of the Sun during future solar cycles and will enable the search for additional temporal features in the emission. Figure~\ref{fig:sun} shows the existing measurements of the Sun in gamma rays along with the projected sensitivity of SWGO.

%% file: Multimessenger.tex
\section[Multi-Messenger and Multi-Wavelength Program]{Multi-Messenger and Multi-Wavelength  Program} 

SWGO will provide the Southern Hemisphere's first wide-field, high-duty-cycle instrument for VHE--UHE time-domain astronomy, complementing the pointed observations of \change{CTAO-South. From the beginning, the SWGO Collaboration has been committed to close scientific coordination with CTAO, recognizing the very strong synergy and complementarity between these instruments}. With its wide FoV, high duty cycle, and continuous monitoring strategy, SWGO will serve as a cornerstone facility for MM/MWL astrophysics. It will offer crucial Southern sky coverage, complementing Northern facilities such as LHAASO, and will play a vital role in the global network of observatories probing the transient universe.

\subsection{\bf Time-Domain Capabilities and Alert Infrastructure}

SWGO's design and operational strategy will address several major challenges in time-domain and MM/MWL astrophysics in the following ways:

{\it High Duty Cycle:} The observatory is expected to operate with a duty cycle exceeding 95\%, enabling near-continuous observation of the sky. Unlike IACTs, which are limited by weather, daylight, and moonlight, SWGO will operate both day and night, regardless of atmospheric conditions. This makes it ideally suited for detecting unpredictable, short-lived, or recurrent astrophysical events.

{\it Large Instantaneous FoV:} With a FoV of approximately 2~steradians, SWGO will monitor a substantial portion of the sky at any given moment. This will increase the likelihood of detecting rare and serendipitous transients across a broad range of spatial and temporal scales. It also enhances SWGO’s potential to discover new classes of variable and transient gamma-ray sources.

{\it All-Sky Monitoring without Repointing:} The SWGO design allows it to continuously observe the entire overhead sky without the need for repointing or scheduling (as opposed to IACTs). This ensures the availability of pre-alert data for retrospective analysis (following external alerts), making SWGO highly effective for identifying not only precursor activity, but also significantly delayed high-energy counterparts.

{\it Real-Time Data Analysis:} SWGO will implement real-time pipelines for both externally triggered follow-ups and blind searches for transient events. These systems will facilitate rapid detection of gamma-ray counterparts to high-energy neutrinos, GWs, and  Fast Radio Bursts (FRBs), as well as unexpected phenomena such as magnetar flares, Galactic transients, and unidentified bursting sources. The ability to respond within seconds to minutes of an external alert will place SWGO at the forefront of rapid-response high-energy astrophysics.

SWGO’s effectiveness in time-domain and MM/MWL astrophysics will rely on how it shares data with the global astrophysics community. As a modern observatory, it must be fully embedded within the global MM network. With this in mind, SWGO will adopt a transparent and community-oriented data policy, designed to maximize accessibility and scientific return:

{\it Real-Time Alert Dissemination:}\label{sec:alertdissemination} Based on the real-time data analysis, the observatory will issue alerts containing key event parameters, such as time, localization (with uncertainty), false alarm rate (FAR), energy range, and flux estimates (or upper limits), using standardized formats like the International Virtual Observatory Alliance (IVOA) VoEvent schema. This will ensure a rapid interpretation of SWGO data and responses from other observatories and automated alert brokers. SWGO will engage with the major alert networks and platforms that constitute this infrastructure, including:

{\it Astronomer’s Telegrams \citep[ATels;][]{1998PASP..110..754R}:} ATels are human-curated reports on variable sources, such as AGN flares or X-ray binaries, which will be monitored by SWGO for VHE gamma-ray activity on timescales of days or longer. SWGO will contribute its own findings back to the community via ATels, ensuring rapid dissemination of significant detections.

{\it General Coordinates Network\footnote{https://gcn.nasa.gov/} (GCN):} SWGO will respond to real-time GCN alerts from instruments detecting GRBs, high-energy neutrinos (e.g., IceCube), and GW events (e.g., LIGO-Virgo-KAGRA). Its ability to instantaneously monitor relevant sky regions will enable prompt detections of potential counterparts (including refined localizations of poorly localized events) or the setting of upper limits. 

{\it Astrophysical Multimessenger Observatory Network \citep[AMON;][]{2013APh....45...56S}:} Building on experience with HAWC, SWGO will contribute real-time data streams of sub-threshold event clusters and selected high-energy photon candidates. This input will help identify statistically significant coincidences across different wavelengths and messengers, facilitating early detection of flaring states or entirely new source classes.

{\it Transient Name Server\footnote{https://www.wis-tns.org} (TNS):} SWGO will monitor reports from the TNS to follow up on optical transients such as supernovae, TDEs, and FRBs, particularly those associated with non-thermal or rebrightening activity.

{\it Vera Rubin/LSST Brokers\footnote{https://www.lsst.org/scientists/alert-brokers}:} With the Rubin Observatory expected to detect millions of transients per night, real-time classification and follow-up will be critical. SWGO will be the only ground-based instrument that will cover the same sky simultaneously with LSST and will therefore be ideally placed to search for gamma-ray counterparts to LSST transients over a wise range of timescales from minutes to weeks and months. As with AMON (see above), SWGO will employ dedicated pipelines to correlate its sub-threshold stream of gamma-ray events with LSST-detected transients.

{\it Astro-COincidence LIBrary for Real-time Inquiry \citep[Astro-COLIBRI;][]{2021ApJS..256....5R}:} This centralized coordination platform for transient astrophysical events, developed within the MM community, will serve as a primary interface for SWGO's burst advocates and shift personnel. It will allow real-time filtering, visualization, and contextualization of alerts across all relevant messengers and wavelengths.

{\it Archival Data and Source Catalogs:} On longer timescales (typically within one year), SWGO will release comprehensive datasets from offline analyses, including transient light curves and catalogs of steady and variable sources. These products will be made available through FAIR-compliant platforms and integrated into established databases such as VizieR, HEASARC, and TeVCat~\citep{tevcat}, to ensure interoperability and long-term discoverability.

\subsection{\bf Prospects for Multi-Messenger Synergies}
\label{sec:MMsynergies}

The alert-driven infrastructure described above already touches several of these messengers directly, for instance, through AMON's cross-messenger coincidence searches. Beyond this real-time role, SWGO's longer-baseline and archival data will also directly inform three messenger-specific science programs: high-energy neutrinos, gravitational waves, and cosmic rays.

{\it High-Energy Neutrinos:} The combination of gamma-rays with high-energy neutrinos is extremely powerful for the identification of high-energy CR accelerators, given the likely association of neutrino signals with hadron acceleration and the larger statistics and precision possible with gamma ray observations. Extensive experimental attempts in recent decades have culminated in the finding of a diffuse neutrino flux by the IceCube Collaboration~\citep{2013PhRvL.111b1103A, IceCube_2021PhRvD.104b2002A, 2022ApJ...928...50A}, as well as the first neutrino sources, including the flaring blazar TXS~0506+056~\citep{Aartsen+18nublazarmma}, the Seyfert galaxy NGC~1068~\citep{IceCube+22_NGC1068}, and the Galactic Plane~\citep{IceCube:2023ame}. Despite the emergence of the first source classes, the origin of the IceCube diffuse flux remains a mystery.

Located in the Southern Hemisphere, SWGO is uniquely placed to follow up IceCube's Southern-sky tracks and Earth-skimming events, as well as KM3NeT's Earth-crossing events, and will provide a map of the TeV--PeV photon sky, which will be invaluable for understanding the Galactic neutrino emission in the same energy range.

A concrete case for this complementarity is provided by the Galactic diffuse emission models discussed in Section~\ref{sec:diffuse}. As shown there, the ``CRINGE'' and ``varmin-rescaled'' interstellar emission models, which encode markedly different assumptions about CR transport, are nearly degenerate in the broad sky regions accessible to current instruments (Figure~\ref{fig:diffuse_lhaaso_mask}), yet SWGO's angular and energy resolution will allow this degeneracy to be broken at the smaller scales of individual molecular clouds in the inner Galaxy (Figure~\ref{fig:iem_detectability}). Since the same CR transport assumptions also set the predicted Galactic diffuse neutrino flux \citep{2024JHEAp..43..140F}, SWGO's gamma-ray-based discrimination between these models will directly narrow the systematic uncertainty currently affecting interpretation of the IceCube Galactic Plane signal. SWGO will also continuously monitor known and flaring blazars (Section~\ref{sec:AGN}) and transient phenomena such as TDEs (Section~\ref{sec:exploratory}), providing candidate electromagnetic counterparts for IceCube and KM3NeT alerts. Possible contributions to starburst-galaxy and GRB neutrino searches remain to be quantified in future work.

{\it Gravitational Wave Sources:} Observations by the LIGO, Virgo, and KAGRA interferometers have established the field of GW astronomy, marked by several key milestones: the first direct GW detection (GW150914), the first binary neutron star (BNS) merger with an electromagnetic counterpart (GW170817), an intermediate-mass BH merger (GW190521), an unusual mass-ratio system (GW190814), and the first neutron star-black hole (NS-BH) merger (GW200115).

Events with electromagnetic counterparts, such as BNS and NS-BH mergers, will remain confined to the local universe ($z \lesssim 0.1$), a redshift range that SWGO can probe despite EBL absorption. Localization regions for GW detections are expected to remain large (110--180~deg$^2$), and SWGO's wide FoV and high duty cycle make it one of the few gamma-ray instruments capable of efficiently covering them without repointing, while its real-time alert system could enable prompt follow-up by facilities like CTAO. A quantitative forecast of SWGO's expected coverage of GW localization regions and the resulting joint detection rate is left to future work.

\textit{Combination with Cosmic-Ray Measurements:} Many SWGO observations relate directly to cosmic particle acceleration, as discussed elsewhere in this document. Direct connections to the CR observatories include SWGO's capability to observe CR anisotropy, discussed in Section~\ref{sec:CR-Anisotropy}, but also complementarity to the UHE CR observatories.

The higher end of the SWGO energy range is limited to the GZK horizon by EBL absorption. Within this range, SWGO's gamma-ray sky maps can directly support searches for ultra-high-energy cosmic-ray (UHECR) sources, complementing the charged-particle data from the Pierre Auger Observatory and the Telescope Array and the neutrino data from IceCube and ANTARES, which have so far found no significant spatial correlation between UHECRs and neutrinos \citep{Albert_2022}. Gamma-ray counterparts identified by SWGO would provide an independent and complementary handle on the accelerators responsible for the locally measured UHECRs.

The Southern Hemisphere location makes SWGO ideally suited to follow up the most significant deviation from isotropy in the UHECR sky: an excess likely associated with the very nearby active galaxy Cen~A. {\bf A SWGO detection of gamma-ray emission correlated with this direction, combined with the existing Auger/Telescope Array excess and ongoing neutrino searches, would offer a three-messenger view of a UHECR accelerator candidate. SWGO's combination of Southern-sky access, sensitivity to many-degree-scale emission, and reach into the PeV range is what makes this gamma-ray contribution possible--- no existing or planned facility combines all three.}

The combination of SWGO mass-resolved anisotropy measurements with a wide range of CR observatories in the TeV--PeV range will be extremely powerful as a probe of local CR transport as well as CR origin.

%% file: Outreach.tex
\section{Outreach \& Communications}


In addition to the strong scientific objectives that constitute the pillars of the SWGO project, the Collaboration has devoted particular attention to the development of a comprehensive outreach and education program. Through our official website \href{http://www.swgo.org}{www.swgo.org} and \href{https://www.youtube.com/@CollaborationSWGO}{YouTube channel},
the Collaboration actively disseminates updates and information about ongoing activities in a way that is accessible to the general public. 

Following the site selection in Pampa La Bola, Chile last year, efforts have been substantially intensified in the host country of Chile to cultivate a strong relationship between the Observatory and the local Atacamenean ethnic communities, with particular focus on engaging educators and students. In this way, the project seeks to not only broaden public understanding of high-energy astrophysics but also nurture a sense of common ownership and pride in scientific discovery that is shared with the Atacamenean worldview and patrimony. This engagement is a vital component of SWGO’s mission as it brings together cutting-edge research and the educational and cultural heritage of the region and country. 

This community engagement extends to the region’s vibrant amateur astronomy network, upon which SWGO will draw for long-term photometric and spectroscopic monitoring of gamma-ray sources. These collaborations, coordinated through platforms such as Astro-COLIBRI (see Section~\ref{sec:alertdissemination}), will expand the observational reach of SWGO while offering valuable training and outreach opportunities. Through partnerships with initiatives such as the United Nations’ Open Universe\footnote{https://openuniverse.cbpf.br} program, SWGO will also contribute to capacity building in data science and astrophysics across Latin America and the Global South, ensuring that its impact is as inclusive as it is scientifically significant.

Our local outreach plans, which will be developed alongside Atacamenean community members, envision a range of educational initiatives that target students, teachers, and the broader public. We aim to collaborate with teachers in the region to develop physics and astronomy concepts that can help inspire a next generation of scientists, building bridges of common growth and understanding in the framework of the Atacamenean worldview. Members of the collaboration have initiated preliminary conversations with high school physics teachers in San Pedro de Atacama and Toconao to explore ways of introducing astroparticle physics topics and the SWGO project. Recently, with support from the AAP, local teachers attended a summer school at Universidad Metropolitana de Ciencias de la Educación (UMCE) focused on the Sun–Earth–Moon system and its role in the Chilean elementary school curriculum. 

An additional goal of the project is to help implement a curriculum-linked science program for elementary and high school students in local towns, providing engaging hands-on experiences that make scientific concepts accessible and exciting. Keeping in line with our commitment to foster scientific curiosity in young people, we also plan to offer technical and scientific training, aiming to ignite interest in pursuing Science, Technology, Engineering, and Mathematics (STEM) careers, particularly in fields such as physics and engineering. Work is underway to translate key content from the SWGO website into Spanish, with the aim of improving accessibility for audiences in the host country and abroad, as well as to support the development of new educational materials. For the broader community, our plans include the creation of the SWGO Scientific Literacy Program, which will offer an open-access online course introducing key scientific concepts such as the scientific method, particle physics, and gamma rays. These efforts will demystify complex topics, address common misconceptions, and improve public understanding of the universe and the science behind it. 

Finally, SWGO is exploring the potential creation of a Community Learning Center, which would serve as an immersive educational hub for scientific tourism, possibly in collaboration with other observatories in the area. This center would offer interactive exhibits and activities, such as hands-on experiments and live demonstrations, to attract visitors eager to learn about the groundbreaking research at the observatory. This dynamic environment would bring SWGO collaborators together to share their work and its impact through presentations, videos, and educational materials. It would also serve as a shared forum where this information and the Atacamenean worldview can be discussed side by side, creating an opportunity to explore and connect both perspectives. While still under consideration, the development of this center represents SWGO’s commitment to increasing community engagement and promoting the importance of science in daily life.

%% file: Summary.tex
\section{Summary} 

The SWGO project is now moving towards construction and will become the second major facility for ground-based gamma-ray astronomy in the Southern Hemisphere, besides the CTAO Southern Array. Here we have shown the broad scientific impact to be expected from SWGO. 

Its initial phase, SWGO-A, will already address key science questions through observations of the Galactic Center, bright Galactic sources including young remnants such as RX\,J1713.7$-$3946 and SN~1006, and massive stellar clusters such as Westerlund~1. Source detections by SWGO-A within its first year of operation are expected to exceed the size of 3HWC and 1LHAASO catalogs. 
It can also provide particle physics constraints comparable to those from HAWC and LHAASO, but from a complementary region of the sky.

Highlights of the science case for SWGO include tightly constraining the thermal relic WIMP paradigm for dark matter, an expected rate of GRB detection of 0.7 per year, and probing Galactic source populations with PeV reach. In all of these cases, the combination of SWGO and CTAO data can greatly enhance the scientific capabilities beyond what can be reached by either observatory alone. As a large array with unprecedented muon tagging capability, SWGO will have not just excellent background rejection power, but also the capability to make the first mass-resolved CR anisotropy measurements in this energy range and help to solve the {\it muon puzzle} in UHE air showers. 

As we finalize the design of SWGO and prepare for construction, we are seeking partners from across the MM/MWL community to further develop the SWGO science case and develop plans for operations.

%% file: acknowledgements.tex

The SWGO Collaboration gratefully acknowledges the support from a number of grants, agencies, and organizations that can be found listed here:
\\ \url{https://www.swgo.org/SWGOWiki/doku.php?id=acknowledgements}.
In addition we thank the two anonymous referees for the their very extensive and extremely useful comments.